\documentclass[11pt,a4paper]{article}
\pdfoutput=1

\usepackage{jheppub}
\usepackage{amsmath,amsfonts,amssymb,mathrsfs,graphicx,bbm}

\DeclareMathAlphabet{\mathpzc}{OT1}{pzc}{m}{it}

\def\V{{\mbox v}}
\def\W{{ \mbox w}}
\def\U{{\mbox u}}
\def\k{{\kappa }}
\def\z{{ z  }}
\newcommand{\bea}{\begin{eqnarray}}
\newcommand{\eea}{\end{eqnarray}}

\newcommand{\alg}[1]{\mathfrak{#1}}
\newcommand{\su}{\alg{su}}

\newcommand{\sls}{\alg{sl}}

\newcommand{\psu}{\alg{psu}}

\newcommand{\el}{\nonumber\\}

\def\ads{{\rm AdS}_5\times {\rm S}^5}
\def\AdS{{\rm AdS}_5\times {\rm S}^5}

\newcommand{\cs}{{\rm cs}}
\def\cso{{\rm cs}_0}
\def\dn{{\rm dn}}
\def\dno{{\rm dn}_0}

\def\sno{{\rm sn}_0}

\def\cno{{\rm cn}_0}
\def\am{{\rm am}}
\def\am0{{\rm am}_0}

\def\mI{\mathbbm{1}}

\def\a{\mathsf{a}}
\def\ad{\mathsf{a}^\dag}
\title{The Quantum Deformed Mirror TBA {\fontfamily{ptm}\selectfont I}}

\author[a,1]{Gleb Arutyunov,}
\note{Correspondent fellow at Steklov
Mathematical Institute, Moscow.}
\author[b]{Marius de Leeuw,}
\author[a]{and Stijn J. van Tongeren}
\affiliation[a]{Institute for Theoretical
Physics and Spinoza Institute, Utrecht University, \\ Leuvenlaan \ 4, 3584 CE
Utrecht, The Netherlands}
\affiliation[b]{ETH Z\"urich, Institut f\"ur Theoretische Physik, \\Wolfgang-Pauli-Str.\ 27, CH-8093 Zurich, Switzerland}

\emailAdd{g.e.arutyunov@uu.nl}
\emailAdd{deleeuwm@phys.ethz.ch}
\emailAdd{s.j.vantongeren@uu.nl}

\abstract{We derive the ground state thermodynamic Bethe ansatz equations for the quantum deformation of the $\AdS$ mirror model, taking the deformation parameter to be a root of unity. By virtue of the deformation, the resulting equations show an interesting  structure between a finite number of Y-functions.}

\begin{document}

\begin{flushright}\small{ITP-UU-12/29\\SPIN-12/27}\end{flushright}

\maketitle

\section{Introduction}

Integrability has and continues to play an important role in the context of the AdS/CFT correspondence \cite{Maldacena}, where it might provide us with the first opportunity ever to exactly solve an interacting field theory at the quantum level. In particular, major achievements have been made in the spectral problem of $\mathcal{N}=4$ SYM in the planar limit \cite{Beisert:2010jr} by its equivalence to a two dimensional integrable quantum field theory; the world-sheet theory of the $\AdS$ superstring in the light-cone gauge \cite{Arutyunov:2009ga}. These developments conceptually culminated in the formulation of the thermodynamic Bethe ansatz equations \cite{AF09b,BFT,GKKV09} describing the spectrum of the world-sheet theory at finite size and finite coupling through the thermodynamics of an accompanying mirror model \cite{Zamolodchikov90,AF07}.

\smallskip

Ultimately the structure of these thermodynamic Bethe ansatz (TBA) equations for the ground state stems from that of  a $\psu(2|2)^2$ invariant
S-matrix \cite{Beisert:2005tm}. Interestingly the centrally extended Lie algebra $\psu(2|2)$, more precisely its universal enveloping algebra,  admits a natural quantum deformation in the sense of quantum groups \cite{Beisert:2008tw,Beisert:2011wq}.
%({\color{red} Marius, could you please extend the references here as you know them}).
We will loosely refer to the quantum deformed algebra $U_q(\psu(2|2))$ simply as $\psu_q(2|2)$.
This algebraic structure is the starting point
to construct a $\psu_q(2|2)^2$ invariant S-matrix, giving a quantum deformation of the $\AdS$ world-sheet S-matrix. The deformation parameter $q$ can in principle be any complex number, but is typically taken to be real or a phase in concrete applications. In the case where $q$ is a phase, more particularly the $k$th principal even root of unity\footnote{It was argued in \cite{Hoare:2011nd} that the S-matrix theory is most naturally defined at these points.} $q=e^{i\frac{\pi}{k}}$, the $\psu_q(2|2)^2$ invariant S-matrix \cite{Hoare:2011wr,Hoare:2012fc} interpolates between the S-matrix of the original string and mirror theory at $q=1$, and in the limit of infinite coupling \cite{Beisert:2010kk} the S-matrix of a relativistic semi-symmetric space sine-Gordon theory \cite{Hoare:2011fj,Hoare:2011nd}. Apart from interpolating between two theories we can anticipate that the model has very interesting intrinsic properties, particularly when the deformation parameter is taken to be a root of unity. Indeed the prototypical example of a $q$-deformed model, the XXZ spin chain, shows a very interesting structure in its TBA equations when $q$ is a root of unity \cite{Takahashi:1972,Takahashi:book}; there are only finitely many Y-functions, some coupling in a non-standard fashion. With this in mind, in the present paper we will derive the TBA equations for the quantum deformed theory at $q=e^{i\frac{\pi}{k}}$.

\smallskip

In the undeformed case of the $\ads$ superstring, the derivation\footnote{For recent reviews in the present context see for example \cite{Kuniba2,Bajnok:2010ke}.}  of the ground state TBA equations \cite{AF09b,BFT,GKKV09}  is based on the  string hypothesis  \cite{AF09a}.  These equations imply the corresponding Y-system \cite{GKV09} with rather intricate analytic properties studied in
\cite{Cavaglia:2010nm,Balog:2011nm}. Excited string states can be described by certain modifications of the ground state TBA equations \cite{KP,DT96,BLZe,Arutyunov:2011uz}, which has been explicitly done for certain string states with real \cite{GKKV09,AFS09,BH10b,Sfondrini:2011rr} and complex momenta \cite{Arutyunov:2011mk, Arutyunov:2012tx}. A particularly nice test of this approach was provided by the TBA equations for the Konishi operator, which were shown to agree numerically \cite{AFS10} and analytically \cite{BH10a} with L\"uscher's perturbative treatment \cite{BJ08,BJ09,LRV09,Janik:2010kd} and with the dual field theory at four loops by explicit computation \cite{Sieg,Velizhanin:2008jd}, and recently at five loops through an approach based on four-point correlation functions of BPS operators \cite{Eden:2012fe}. Furthermore, the TBA approach has been successfully applied to the computation of the cusp anomalous dimension in the context of Wilson loops \cite{Correa:2012hh,Drukker:2012de}, and there have been interesting developments   in describing the spectral problem through a finite set of non-linear integral equations \cite{Gromov:2011cx,Balog:2012zt}, see also  \cite{Suzuki:2011dj}, complementary to the TBA equations. These last two topics came together nicely very recently in \cite{Gromov:2012eu}.

\smallskip

Deformations in the form of twisted boundary conditions have been previously considered in the AdS/CFT TBA setting \cite{arXiv:1009.4118,Gromov:2010dy,Beccaria:2011qd,deLeeuw:2011rw,arXiv:1108.4914,deLeeuw:2012hp}. However, since twisted boundary conditions do not affect the mirror Bethe equations, the TBA equations are basically identical. Here on the other hand, the deformation affects the theory at the fundamental level of the dispersion relation already, and hence we have to go through the entire procedure outlined just above to derive the TBA equations for the ground state of the deformed model at roots of unity. To go beyond the ground state we need to have a candidate asymptotic solution for excited states, and we will address this question in an upcoming publication \cite{qdeftbapart2}.

\smallskip

The $\psu_q(2|2)^2$ invariant S-matrix which lies at the basis of our derivation of course satisfies all usual requirements of integrability. However, its physical unitarity has not been fully investigated before and as such remained an open question. We have found that this S-matrix (as well as the $\psu_q(2|2)$ invariant $R$-matrix of course) is not physically unitary but rather physically pseudo-unitary\footnote{The notion of pseudo-unitarity is defined in appendix \ref{app:matrixSmatrix}, see eqn. \eqref{eq:pseudounitarity}. For the notion of pseudo-Hermitian quantum mechanics, see e.g.  \cite{Mostafazadeh:2001A}.}, in particular in a local fashion\footnote{In other words, the Hermitian automorphism involved in the pseudo-unitarity factors over the one particle states, \emph{cf.} eqs. \eqref{eq:pseudounitarity} and \eqref{eq:pseudounitaritylocal}.}. Now in general pseudo-unitary (pseudo-Hermitian) models come in two classes; the Hamiltonian has a self-conjugate or a real spectrum. In the latter case the model is quasi-unitary (quasi-Hermitian) \cite{Mostafazadeh:2001B}. Our uniformized two body S-matrix generically satisfies generalized pseudo-unitary for complex arguments, and has a unitary spectrum for both the string and mirror theory. This quasi-unitary structure does not appear to be compatible with locality however, and as such does not necessarily extend to the many-body scattering theory. In fact, the many-body S-matrix makes a clear distinction between the string and mirror theory; the string theory many body S-matrix has non-unitary eigenvalues and is therefore only pseudo-unitary, while the mirror theory many body S-matrix appears to remain quasi-unitary. This difference means that we can expect the mirror theory to have a stronger sense of reality than the string theory, and in fact this is what we observe as we will discuss soon; these properties translate to the string hypothesis and TBA equations.

\smallskip

In general the question of a $q$-deformed TBA at roots of unity is notoriously difficult to answer. It might be surprising to learn that the TBA for a general $\su_q(N)$ spin chain is in fact not known. One of the difficulties is related to the fact that the associated local Hamiltonian is non-Hermitian\footnote{By a quick investigation they (naturally) appear to be pseudo-Hermitian, but not quasi-Hermitian, perfectly in line with our discussion.} for $N>2$, highlighting the special status of the XXZ model. For $\su_q(3)$ work has been done on the TBA for complex Toda theories \cite{Saleur:2000bq}, involving interesting but rather unusual non-unitary string complexes; still the general story remains unknown to the authors' current knowledge. Nevertheless, likely due to the quasi-unitarity of the mirror theory, we appear to be able to correctly account for the thermodynamics of the quantum deformed Hubbard model\footnote{The $\psu_q(2|2)$ invariant R-matrix of \cite{Beisert:2008tw} gives models which are closely related to the Alcaraz-Bariev model \cite{Alcaraz:1999}, however the detailed relation is not yet fully understood in all cases. The Alcaraz-Bariev model can be viewed as a quantum deformed version of the Hubbard model and is sometimes referred to as such. We loosely refer to our $\psu_q(2|2)$ invariant model as the quantum deformed Hubbard model here.} with a `real' string hypothesis, giving an elegant structure highly analogous to the XXZ case. Thus it appears that the \emph{mirror} $\su_q(2|2)$ theory is the next simplest case beyond the XXZ model. The coupling between the two copies of the deformed Hubbard model merges with this structure very neatly.

\smallskip

As we will show below, the spectrum of excitations of the deformed mirror model in the thermodynamic limit is severely constrained. This is analogous to what happens in the XXZ spin chain at roots of unity, though more involved. As a result, we obtain a finite set of $5k$ TBA equations\footnote{$3k$ for left-right symmetric states.} which couple in an intricate way.

\begin{figure}[h]
\begin{center}
\includegraphics[width=0.4\textwidth]{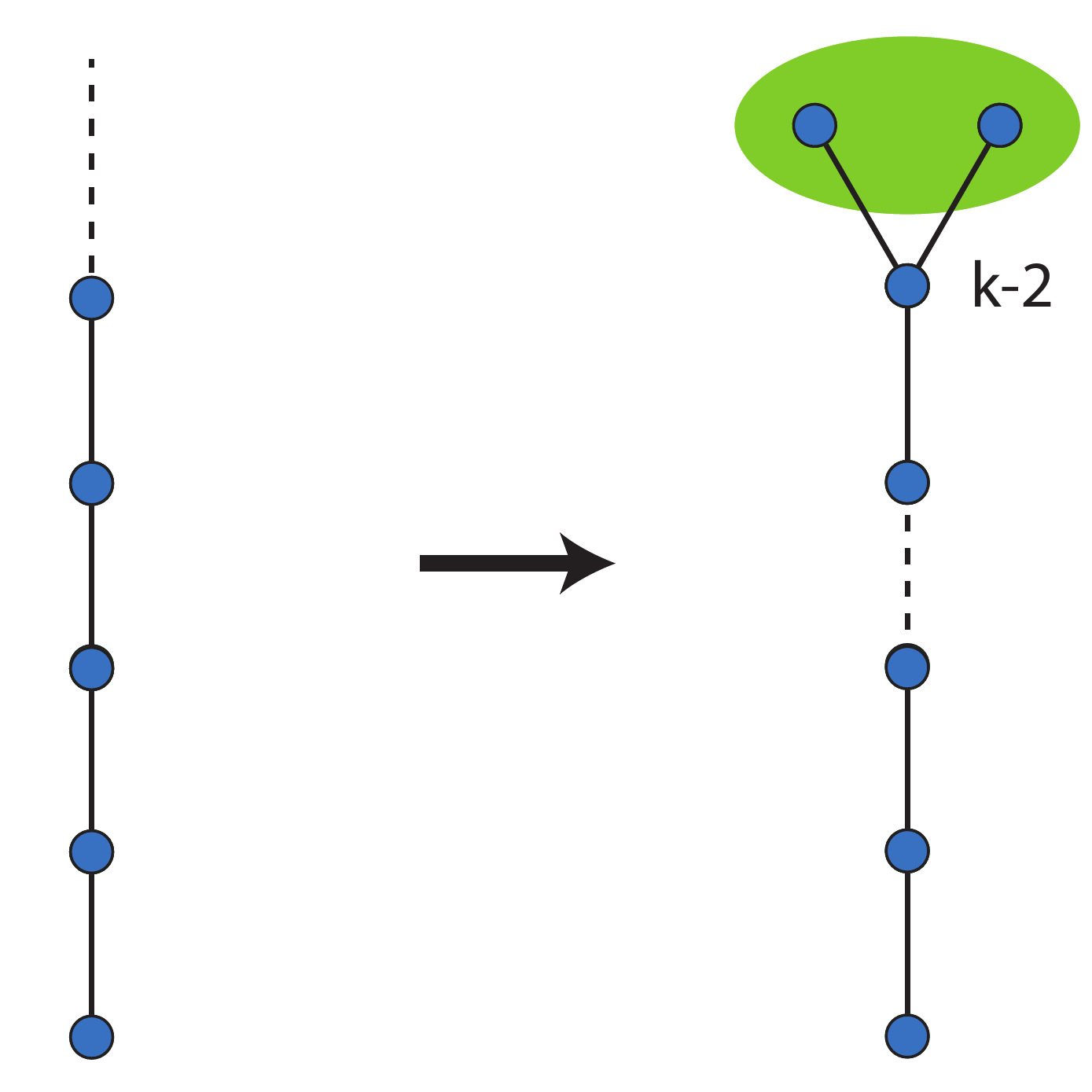}
\caption{The TBA structure of the XXX model and the XXZ model at roots of unity. The XXX model has a standard nearest neighbour coupling between infinitely many Y-functions (the vertices in the diagram), while the XXZ model has only a finite number of them. In the XXZ model the last two Y-functions are related directly algebraically (represented by the lime-green ellipse) and only couple back to the next to last regular Y-function.}
\label{fig:XXXtoXXZ}
\end{center}
\end{figure}

The XXZ model with $k=2$ is equivalent to free fermions, and also in our model the point $k=2$ is degenerate\footnote{For instance, on top of the obvious degeneration of the Bethe-Yang equations, the shape of the physical region on the torus degenerates dramatically.}, therefore we will consider $k>2$ in the present paper\footnote{There is no principal objection to treating the case where $q$ is a generic phase along the lines of the XXZ model \cite{Takahashi:1972,Takahashi:book}, but already there the technicalities become somewhat involved which would lead to a truly undesirable level of technical details here.}. With $q=e^{i\frac{\pi}{k}}$ and $k>2$, the string hypothesis for the XXZ spin chain allows only $k$ different types of string solutions, and the standard TBA equations are cut at level $k-2$ while adding an end structure between two algebraically related Y-functions and the last regular one \cite{Takahashi:1972,Takahashi:book}, see figure \ref{fig:XXXtoXXZ}. Of course the algebraic relation can be used to eliminate one of the last two Y-functions in exchange for a nonstandard coupling; this pictorial representation is chosen to emphasize their origin as two separate Y-functions.

\begin{figure}[h]
\begin{center}
\includegraphics[width=0.95\textwidth]{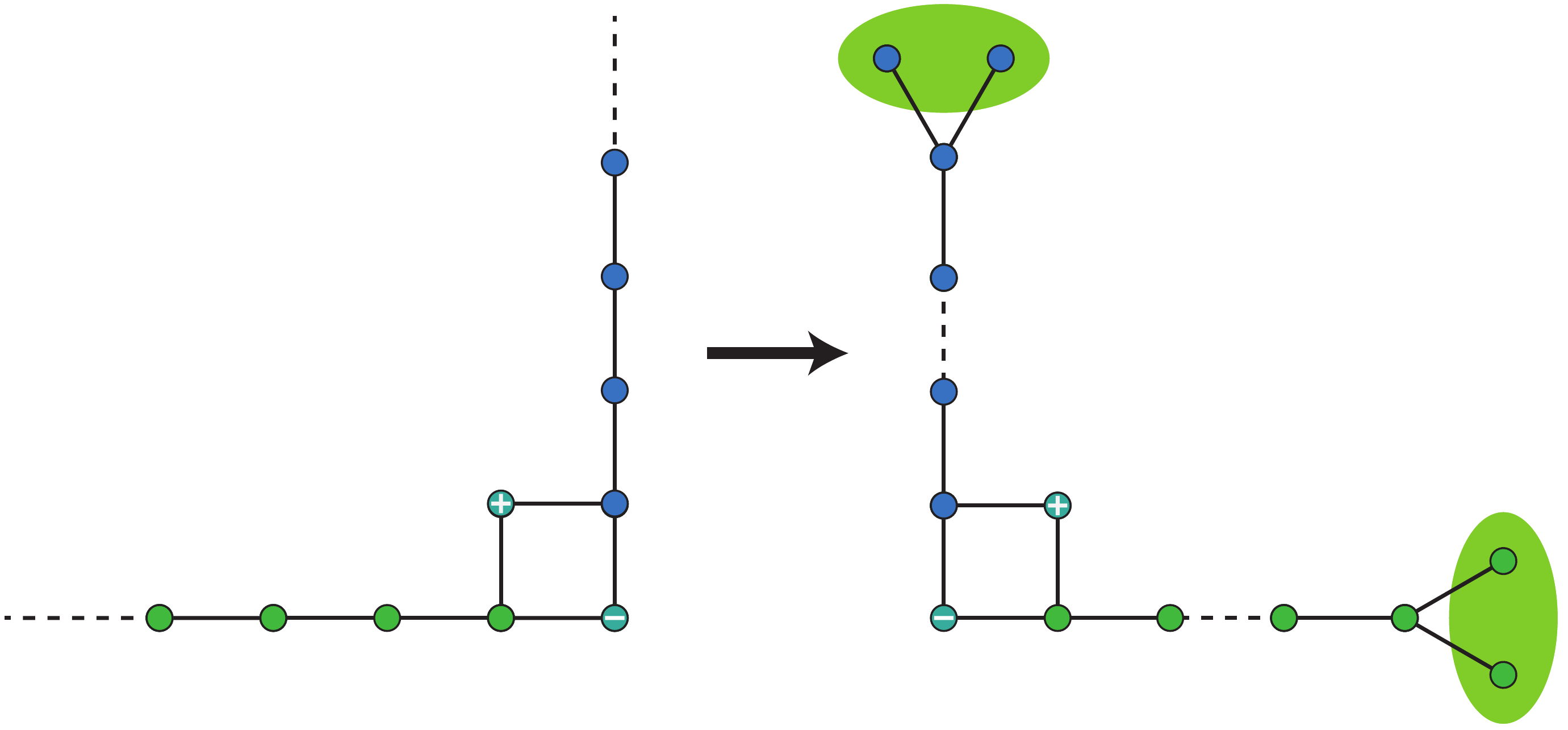}
\caption{The TBA structure of the Hubbard model (left) and the deformed Hubbard model at roots of unity (right). The green, teal (with $\pm$) and blue dots indicate what we call $Y_{M|w}$, $Y_{\pm}$, and $Y_{M|vw}$ functions respectively; their coupling is nearest neighbour apart from the indicated coupling near the corner. The modification of this structure in the deformed case is analogous to the XXZ case on each of the $\su_q(2)$ wings. (The reflection of the diagram is for purely aesthetic purposes.)}
\label{fig:HubbtoDefHubb}
\end{center}
\end{figure}

For our model we will firstly need to identify the TBA structure of the $q$-deformed Hubbard subsystem. The undeformed Hubbard TBA structure is that of two $\mathfrak{su}(2)$ wings coupled via two extra Y-functions. Quite naturally, the deformation results in an XXZ-like modification of each of these wings independently, giving the structure illustrated in figure \ref{fig:HubbtoDefHubb} at roots of unity\footnote{The TBA diagram for the generically deformed Hubbard model would have infinite extent like the undeformed case.}. This means that for the deformed Hubbard model at roots of unity we have $2k+2$ Y-functions with two algebraic relations between them, resulting in $2k$ independent equations.

\begin{figure}[h]
\begin{center}
\includegraphics[width=0.95\textwidth]{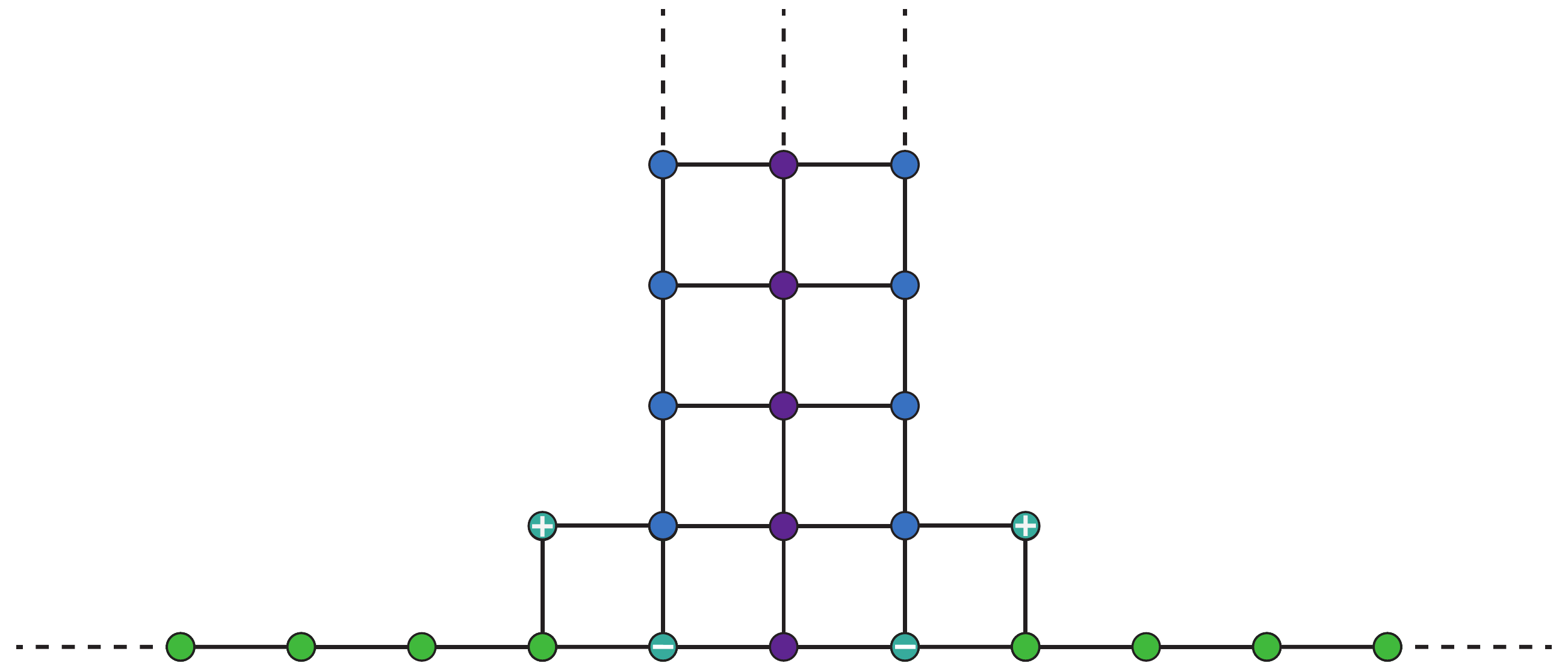}
\caption{The structure of the $\AdS$ TBA. Two copies of the Hubbard model are connected via the momentum carrying nodes (purple) of the mirror theory, corresponding to $Y_Q$-functions in our notation. Note that $Y_1$ couples to $Y_-$, but no $Y_Q$ couples to $Y_+$ in a local fashion.}
\label{fig:NormalYsys}
\end{center}
\end{figure}

Coming to the full $\AdS$ model then, in the undeformed case the Y-system \cite{GKV09} is given by coupling the Y-systems of two Hubbard models through physical mirror excitations and implies the TBA structure indicated in figure \ref{fig:NormalYsys}.

\smallskip

\begin{figure}[h]
\begin{center}
\includegraphics[width=0.95\textwidth]{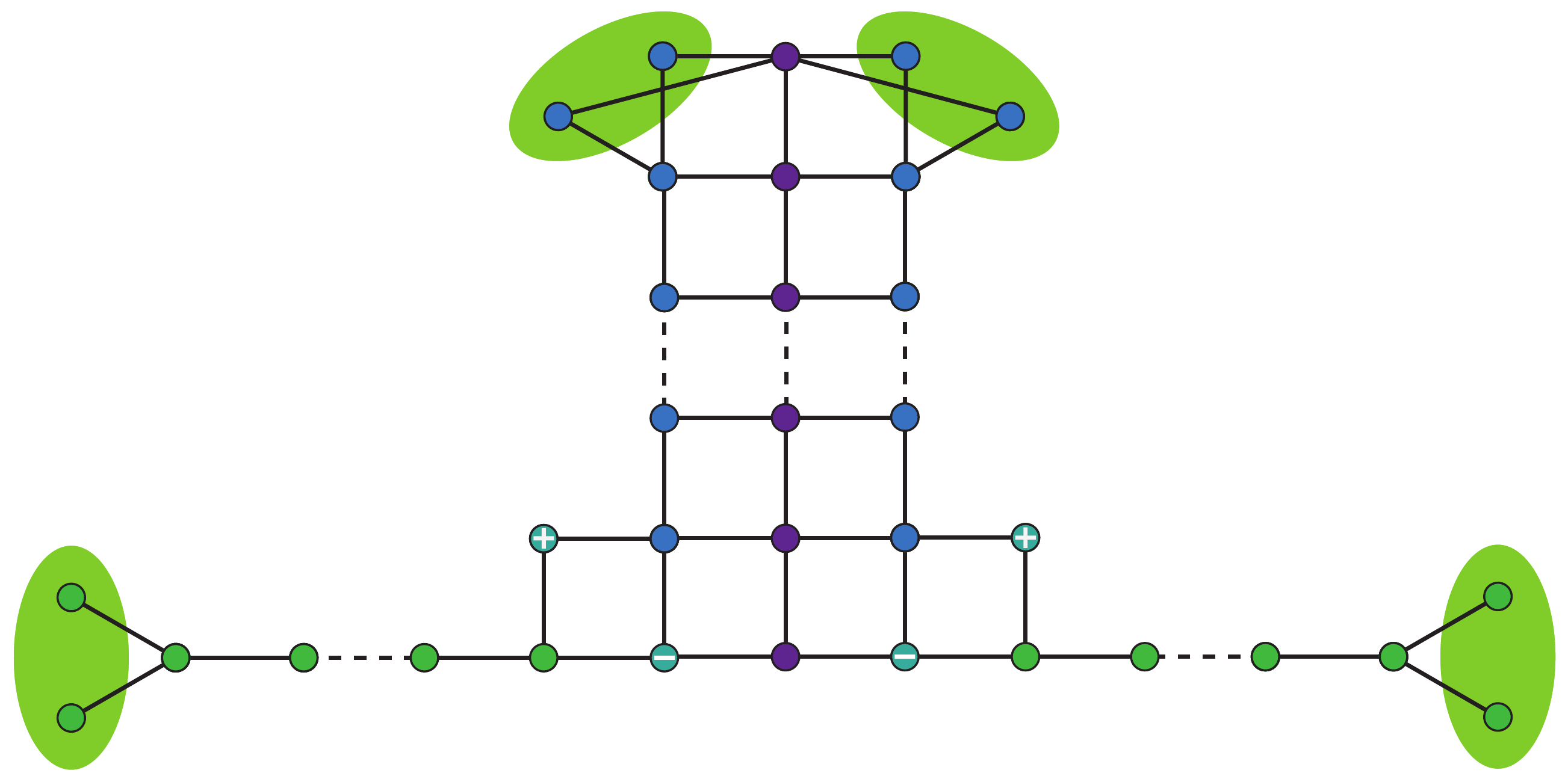}
\caption{The structure of the full deformed Y-system. The coupling between the two deformed Hubbard systems is via $k$ momentum carrying $Y_Q$-functions, where the $k$th one precisely couples to the end structure of the deformed Hubbard model, notably in a local fashion.}
\label{fig:DefYsys}
\end{center}
\end{figure}

As we will show in this paper, when we analogously couple the two $\psu_q(2|2)$ systems the spectrum of momentum carrying particles of the deformed mirror theory naturally terminates at bound states of length $k$. These bound states and their scattering properties arrange themselves perfectly with the subsystems, resulting in the TBA structure illustrated in figure \ref{fig:DefYsys}. There are $5k+4$ Y-functions with four algebraic relations between four pairs of them, resulting in $5k$ independent TBA equations.

\smallskip

The deformed model shows various interesting features on top of the fact that it can be described through a finite number of TBA equations. To start with, as already indicated the S-matrix upon which the TBA equations are based is not unitary but pseudo-unitary. Nonetheless the TBA equations are real, which is no doubt due to the not only pseudo but apparent quasi-unitarity of the mirror theory, briefly discussed in appendix \ref{app:matrixSmatrix}. At the level of kinematics, the physical region of complex rapidities of the mirror theory is deformed and in particular the real mirror line on the rapidity torus is now a true interval, as explained in the text around  figure \ref{fig:torusvsplane}. Next, as a cute result of the periodicity introduced by the deformation it is possible to define a mirror $x$ function which maps a single complex $u$-plane onto half of the torus in one go, compared to the quarter of the torus covered in the undeformed case, \emph{cf.} eqs. \eqref{eq:xmirror} and \eqref{shift} and figure \ref{fig:torusvsplane}. This allows for a description of the crossing transformation on the $u$-plane directly, if desired. Furthermore, the end structure at the level of the auxiliary TBA equations alluded to above shows an interesting coupling. The last two Y-functions introduced on either of the $\mathfrak{su}_q(2)$ wings turn out to be exactly inverse to one another (eqs. \eqref{eq:Y0inverse} and \eqref{eq:inverseYfunctions}), and result in a doubled contribution to the last regular simplified TBA equations and associated Y-system as shown in eqs. (\ref{eq:TBAvwbndry1},\ref{eq:TBAvwbndry2},\ref{eq:sTBAQk}) and (\ref{eq:YsysMw},\ref{eq:YsysMvw},\ref{YsysYk}), respectively. The fact that it is possible to obtain a  Y-system equation for length $k$ mirror bound states is due to quite nontrivial cancellations around this boundary, in particularly involving the $q$-deformed dressing phase as discussed in section \ref{sec:sTBA}. These cancellations are moreover precisely such that in the end we have a doubled contribution from the $vw$ functions resulting in a boundary Y-system equation of the form
\begin{align}
\nonumber
\frac{Y_k^+ Y_k^-}{Y_{k-1}^2} = & \frac{\displaystyle{\prod_{\alpha=1,2}}\left(1+\tfrac{1}{Y_{k-1|vw}^{(\alpha)}}\right)^2}{1+Y_{k-1}}\, .
\end{align}
The solution of the ground state TBA equations shows that the ground state of the deformed `string' theory has zero energy.

\smallskip
This paper is organized as follows. In section \ref{sec:kinematics} we introduce our main variables $x^\pm$ which parametrize the fundamental representation of the $q$-deformed algebra, and discuss the structure of the underlying rapidity torus and the dispersion relation of the $q$-deformed model and its mirror. Then in section \ref{sec:BetheYang} we present the Bethe-Yang equations of the mirror model, followed by a discussion of two- and multi-particle mirror bound states in section \ref{sec:boundstates}. Based on this analysis and the analysis done in appendix \ref{app:Stringhypo} we proceed to formulate the string hypothesis for our model in section \ref{sec:string hypothesis} and derive the canonical TBA equations. We find their simplified form and associated Y-system in section \ref{sec:sTBAandYsystem}. Finally, in section \ref{sec:groundsolution} we use the TBA equations to compute the energy of the ground state and show that it vanishes in complete analogy with the undeformed case. These are the main results of our paper. In the conclusion we recapitulate our findings and indicate interesting open questions as well as some preliminary findings not addressed here. We would like to emphasize that our string hypothesis in principle contains the string hypothesis of the quantum deformed Hubbard model at roots of unity, which has not been investigated before.
Technical details regarding the S-matrix (including its pseudo-unitarity), string hypothesis, TBA kernels, the simplified TBA equation for $Y_k$, and representation theory of $\psu_q(2|2)$ have been collected in five appendices.

\section{Kinematics of the $q$-deformed model and its mirror}

\label{sec:kinematics}
Here we present various useful parametrizations of the fundamental representation of the centrally extended $\psu_q(2|2)$ algebra
as well as the dispersion relations of the $q$-deformed model and its mirror.  For a detailed discussion of (atypical)
representations of $\psu_q(2|2)$ with $q$ a root of unity we refer the reader to appendix  \ref{app:reptheory}.

\subsection{Rapidity torus, u- and $u$-planes}
\begin{figure}[t]
\begin{center}
\includegraphics[width=1.0\textwidth]{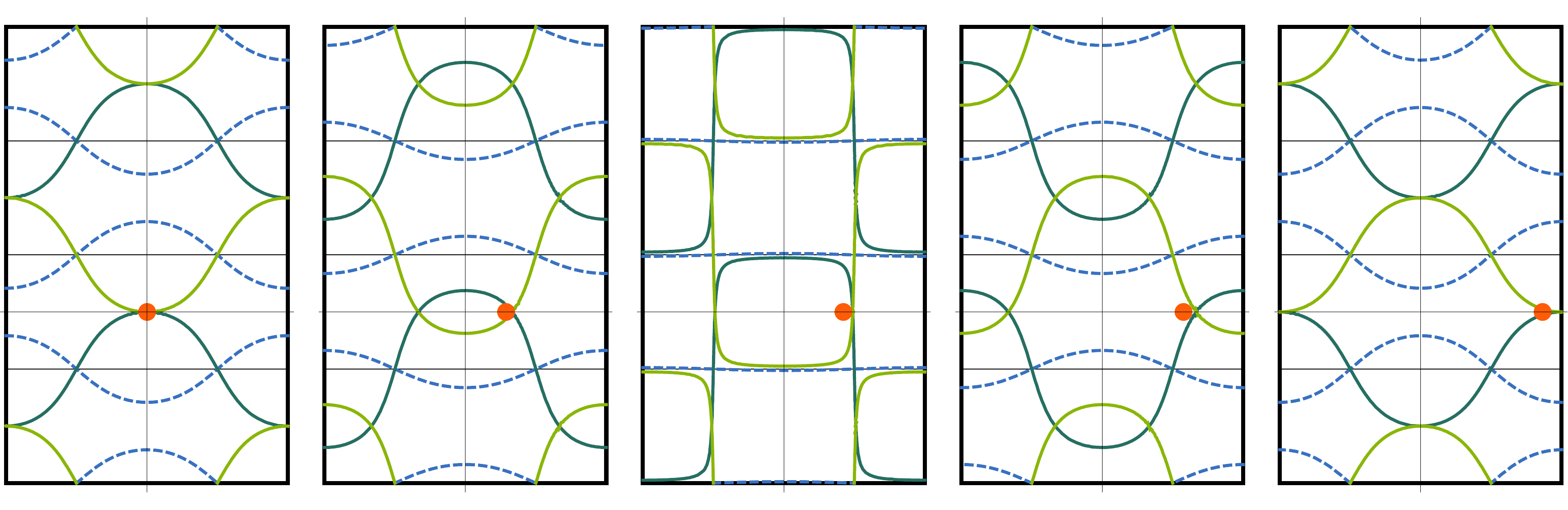}
\caption{
The division of the torus by the curves $|x^{\pm}|=1$ (blue, dashed) and $\mbox{Im}(x^{\pm})=0$ (two greens, solid) for five different values of $q$ on the unit circle. The orange dot on each plot indicates the position of the corresponding $\z_0$; the middle plot corresponds to $\z_0$ just below the value for which $q=i$. The shape of the `string' region ($|x^{\pm}|=1$) smoothly changes from the ``fish" to the ``hour-glass" shape as $\z_0$ runs from $0$ to $\omega_1(\k)/2$, {\it i.e.} $q$ from $q=1$ to $q=-1$, but it does interchange the location of its crests and throughs at $q=i$. For the `mirror' region ($\mbox{Im}(x^{\pm})=0$) the change around and the situation at $q=i$ is much more dramatic as the figure shows.}
\label{fig:stringtorusqvariable}
\end{center}
\end{figure}

Let us start by recalling that the basic variables $x^{\pm}$ parametrizing a fundamental representation of the $q$-deformed algebra satisfy the following constraint \cite{Beisert:2008tw,Beisert:2011wq}
\bea
\label{fc}
\frac{1}{q}\left(x^++\frac{1}{x^+}\right)-q\left(x^-+\frac{1}{x^-}\right)=\left(q-\frac{1}{q}\right)\left(\xi+\frac{1}{\xi}\right)\, ,
\eea
where the parameter $\xi$ is related the coupling constant $g$ as
\bea
\xi=-\frac{i}{2}\frac{g(q-q^{-1})}{\sqrt{1-\frac{g^2}{4}(q-q^{-1})^2}}\, .
\eea
As in the undeformed case the fundamental variables $x^{\pm}$ can be uniformized on an elliptic curve. This elliptic curve has real period $2\omega_1(\k)=4{\rm K}(m)$ and imaginary period $2\omega_2(\k)=4i{\rm K}(1-m)-4{\rm K}(m)$, where ${\rm K}(m)$ is the elliptic integral of the first kind considered as a function of the elliptic modulus $m=\k^2$. It turns out convenient to introduce a variable $\z_0$ in place of the deformation parameter $q$, related to it as
\bea
q=e^{i {\rm am}(2\z_0)}=\frac{\cso+i\dno}{\cso-i\dno}\, .
\eea
Here and below we use a concise notation for Jacobi elliptic functions; no subscript denotes the free variable in the equation, while the zero subscript indicates evaluation at $z_0$. For instance, ${\rm cs}(\z_0)\equiv {\rm cs}_0$.
The modulus $\k$ is related to the coupling constant $g$ as
\bea
g=-\frac{i\k}{2 \dno}\sqrt{1-\k^2 {\rm sn}^4_0}\, .
\eea
In the limit $q\to 1$, {\it i.e.} $\z_0\to 0$, we recover the familiar relation $\k^2=-4g^2$. With these conventions, the variables $x^{\pm}$ are the following meromorphic functions on the $\z$-torus
\bea
x^{+}(\z)=\k\, {\rm sn}_0^2 \frac{\cs +\cso}{\dn -\dno}
\frac{\dn -i\cso}{\cs -i\dno} \, , ~~~~
x^{-}(\z)=\frac{1}{\k\, {\rm sn}_0^2}\frac{\dn+\dno}{\cs-\cso}
\frac{\cs-i\dno}{\dn-i\cso} \, ,
\eea
while $\xi$ is given by
\bea
\xi=-i\k\frac{\sno\cno}{\dno}\, .
\eea
In this paper we want to take $q$ to be a root of unity and therefore restrict ourselves to real $\z_0$. Moreover, for $q$ to cover the unit circle once, the corresponding values of $\z_0$ can be restricted to the domain
\bea
-\omega_1(\k)/2 \leq \z_0 \leq \omega_1(\k)/2\, .
\eea
We further take the modulus $\k$ to be purely imaginary with $\mbox{Im}\, \k>0$, so that the coupling constant $g$ is positive. Accordingly, $\xi$ lies between zero and one. Note that for real $\z_0$ conjugation of the $x^\pm$ functions on the torus takes the form
\begin{equation}
x^{\pm}(\z)^*=x^{\mp}(\z^*)\,.
\end{equation}
For reasons already indicated in the introduction, we will take $q=e^{i \frac{\pi}{k}}$ with $k$ an integer greater than two. The special nature of $k=2$ can be seen in figure \ref{fig:stringtorusqvariable}, which show that the `string' and `mirror' type regions `flip their character' precisely at $q = i$.
The central charges of the fundamental representation are
\bea
U^2&=& \frac{1}{q} \frac{x^+ + \xi}{x^- + \xi}=\frac{\cs+i\dno}{\cs-i\dno}=e^{i({\rm am} (\z+\z_0)+{\rm am}(\z-\z_0))}\, , \\
V^2&=& q \frac{x^+}{x^-} \frac{x^- + \xi}{x^+ + \xi}=\frac{\cso+i\dn}{\cso-i\dn}=e^{i({\rm am} (\z+\z_0)-{\rm am}(\z-\z_0))}\,
\eea
and the shortening condition reads
\bea
\Big(\frac{V-V^{-1}}{q-q^{-1}}\Big)^2-\frac{g^2}{4}(1-U^2V^2)(V^{-2}-U^{-2})=1\, .
\eea

\begin{figure}[h]
\begin{center}
\includegraphics[width=1.0\textwidth]{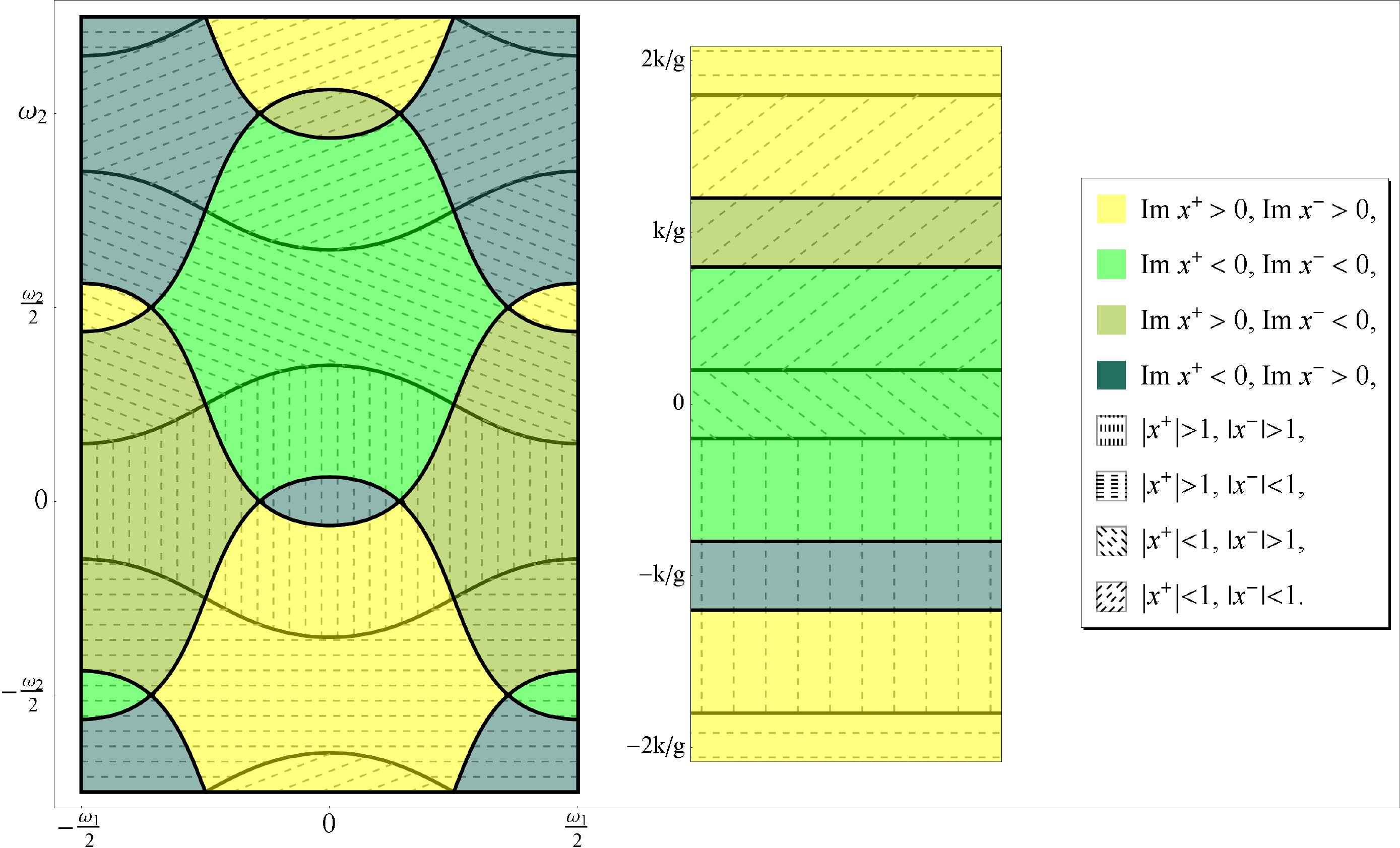}
\caption{The torus and the mirror $u$-plane. The left figure shows the torus with the distinguishing properties of the $x^\pm$ functions defined on it. The right figure depicts the mirror $u$-plane, with the same indications regarding $x^\pm$, illustrating the map between the mirror $u$-plane and the torus.}
\label{fig:torusvsplane}
\end{center}
\end{figure}

To proceed, let us introduce a multiplicative evaluation parameter $\U\equiv \U(x)$ defined as
\begin{equation}
\label{ev}
\U =-\frac{x+\frac{1}{x}+\xi+\frac{1}{\xi}}{\xi-\frac{1}{\xi}}\, ,
\end{equation}
so that $\U(x^+)=q^2\,  \U(x^-)$ and $x^{\pm}$ can be determined in terms of a single variable $x$ as
\bea
\U(x^+)=q\,\U(x)\, , ~~~~~\U(x^-)=q^{-1}\U(x)\, .
\eea
This evaluation parameter has the following expression via the torus variable $\z$
\bea
\U=\frac{\cso\dno+\cs\,\dn }{\cso\dno-\cs\,\dn}\, .
\eea
From eq.(\ref{ev}) we determine the inverse function $x(\U)$ as
\begin{align}
\label{eq:multxmirror}
x(\U) = \frac{\xi-\tfrac{1}{\xi}}{2}\Big[\frac{1+\xi^2}{1-\xi^2}-\U+i\sqrt{\Big(\U-\frac{1-\xi}{1+\xi}\Big)\Big(\frac{1+\xi}{1-\xi}-\U\Big)}\Big]\, .
\end{align}
There are two branch points $0<\frac{1-\xi}{1+\xi}<1$ and $1<\frac{1+\xi}{1-\xi}$ since $\xi$ is positive and less than $1$. The map $x(\U)$ is chosen in such a way that the cuts are $]-\infty,\frac{1-\xi}{1+\xi}]\cup [\frac{1+\xi}{1-\xi},+\infty[$.  As $\U$ runs over the interval $\frac{1-\xi}{1+\xi}\leq \U\leq \frac{1+\xi}{1-\xi}$ its image $x(\U)$ spans the unit half-circle in the lower half plane. Correspondingly, $1/x(\U)$ spans the upper half-circle. Note that the variables $x^{\pm}$ are expressed via $x(\U)$ in a simple manner\footnote{The parameters $x^{\pm}$ are two complex variables obeying one (complex) relation (\ref{fc}). Expressing $x^{\pm}$ in terms of $\U$, this relation is explicitly resolved.}
\bea
x^{\pm}=x(q^{\pm 1} \U)\, .
\eea
This $x$ function maps the $\U$-plane onto the green regions of the torus indicated in figure \ref{fig:torusvsplane}. As the branch cuts of $x(\U)$ are straight lines, the branch cuts of $x^\pm (\U)$ necessarily intersect and in this fashion cut of part of the $\U$-plane. It is precisely the disconnected green region on the torus which corresponds to this disconnected region of the $\U$-plane.

\smallskip

For the work that follows it will provide fruitful to work in an additive setting\footnote{We will shortly see that this is exactly the analogue of the conventional hyperbolic parametrization of the XXZ model with $|\Delta|<1$.}, which is obtained by the $u$-plane parametrization $\U=q^{-igu}=e^{\frac{\pi g u}{k}}$. Since the shift $u\to u+\frac{2ik}{g}$ leads to the same value of $\U$, the $u$-plane is an infinitely-sheeted cover of the $\U$-plane. For this parametrization the map $x(\U)$ turns into
\bea
\label{eq:xmirror}
x(u)=\frac{e^{\frac{\pi g u}{2k}}\Big(\sinh\frac{\pi g u}{2k}-i\,  \sqrt{g^2\sin^2\frac{\pi}{k}-\sinh^2\frac{g\pi u}{2k} }\Big)-g^2\sin^2\frac{\pi}{k} }{g\sin\frac{\pi}{k}\sqrt{1+g^2\sin^2\frac{\pi}{k}}}\, ,
\eea
where we have replaced $e^{\frac{\pi g u}{k}}$ which originally appears under the square root by $e^{\frac{\pi g u}{2k}}$ in front of it, removing a square root ambiguity. Because of this the function is no longer periodic with period $2ik/g$, but rather
\bea
\label{shift}
x\left(u+\tfrac{2ik}{g}\right)=\frac{1}{x(u)}\,
\eea
meaning $x(u)$ is periodic on the $u$-plane with period $\frac{4 i k}{g}$. In other words, by resolving a square root ambiguity, we have extended our $x$ function beyond the original mirror theory $\U$-plane.
Extended in this fashion, the mirror $x$ function covers a full vertical band of the torus, as illustrated in figure \ref{fig:torusvsplane}.
In fact, eq.(\ref{shift}) is nothing but the crossing transformation.
Moreover, the scattering matrix of the $q$-deformed model, see appendix  \ref{app:matrixSmatrix}, can be put on two copies of the $u$-plane, where it is compatible with crossing symmetry and its matrix part is periodic with period $4ik/g$ in either argument.

\begin{figure}[h]
\begin{center}
\includegraphics[width=0.45\textwidth]{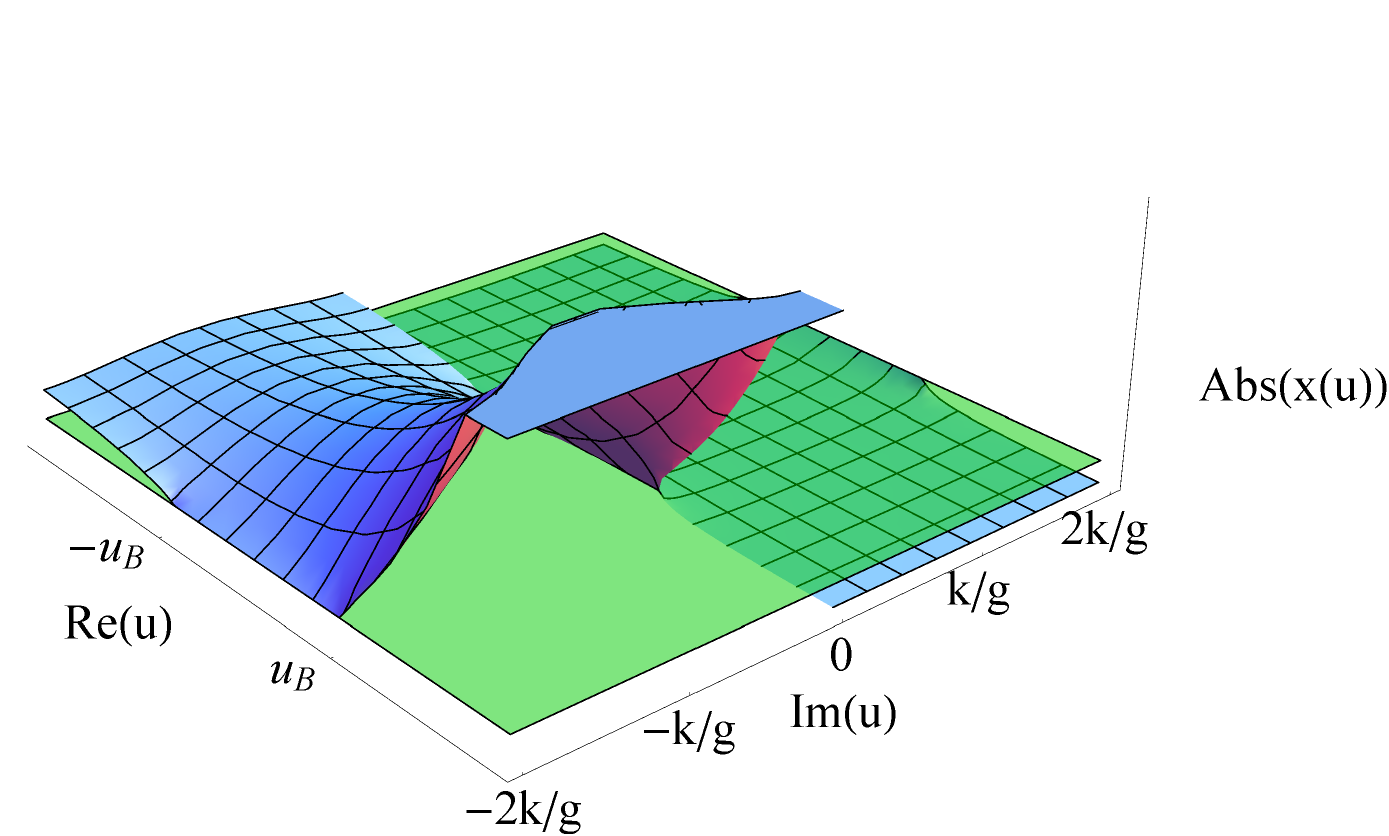}
\includegraphics[width=0.45\textwidth]{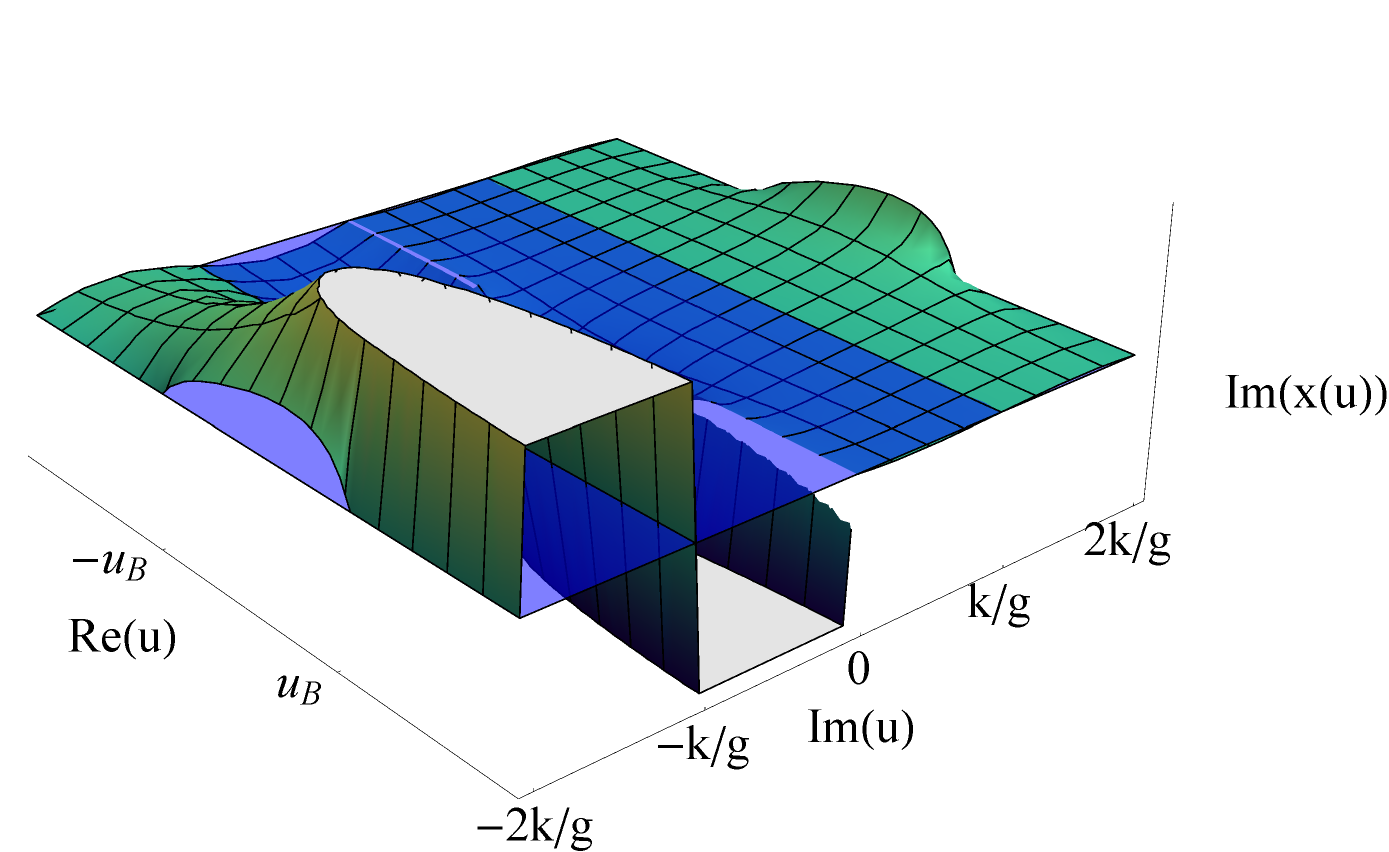}
\caption{The absolute value and imaginary part of $x(u)$ on the $u$-plane. The green surface distinguishes $|x|>1$ from $|x|<1$ in the left plot, while the blue surface distinguishes the sign of the imaginary part in the right plot.}
\label{fig:xmirror}
\end{center}
\end{figure}

The $x$ function has branch cuts on the $u$-plane running outward along the real line from $\pm u_b$ where
\begin{equation}
u_b = \frac{k}{\pi g} \log\frac{1+\xi}{1-\xi}=\frac{2k}{\pi g}{\rm arcsinh}\Big(g\sin\frac{\pi}{k}\Big)\, ,
\end{equation}
as well as outward from the points $\pm u_b + 2ik/g$ as illustrated in figure \ref{fig:xmirror}.
In the limit $k\to\infty$, the variable $x(u)$ tends to the standard mirror variable of the undeformed theory
\bea
x(u)=\frac{1}{2}(u-i\sqrt{4-u^2})\, ,
\eea
and naturally has the conjugation property of the standard mirror variable; $x(u)^* = \frac{1}{x(u^*)}$.

\subsection{The dispersion relations}

It is natural to make the following identification
\bea
\label{eq:UandVtoEandP}
V=q^{\frac{H}{2}}\, , ~~~~~~U=e^{i\frac{p}{2}}\, ,
\eea
where $H$ and $p$ are the energy and momentum of the model. For real $\z$, the energy is a non-negative periodic function, while the momentum takes values in the interval $(-\pi,\pi)$ as $\z$ runs over $(-\tfrac{\omega_1}{2},\tfrac{\omega_1}{2})$ as in the undeformed case. In the limit $q\to 1$ we smoothly obtain the undeformed string theory result  $p=2\, {\rm am}\, \z$ and $H=\dn\, \z$. Note that in terms of $H$ and $p$ thus introduced,
the shortening condition turns into the following dispersion relation
%\bea
%\label{Sdispersion}
%e^2 \frac{[H/2]^2_{q}}{\left[1/2\right]^2_q}- \frac{g^2}{\left[1/2\right]_q^{2}}\sin^2\frac{p}{2}=1\, ,
%\eea
\bea
\label{Sdispersion}
e^2 [H/2]^2_{q}- g^2\sin^2\frac{p}{2}=\left[1/2\right]_q^2\, ,
\eea
where we have introduced a manifestly positive ``coupling constant" $e$
\bea
e^2\equiv1-\frac{g^2}{4}(q-q^{-1})^2 = 1+ g^2 \sin^2 \frac{\pi}{k}\, ,
\eea
and the notation
\begin{equation}
\left[ x \right]_q \equiv \frac{q^x - q^{-x}}{q-q^{-1}}\, .
\end{equation}
To obtain the dispersion relation for the mirror theory, we perform the double Wick rotation, which just as in the undeformed case amounts to the following replacement in (\ref{Sdispersion})
\bea
\label{eq:mirrortf}
H\to i\tilde{p}\, , ~~~~p\to i\tilde{H}\, ,
\eea
where $\tilde{p}$ and $\tilde{H}$ are the momentum and energy of the mirror model. In this way we find
\bea
\tilde{H}=2\, {\rm arcsinh}\frac{[1/2]_q}{g}\sqrt{1-e^2[i\tilde{p}]^2_{q^{1/2}}}\, .
\eea
In particular, for  $q=e^{\frac{i\pi}{k}}$ the formula takes the form
\bea
\tilde{H}=2\, {\rm arcsinh}\left(\frac{1}{g}\frac{\sin \frac{\pi}{2k}}{\sin\frac{\pi}{k}}\sqrt{1+e^2\, \frac{\sinh^2\frac{\pi}{2k}\tilde{p}}{\sin^2\frac{\pi}{2k}}}\right)\, .
\eea
Very importantly, the mirror momentum is real on the line at $\tfrac{\omega_2}{2}$ in the green region of the torus, but \emph{not} on this line in the yellow region. This is perhaps surprising, but most certainly not a problem since this interval on the torus already covers the whole real line of mirror momenta; this interval corresponds to the whole real line of the $u$-plane. On this same interval the energy is positive and bounded from below. There are two natural ways to consider the limit $g\to \infty$. We can take the limit $g\to \infty$ with $k$ and $\tilde{p}$ fixed, resulting in a linear (phononic) dispersion
\bea
{\tilde H}=\frac{\pi}{k}|\tilde{p}|\, .
\eea
Alternatively if we first rescale the energy and momentum as $\tilde{H} \rightarrow  \tfrac{\tilde{H}}{g}$ and $\tilde{p} \rightarrow \tfrac{k}{\pi} \tfrac{\tilde{p}}{g}$ as in \cite{Hoare:2012fc}, we obtain the relativistic dispersion
\begin{equation}
\tilde{H}^2 - \tilde{p}^2 = \frac{1}{\cos^2{\tfrac{\pi}{2k}}}\,
\end{equation}
up to a rescaling of $g$ by a factor of two. Finally, let us note that the mirror transformation on the rapidity torus is given by a shift of the $\z$ variable by a quarter of the imaginary period, as in the undeformed case.

\section{Bethe-Yang equations for the mirror model}

\label{sec:BetheYang}

%deleted sections moved below \end{document}%
We are interested in constructing  the $q$-deformation of the TBA equations for strings on ${\rm AdS}_5\times {\rm S}^5$. Thus we assume that the corresponding S-matrix has $\psu_q(2|2)\oplus \psu_q(2|2)$ symmetry and therefore factorizes into two copies each invariant under $\psu_q(2|2)$. As in the undeformed case, this uniquely fixes the S-matrix up to a scalar factor. Of course this scalar factor is not uniquely determined by the requirements of unitarity and crossing,  but a very natural generalization\footnote{Taking a different scalar factor corresponding to a different solution to the crossing equation would change the S-matrix. However, insisting on the proper $q \rightarrow 1$ limit we expect the modification not to change the bound state picture we describe below. Therefore the only modification would be a simple prefactor in the $\mathfrak{sl}(2)$ S-matrix which should not affect our results in any further way.} of the undeformed scalar factor that satisfies these requirements has been found \cite{Hoare:2011wr}. We will take this as the S-matrix defining our theory and proceed from there.

\smallskip

The auxiliary problem corresponds to two copies of the quantum deformed Hubbard model, and analyticity of the associated transfer matrix\footnote{This transfer matrix will be explicitly presented in an upcoming paper.} implies the following Bethe-Yang equations for a single copy of the auxiliary roots $y_m$ and $\W_n$
\begin{align}
\label{eq:auxy}
1&= \prod_{i=1}^{K^{\mathrm{I}}}\sqrt{q}\frac{y_m - x^-_i}{y_m - x^+_i}\sqrt{\frac{x^+_i}{x^-_i}}
\prod_{j=1}^{K^{\mathrm{III}}}\frac{\V_m - q \W_j}{q \V_m - \W_j}\, ,\\
\label{eq:auxw}
-1&= \prod_{i=1}^{K^{\mathrm{II}}} \frac{q \W_n - \V_i}{\W_n - q \V_i} \prod_{j=1}^{K^{\mathrm{III}}}\frac{\W_n - q^2 \W_j}{ q^2 \W_n - \W_j }\, .
\end{align}
In the $u$-plane parametrization
\bea
\U=e^{\frac{\pi g u}{k}}\, , ~~~~\V=e^{\frac{\pi g v}{k}}\, , ~~~~\W=e^{\frac{\pi g w}{k}}\, , \eea
the auxiliary Bethe equations take the form
\begin{align}
\label{eq:auxy2}
1&= \prod_{i=1}^{K^{\mathrm{I}}}\sqrt{q}\frac{y_m - x^-_i}{y_m - x^+_i}\sqrt{\frac{x^+_i}{x^-_i}}
\prod_{i=1}^{K^{\mathrm{III}}}\frac{\sinh{\frac{\pi g}{2k}\big(v_m - w_i-\frac{i}{g}\big)}}{\sinh{\frac{\pi g}{2k}\big(v_m - w_i+\frac{i}{g}\big)}}\, ,\\
\label{eq:auxw2}
-1&= \prod_{i=1}^{K^{\mathrm{II}}} \frac{\sinh{\frac{\pi g}{2k}\big(w_n - v_i + \frac{i}{g}\big)}}{\sinh{\frac{\pi g}{2k}\big(w_n - v_i - \frac{i}{g}\big)}}\prod_{j=1}^{K^{\mathrm{III}}}\frac{\sinh{\frac{\pi g}{2k}\big(w_n - w_j - \frac{2i}{g}\big)}}{\sinh{\frac{\pi g}{2k}\big(w_n - w_j + \frac{2i}{g}\big)}}\, ,
\end{align}
where
\begin{equation}
e^{\frac{\pi g v}{k}}= \V = -\frac{y+\tfrac{1}{y} + \xi +\tfrac{1}{\xi}}{\xi - \tfrac{1}{\xi}}\,.
\end{equation}
As mentioned above, the second set of auxiliary equations is identical to those of the inhomogeneous Heisenberg XXZ spin chain; the limit $v_i \rightarrow 0$ gives the homogeneous XXZ spin chain. In the limit $k \to \infty$ with Bethe roots kept finite, by construction we get the auxiliary Bethe-Yang equations of the undeformed model
\begin{align}
1 & = \prod_{i=1}^{K^{\rm I}} \frac{y_m -x^-_i}{y_m -x^+_i}\sqrt{\frac{x^+_i}{x^-_i}} \prod_{j=1}^{K^{\rm III}} \frac{v_m - w_j -\frac{i}{g}}{v_m - w_j +\frac{i}{g}} \, ,\\
-1 & = \prod_{i=1}^{K^{\rm II}}  \frac{w_n - v_i +\frac{i}{g}}{ w_n - v_i-\frac{i}{g}}\prod_{j=1}^{K^{\rm III}} \frac{ w_n - w_j -\frac{2i}{g}}{w_n - w_j +\frac{2i}{g}} \, ,
\end{align}
where $v = y + \frac{1}{y}$.
With the definition of momentum (\ref{eq:UandVtoEandP},\ref{eq:mirrortf}) of the previous section and the scalar factor of the S-matrix, we can write down the full set of Bethe-Yang equations of the quantum deformed mirror model
\begin{align}
\label{eq:main}
1& = e^{i\tilde{p}_l R} \prod_{i\neq l}^{K^{\mathrm{I}}} S_{\sls(2)}(x_l,x_i)\prod_{\alpha=1}^2\prod_{i=1}^{K^{\mathrm{II}}_{(\alpha)}} \sqrt{q}\frac{y_i^{(\alpha)} - x_l^-}{y_i^{(\alpha)} - x_l^+}\sqrt{\frac{x_l^+}{x_l^-}},\\
\label{eq:auxyfinal}
1&= \prod_{i=1}^{K^{\mathrm{I}}}\sqrt{q}\frac{y_m^{(\alpha)} - x^-_i}{y_m^{(\alpha)} - x^+_i}\sqrt{\frac{x^+_i}{x^-_i}}
\prod_{i=1}^{K^{\mathrm{III}}_{(\alpha)}}\frac{\sinh{\frac{\pi g}{2k}\big(v_m^{(\alpha)} - w_i^{(\alpha)}-\frac{i}{g}\big)}}{\sinh{\frac{\pi g}{2k}\big(v_m^{(\alpha)} - w_i^{(\alpha)}+\frac{i}{g}\big)}}\, ,\\
\label{eq:auxwfinal}
-1&= \prod_{i=1}^{K^{\mathrm{II}}_{(\alpha)}} \frac{\sinh{\frac{\pi g}{2k}\big(w_n^{(\alpha)} - v_i^{(\alpha)} + \frac{i}{g}\big)}}{\sinh{\frac{\pi g}{2k}\big(w_n^{(\alpha)} - v_i^{(\alpha)} - \frac{i}{g}\big)}}\prod_{j=1}^{K^{\mathrm{III}}_{(\alpha)}}\frac{\sinh{\frac{\pi g}{2k}\big(w_n^{(\alpha)} - w_j^{(\alpha)} - \frac{2i}{g}\big)}}{\sinh{\frac{\pi g}{2k}\big(w_n^{(\alpha)} - w_j^{(\alpha)} + \frac{2i}{g}\big)}}\, ,
\end{align}
where $\alpha=1,2$. Here, as usual, $\tilde{p}_l$ stands for the mirror momentum of $l$th particle and $R$ is a length of the mirror circle. We will come back to the scalar factor $S_{\sls(2)}$ when discussing bound states in the next section, and have summarized the full deformed S-matrix and its properties in appendix \ref{app:matrixSmatrix}.

\section{Bound states of the mirror theory}
\label{sec:boundstates}

In order to discuss the thermodynamics of the mirror theory in the infinite volume limit, we need to determine the spectrum of excitations that make up the thermodynamic ensemble in infinite volume. In this section we discuss the spectrum of physical excitations of the infinite volume mirror model, while in the next section we join these with the spectrum of auxiliary excitations whose analysis we have carried out in appendix \ref{app:Stringhypo}. Together this goes under the name of the string hypothesis.

\subsection{Two-particle bound states}

To discuss the physical bound states of the mirror theory we need to analyze consistency of the Bethe equations in the limit $R\to \infty$. In the absence of auxiliary roots, the main Bethe equation (\ref{eq:main}) takes the form
\begin{align}
\label{eq:mainagain}
1& = e^{i\tilde{p}_lR} \prod_{i\neq l}^{K^{\mathrm{I}}} S_{\sls(2)}(x_l,x_i)\, ,
\end{align}
where the S-matrix corresponding to the $q$-deformed analogue of the $\sls(2)$ sector of string theory is given by
\bea
\label{sl2}
S_{\sls(2)}(x_1,x_2)=\sigma^{-2}\frac{x_1^+-x_2^-}{x_1^--x_2^+}\frac{1-\frac{1}{x_1^-x_2^+}}{1-\frac{1}{x_1^+x_2^-}}=
\frac{\sinh\frac{\pi g}{2k} (u_1-u_2+\frac{2i}{g})}{\sinh\frac{\pi g}{2k}(u_1-u_2-\frac{2i}{g})}\left(\frac{1-\frac{1}{x_1^-x_2^+}}{1-\frac{1}{x_1^+x_2^-}}\sigma^{-1}\right)^2\, ,~~
\eea
and $\sigma$ is the dressing factor. An explicit formula for $\tilde{p}(u)$ is given below in eq. \eqref{eq:ptildeofu}. For complex values of momenta the S-matrix exhibits a pole at $x_1^-=x_2^+$. At the level of the Bethe equations this pole is accompanied by a divergence of the momentum factor in the limit $R\to\infty$. Indeed, if the first particle has a momentum with positive imaginary part, the factor $e^{i\tilde{p}_1R}$ in \eqref{eq:mainagain} goes to zero, but the total equation remains finite when accompanied by a pole at $x_1^-=x_2^+$. If the second particle has a momentum with negative imaginary part such that the resulting total momentum $\tilde{p}_1+\tilde{p}_2$ is real, we obtain a two-particle bound state.

\smallskip

\begin{figure}[h]
\begin{center}
\includegraphics[width=0.7\textwidth]{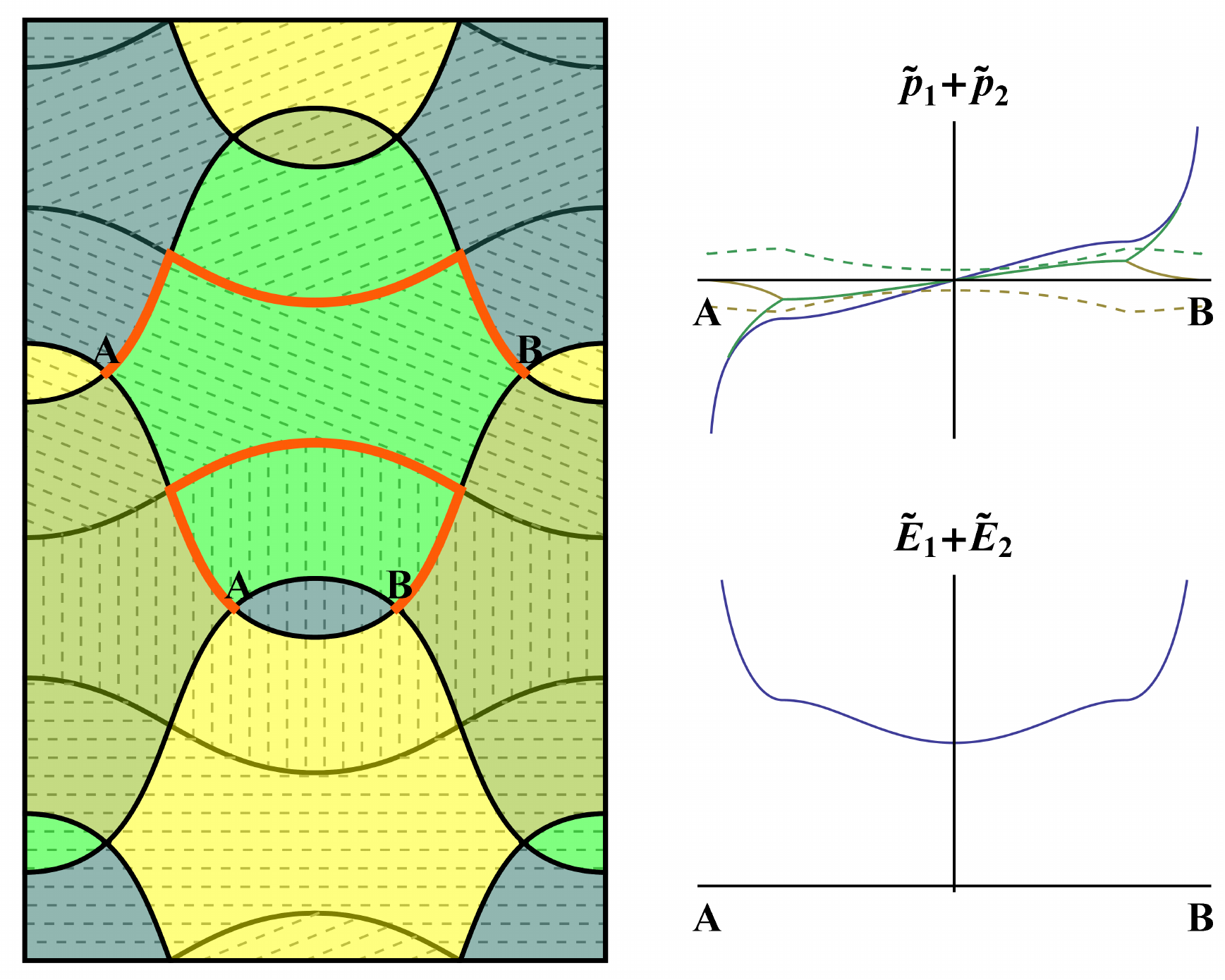}
\caption{The two-particle bound state on the torus with its momentum and energy. The constituents of the bound state lie along the orange curves, where $\tilde{p}=\tilde{p}_1+\tilde{p}_2$ increases from $-\infty$ to $\infty$ as we move from point $A$ to $B$. The kinks corresponds to critical values of the momentum, beyond which the two constituent momenta are no longer complex conjugate. This is shown in the upper right graph where the green and yellow lines show the real (solid) and imaginary (dashed) parts of the constituent momenta.}
\label{fig:2partboundstate}
\end{center}
\end{figure}

The two-particle bound state condition can be solved in terms of the torus $\z$-variable and the solution corresponding to the simplest two-particle bound state is shown in figure \ref{fig:2partboundstate}. As in the undeformed model, there is a critical value of the particle momenta; below the critical value  the rapidities of the constituent particles are conjugate to each other (with respect to the real line of the mirror theory) while beyond the critical value the conjugation property is lost. Because of this, in addition to the solution indicated in figure  \ref{fig:2partboundstate} there is another solution corresponding to
reflecting the picture in the mirror line.  This behavior is completely analogous to the undeformed case \cite{AF07},
and has been observed in the present context in \cite{Hoare:2012fc}.

\smallskip

The critical value of the momenta for the two-particle bound state is given by $\tilde{p}_{cr}=\tilde{p}_2(u_b)$. Explicit computation gives
\bea
\tilde{p}_{cr}=\frac{k}{\pi}\log\frac{\cos^2\frac{\pi}{k}+g^2\sin^2\frac{\pi}{k} +\rho_1\sin^2\frac{\pi}{k}+\rho_2\sin\frac{\pi}{k} }{1+g^2\sin^2\frac{\pi}{k}}\, ,
\eea
where
\bea
\rho_1&=&\sqrt{1+4g^2+4g^4\sin^2\frac{\pi}{k}}\, ,  \\
\rho_2&=&\sqrt{-2\cos^2\frac{\pi}{k} +2g^2\sin^2\frac{\pi}{k} +4g^4\sin^4\frac{\pi}{k} +2(\cos^2\frac{\pi}{k}+g^2\sin^2\frac{\pi}{k})\rho_1 }  \, .
\eea
In the limit $k\to \infty$ we get $\tilde{p}_{cr}=\sqrt{-2+2\sqrt{1+4g^2}}$ which coincides with the result obtained in \cite{AF07}.

Note that the two-particle mirror bound state trajectory, together with part of the two-particle string bound state trajectory which lies along the boundary of the small blue region in figure \ref{fig:2partboundstate} isolates a region of the torus which can be naturally called the physical mirror region of the theory.

\subsection{Multi-particle bound states}

In case the momentum of the two-particle bound state is not real we will necessarily have to involve a third particle to render the first two equations finite; the bound state grows. Without loss of generality we can take $\tilde{p}_1+\tilde{p}_2$ to have positive imaginary part\footnote{Otherwise we could have equivalently analyzed the poles of $S^{-1}$ starting with the Bethe-Yang equation for the second particle.}, giving a zero that is cancelled by the pole $x_2^-=x_{3}^+$.  At this point either the total momentum is real again, giving a three particle bound state, or we add a fourth particle and continue the process. In this way, we can obtain a $Q$-particle bound state defined by the conditions
\bea
x_1^-=x_2^+\, , ~~~x_2^-=x_3^+\, , \,\, \ldots \,\, , x_{Q-1}^-=x_Q^+\, .
\label{eq:bs}
\eea
The question now becomes which bound states are physical.
\smallskip

As in the undeformed case, we can insist that the bound state condition has a unique solution in the physical region of our theory. The physical mirror region referred to just above, the big green region of figures \ref{fig:torusvsplane} and \ref{fig:2partboundstate}, has this property, and contains bound states length of up to and including $k$. Were we to go beyond length $k$ we would first enter the lower blue and upper olive-green regions of the torus, as will become clear shortly when we consider bound states on the $u$-plane (\emph{cf.} eq. \eqref{uj} just below). These regions contain in particular fundamental particles and on their boundaries two particle bound states of the string and anti-string theory, making it undesirable to consider proceeding into them. Were we to continue even further, we would enter what we can by now refer to as the anti-mirror region and we would manifestly lose the uniqueness of the solution to the bound state equation. Moreover, this would correspond to constructing bound states containing both particles and anti-particles.
\smallskip

For these reasons we define the physical region of the mirror theory as the large green region on the torus of figures \ref{fig:torusvsplane} and \ref{fig:2partboundstate}. On the $u$-plane this corresponds to the strip $|\mbox{Im}(u)|\leq k-1$, as is clear from figure \ref{fig:torusvsplane}.

\begin{figure}[h]
\begin{center}
\includegraphics[width=0.48\textwidth]{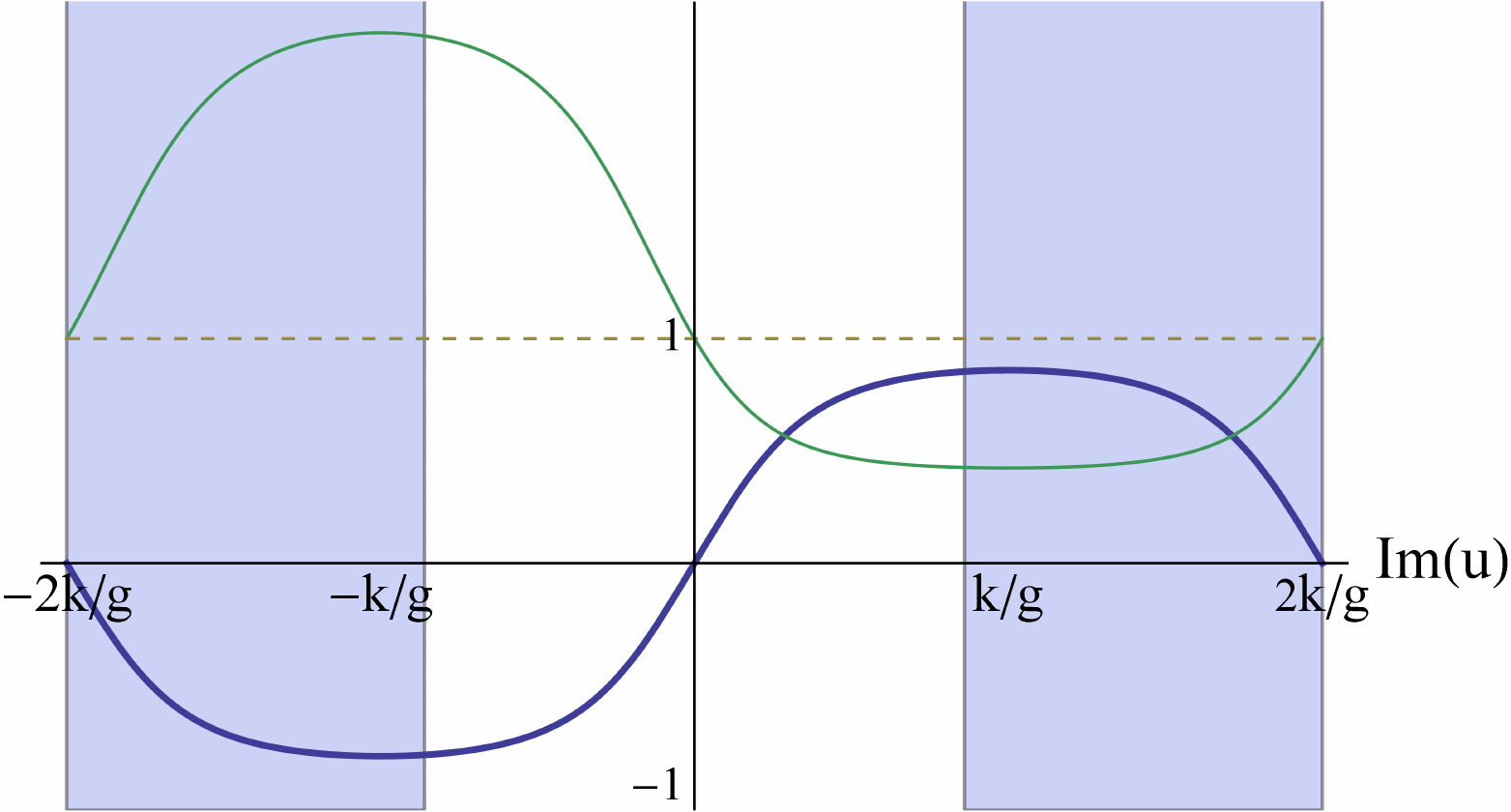}\hspace{5pt}\includegraphics[width=0.48\textwidth]{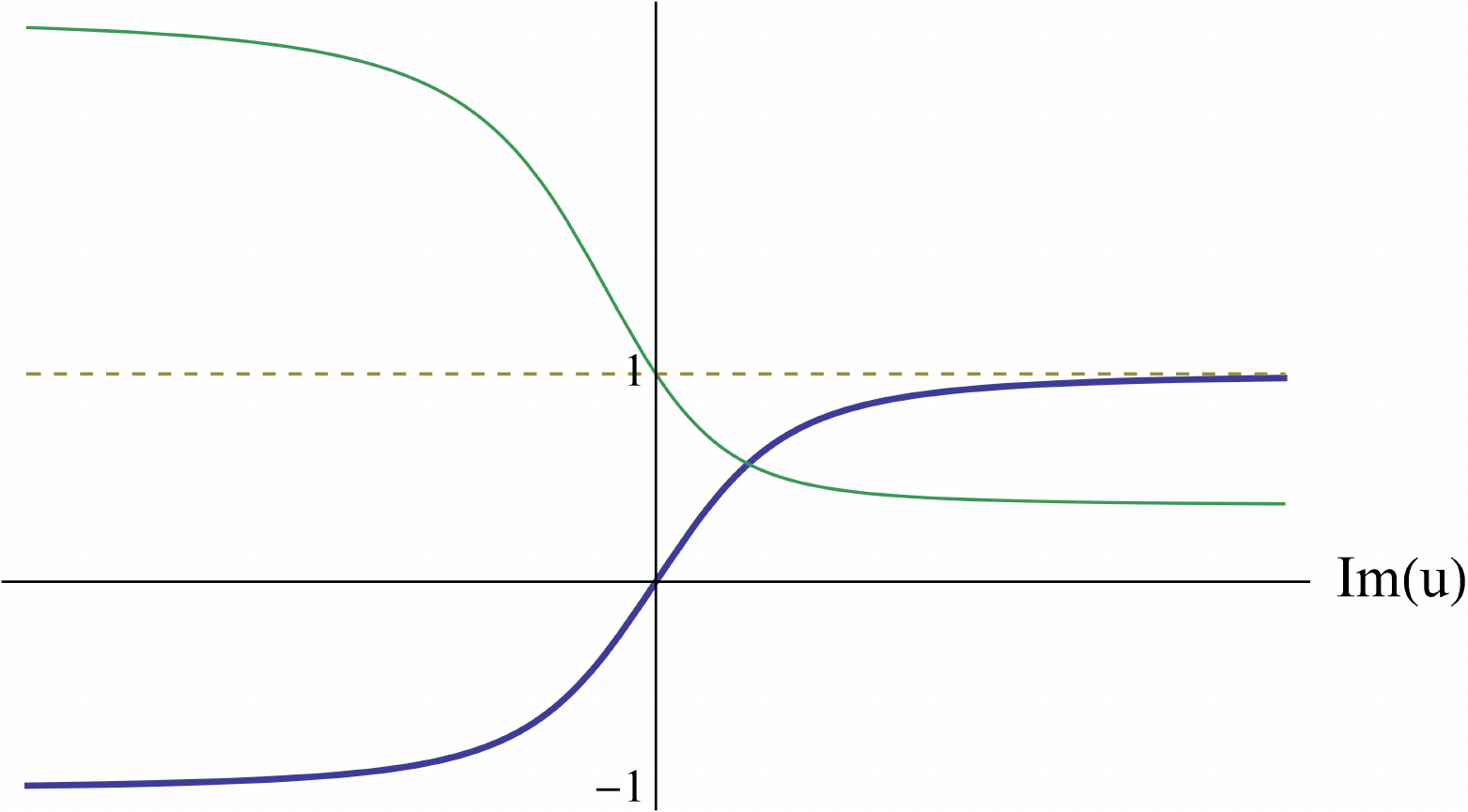}
\caption{The left plot shows the imaginary part of $\tilde{p}(u)$ in blue for some representative value of $\mbox{Re}(u)$, the thin green line shows $|e^{i \tilde{p}}|$. The shaded area of the plot lies outside the physical mirror region. The right plot shows the same functions for the undeformed model. The maximal value of ${\rm Im}\, \tilde{p}$ in reached at $\mbox{Im}(u)=k/g$ in the deformed model and at $\mbox{Im}(u)=\infty$ in the undeformed one.}
\label{fig:immom}
\end{center}
\end{figure}

For a more concrete discussion, let us put our bound states on the $u$-plane. There the set of roots making up a $Q$-particle bound state takes the standard form of the Bethe string
\bea
\label{uj}
u_j=u+\frac{i}{g}(Q+1-2j)\, , ~~~j=1,\ldots, Q\, .
\eea
The pole structure of these bound states is compatible with the Bethe-Yang equations since the mirror momentum $\tilde{p}(u)$ has positive imaginary part in particular for $0<\mbox{Im}(u)<k-1$ as illustrated in figure \ref{fig:immom}. This means that the imaginary part of the momentum of the first half of the particles is positive for all bound states we consider\footnote{At this point the attentive reader might be slightly concerned that by these considerations higher length bound states also appear to be allowed, at least as far as the imaginary part of the mirror momentum is concerned, while we just argued they should not be. The \emph{only} reason we are even pondering this question at this point is that due to the deformation the $u$-plane can naturally cover both the mirror and anti-mirror region, making it possible to smoothly consider the mirror momentum along the full imaginary direction of the torus in one go. In the undeformed limit, we could construct a very similar picture, but (un)fortunately it can only be obtained by gluing two entire planes together. Colloquially speaking, fusing through the line $ik/g$ is like fusing through infinity in the undeformed case.}.
\smallskip

Hence in the limit $R\to\infty$ we have $Q$-particle bound states of length up to and including $k$ defined by the following equations\footnote{Note that due to the periodicity introduced by the deformation, this type of string configuration for $Q = k$ necessarily implies $x_1^+ = 1/x_k^-$. This appears to contradict the string hypothesis since the improved dressing phase $\Sigma(x_1,x_k)\equiv \tfrac{1-1/{x_1^+x_k^-}}{1-1/{x_1^-x_k^+}}\sigma(x_1,x_2)$ has an apparent zero there. However, as the name suggests the improved dressing phase is perfectly finite at these points; the zero is precisely cancelled by a pole in $\sigma(x_1,x_2)$.}
\bea
x_1^-=x_2^+\, , ~~~x_2^-=x_3^+\, , ~~ \ldots\, , x_{Q-1}^-=x_Q^+\, .
\label{eq:bs}
\eea
These $Q$-particle bound states transform in short representations of the symmetry algebra with central charges
\bea
U_Q^2 =  \frac{1}{q^Q} \frac{x^+ + \xi}{x^- + \xi}\, , ~~~~~~~~
V_Q^2 =  q^Q \frac{x^+}{x^-} \frac{x^- + \xi}{x^+ + \xi}\, .
\eea
Here the variables $x^\pm = x(u\pm i Q/g)$ satisfy the relation
\bea
\label{fcb}
\frac{1}{q^Q}\left(x^++\frac{1}{x^+}\right)-q^Q\left(x^-+\frac{1}{x^-}\right)=\left(q^Q-\frac{1}{q^Q}\right)\left(\xi+\frac{1}{\xi}\right)\, .
\eea
The associated mirror momentum and energy are given by
\bea
\tilde{p}_Q = & -\frac{k}{\pi} \log V_Q^2\, ,~~~~~
\tilde{\mathcal{E}}_Q = & - \log U_Q^2\, .
\eea
Both the energy and momentum are real quantities for real values of $u$, as follows from the conjugation property of $x(u)$
\bea\label{cr}
\Big[x^+(u)\Big]^*=\frac{1}{x^-(u^*)}\, .
\eea
On the $u$-plane they are explicitly given by
\begin{align}
\label{eq:ptildeofu}
\tilde{p}_Q=&\,-\frac{k}{\pi}\log\frac{\cosh\frac{\pi g}{2k}(u+\tfrac{iQ}{g})-i\sqrt{g^2\sin^2\frac{\pi}{k} -\sinh^2\frac{\pi g}{2k}(u+\tfrac{iQ}{g})}}
{\cosh\frac{\pi g}{2k}(u-\tfrac{iQ}{g})-i\sqrt{g^2\sin^2\frac{\pi}{k} -\sinh^2\frac{\pi g}{2k}(u-\tfrac{iQ}{g})}} \, ,\\
\label{eq:Etildeofu}
\tilde{\cal E}_Q=&\,-\log\frac{\sinh\frac{\pi g}{2k}(u+\tfrac{iQ}{g})-i\sqrt{g^2\sin^2\frac{\pi}{k} -\sinh^2\frac{\pi g}{2k}(u+\tfrac{iQ}{g})}}
{\sinh\frac{\pi g}{2k}(u-\tfrac{iQ}{g})-i\sqrt{g^2\sin^2\frac{\pi}{k} -\sinh^2\frac{\pi g}{2k}(u-\tfrac{iQ}{g})}} \, .
\end{align}

\section{The string hypothesis and TBA equations}
\label{sec:string hypothesis}

In this section we describe the spectrum of the infinite volume mirror model and derive the TBA equations describing thermodynamics of the mirror theory.

\subsection{Physical excitations of the mirror model}

Based on the above analysis of bound states of the mirror theory and the analysis carried out in appendix \ref{app:Stringhypo}, we propose that the spectrum of excitations of the $q$-deformed mirror model in the thermodynamic limit is given by
\begin{enumerate}
\item{Bound states of fundamental particles; $Q$-particles,}
\item{Complexes of $y$ and $w$-roots; $M|vw$-strings,}
\item{Complexes of $w$-roots; $M|w$-strings,}
\item{Single $y$-roots; $y$-particles.}
\end{enumerate}
The significant difference with the string hypothesis for the undeformed string mirror model \cite{AF09a} is that the length and type of such complexes is considerably constrained; we have put the technical discussion appendix \ref{app:Stringhypo} and present an overview here.

\paragraph{\underline{$Q$-particles}} are bound states of fundamental particles, which in the present context can have a maximal length of $Q=k$. The rapidities representing possible $Q$-particle bound state are given by
\begin{equation}
\{ u \} = \{u +\frac{i}{g}(Q+1-2j) \, | \, j=1,\ldots,Q \} \, , \, \, \, Q =1,\ldots,k \, .
\end{equation}

\paragraph{\underline{$vw$-strings}} are most naturally classified in as two types; string of length less than $k$ and strings of length one with a particular type of complex rapidity. We will refer to these as strings with positive and negative parity respectively\footnote{This is the nomenclature in the XXZ model \cite{Takahashi:1972}.}.
The roots constituting possible positive parity $M|vw$ string are given by
\begin{eqnarray}
\{w\}&=&\{ v + \frac{i}{g}(M+1-2j)\, | \, j=1,\ldots,M\}\, ,
\nonumber \\
\{v\}&=&\{v + \frac{i}{g}(M-2j)\, | \, j=1,\ldots,M-1\}\cup\{v + \frac{i}{g} M , v - \frac{i}{g} M \}\, ,
\end{eqnarray}
for $M=1, \ldots, k-1$ and $v$ real, whereas the negative parity length one string has its center on the line $i k/g$\footnote{This is not a strange type of solution, merely a consequence of parametrization. In the $\U$-type variables, this simply corresponds to negative roots which are perfectly allowed in the Bethe equations.}. To each of the $v$ rapidities in the first set of the second line we associate two $y$-roots, one with $|y|>1$ and one with $|y|<1$, while the single $y$-roots associated to $v+ i M/g$ and $v- i M/g$ have $|y|<1$ and $|y|>1$ respectively. In terms of the mirror $x$ function we have $y_{\pm M} = x(v \pm i M/g)$ for positive parity strings, while $y_{1} = x(v+i(k+1)/g)$ and $y_{-1} = \frac{1}{x(v+i(k-1)/g)}$ with $v$ real for negative parity strings.

\paragraph{\underline{$w$-strings}} come in the same types as $vw$ strings, which is no coincidence as we explain in the appendix. The set of $w$-roots making up a $M|w$ string is given by
\begin{equation}
\{w\}=\{w + i (M+1-2j)/g\}\, ,~~~~j=1,\ldots, M\, ,
\end{equation}
where $M=1, \ldots, k-1$ and $w$ real or $M=1$ with $w$ on the line $i k/g$ for positive respectively negative parity strings.

\paragraph{\underline{$y$-particles}} are characterized by the property that $|y|=1$.  This means their associated rapidities $v$ run over the interval $(-u_b,u_b)$. In terms of the mirror $x$ function we have $y^- = x(v)$ while $y^+ = \frac{1}{x(v)}$, where the $\pm$ denotes the sign of the imaginary part of $y$.

\paragraph{\underline{Negative parity strings}} turn out to disappear rather naturally from the problem; their Y-functions are inverse to those of strings of length $k-1$. We will derive this carefully below, though this fact already manifests itself at the level of the string hypothesis; by fusing a length one negative parity $(v)w$-string with a length $k-1$ $(v)w$ string we get a configuration with momentum $\pi$, which effectively essentially means the associated Y-functions must be inverse in the thermodynamic limit.

\smallskip

Given a string hypothesis and the corresponding set of Bethe-Yang equations the derivation of the TBA equations becomes a text book story. There is a subtlety regarding the negative parity states, so for this reason and general completeness we go through the details carefully for the interested reader in appendix \ref{app:TBAderivation}. This gives a set of coupled integral equations between so-called Y-functions, one for each particle type in the spectrum we just discussed; $k$ $Y_Q$ functions for $Q$-particles, $Y_\pm$ functions for $y$ particles with positive, respectively negative imaginary part, and $k$ $Y_{M|(v)w}$ functions for $(v)w$ strings. Of course we have a set of $Y_\pm$ and $Y_{M|(v)w}$ functions for both values of $\alpha$. The index on the $Y_{M|(v)w}$ functions runs from \underline{zero} to $k-1$, where we denote negative parity strings separately as the \underline{zeroth} type of strings. However, as indicated above and derived explicitly in the appendix (\emph{cf.} eq. \eqref{eq:inverseYfunctions}) the Y-functions for negative parity strings are inverse to those of length $k-1$ positive parity ones
\begin{equation}
\label{eq:Y0inverse}
Y^{(\alpha)}_{0|(v)w}=\frac{1}{Y^{(\alpha)}_{k-1|(v)w}}\, .
\end{equation}
We eliminate $Y^{(\alpha)}_{0|(v)w}$ from the equations in this fashion, and consequently we introduce the following notation. The indices $M$,$N$ and $L$ run from one to $k-1$, while $Q$,$P$ and $R$ run from one to $k$; repeated indices indicate a standard sum. The $\star$ denotes convolution, the different types are explicitly defined in eqs (\ref{eq:stdconv}-\ref{eq:checkconv}). This gives the following equations.

\subsection{The canonical TBA equations and free energy}

\paragraph{The TBA equations for $Q$-particles} are

\begin{align}
\label{eq:cTBAQ0}
\log Y_Q = \, &  -L\, \tilde{\mathcal{E}}_{Q} + \log \left(1+Y_P \right)\star K_{\mathfrak{sl}(2)}^{PQ} + \log \left(1+\tfrac{1}{Y_{M|vw}^{(\alpha)}} \right)\star  K^{MQ}_{vwx} \\ \nonumber
&  + \log\left(1+Y_{k-1|vw}^{(\alpha)} \right) \star  K^{k-1,Q}_{vwx} + \log\left(1-\tfrac{1}{Y_{\beta}^{(\alpha)}} \right)\, \hat{\star}\, K^{yQ}_\beta\, ,
\end{align}
where we have an implicit sum over $\alpha=1,2$ and $\beta=\pm$.

\paragraph{The TBA equations for $y$-particles} are
\begin{equation}
\label{eq:cTBAypm0}
\log{Y_{\pm}^{(\alpha)}} = -\log\left(1+Y_{Q}\right)\star K^{Qy}_{\pm} + \log\frac{1+\frac{1}{Y_{M|vw}^{(\alpha)}}}{1+\frac{1}{Y_{M|w}^{(\alpha)}}}\star K_{M}+ \log\frac{\left(1+Y_{k-1|vw}^{(\alpha)}\right)}{\left(1+Y_{k-1|w}^{(\alpha)}\right)}\star K_{k-1} \, .
\end{equation}

\paragraph{The TBA equations for $w$-strings} are
\begin{align}
\label{eq:cTBAw0}
\log{Y_{M|w}^{(\alpha)}}=& \,\log\left(1+\tfrac{1}{Y_{N|w}^{(\alpha)}}\right)\star K_{NM}+\log\left(1+Y_{k-1|w}^{(\alpha)}\right)\star K_{k-1,M}\\
& \, +\log\frac{1-\frac{1}{Y_-^{(\alpha)}}}{1-\frac{1}{Y_+^{(\alpha)}}}\,\hat{\star}\,K_M\, .\nonumber
\end{align}

\paragraph{The TBA equations for $vw$-strings} are
\begin{align}
\label{eq:cTBAvw0}
\log{Y_{M|vw}^{(\alpha)}}
& =\log\left(1+\tfrac{1}{Y_{N|vw}^{(\alpha)}}\right)\star K_{NM} + \log\left(1+Y_{k-1|vw}^{(\alpha)}\right)\star K_{k-1,M} \\
& \quad \quad +\log\frac{1-\frac{1}{Y_-^{(\alpha)}}}{1-\frac{1}{Y_+^{(\alpha)}}}\,\hat{\star}\,K_M - \log\left(1+Y_Q\right)\star K^{QM}_{xv}\,. \nonumber
\end{align}
Clearly $Y_{k-1|vw}$ and $Y_{k-1|w}$ enter the above equations in a special way due to eq. \eqref{eq:Y0inverse}. As usual these equations can be simplified and subsequently used to derive the Y-system; this requires identities satisfied by the kernels which we show in appendix \ref{app:kernels}.

\paragraph{The free energy} of the mirror model allows us to find the energy of the deformed `string' theory as
\bea
\label{eq:Energy}
E(L) &=&-\int {\rm d}u\, \sum_{Q=1}^{k}\frac{1}{2\pi}\frac{d\tilde{p}^Q}{du}\log\left(1+Y_Q \right)\,.
\eea
Note that we have only a finite number of TBA equations and a finite number of Y-functions contributing to the energy.

\section{Simplified TBA equations and Y-system}

\label{sec:sTBAandYsystem}
In this section we simplify the canonical TBA equations of the previous section and derive the corresponding Y-system.

\subsection{Simplified TBA equations}
\label{sec:sTBA}

\subsubsection*{$w$ and $vw$-strings}

Using \eqref{eq:simpKM}, \eqref{eq:simpKMN} and \eqref{eq:simpKQMxv} and the canonical equations for $\log{\frac{Y_+}{Y_-}}$ we obtain the following simplified TBA equations for $vw$-strings
\begin{align}
\log{Y_{M|vw}} = \, & \log{(1+Y_{M+1|vw})(1+Y_{M-1|vw})} \star s - \log\left(1+Y_{M+1}\right)\star s \nonumber\\
& + \delta_{M,1} \log{\left(\frac{1-Y_-}{1-Y_+}\right)}  \hat{\star}  s \, , \label{eq:TBAvwbulk}\\
\log{Y_{k-2|vw}}  = \, & \log{(1+Y_{k-3|vw})(1+Y_{k-1|vw})^2} \star s  - \log\left(1+Y_{k-1}\right)\star s\, , \label{eq:TBAvwbndry1}\\
\log{Y_{k-1|vw}}  = \, & \log{(1+Y_{k-2|vw})} \star s - \log\left(1+Y_{k}\right)\star s\, , \label{eq:TBAvwbndry2}
\end{align}
where $Y_{0|vw}=0$. Note how the contribution of $Y_{k-1|vw}$ is still special, but now definitely more elegant. For $w$-strings we similarly find
\begin{align}
\log{Y_{M|w}} & = \log{(1+Y_{M+1|w})(1+Y_{M-1|w})} \star s + \delta_{M,1} \log{\left(\frac{1-\frac{1}{Y_-}}{1-\frac{1}{Y_+}}\right)}  \hat{\star}  s\, , \label{eq:TBAwbulk}\\
\log{Y_{k-2|w}} & = \log{(1+Y_{k-3|w})(1+Y_{k-1|w})^2} \star s\, , \label{eq:TBAwbndry1}\\
\log{Y_{k-1|w}} & = \log{(1+Y_{k-2|w})} \star s \, , \label{eq:TBAwbndry2}
\end{align}
where $Y_{0|w}=0$.

\subsubsection*{$Q$-particles}

Applying $(K+1)^{-1}$ defined in \eqref{eq:Kp1inv} and using the identities\footnote{We have \emph{assumed} identities involving the $q$-deformed dressing phase to be analogous to the undeformed case, but have not proven this.} in appendix \ref{app:kernels}, we find the simplified TBA equation for $Y_1$
\bea
\log Y_1=\log\frac{\left(1-\tfrac{1}{Y_-^{(\alpha)}}\right)}{1+\tfrac{1}{Y_2}}\star s -\check{\Delta}\check{\star} s\, ,
\eea
where
\bea
\check{\Delta}&=&L\check{\cal E}+\log \left(1-\tfrac{1}{Y_-^{(\alpha)}}\right)\left(1-\tfrac{1}{Y_+^{(\alpha)}}\right)\star \check{K}+2\log(1+Y_Q)\star
\check{K}^{\Sigma}_Q\\
\nonumber
&&\quad\quad\quad +\log\left(1-\tfrac{1}{Y_{M|vw}^{(\alpha)}}\right)\star \check{K}_M+ \log\left(1-Y_{k-1|vw}^{(\alpha)}\right)\star \check{K}_{k-1}\, .\eea
The kernels appearing in the expression above are defined in appendix \ref{app:kernels}. Also here and below the sum over $\alpha$ is left implicit. Analogously, for $Q=2,\ldots,k-1,$ we find that the simplified TBA equations are given by
\begin{align}
\log Y_Q = \, & \log\frac{{Y_{Q+1}Y_{Q-1}}}{(1+Y_{Q-1})(1+Y_{Q+1})}\star s + \log \left(1+\tfrac{1}{Y_{Q-1|vw}^{(\alpha)}}\right) \star s\, .\label{eq:sTBAQg1}
\end{align}
For $Q=k$ we cannot simply apply the usual operator
\bea
(K+1)^{-1}_{Pk}=\delta_{Pk} - s (\delta_{Pk-1}+\delta_{Pk+1})\, ,
\eea  since there is no $k+1$st equation. Instead, we apply the operator $\delta_{Q,k} - 2 \delta_{Q,k-1} s$. As we discuss in appendix \ref{app:sTBAYk}, using special identities and the TBA equations for $vw$ strings we obtain
\begin{equation}
\label{eq:sTBAQk}
\log Y_k =\, 2\log{Y_{k-1}}\star s -\log(1+Y_{k-1})\star s + \log\left(1+\tfrac{1}{Y_{k-1|vw}^{(\alpha)}}\right)^2 \star s \, .
\end{equation}
Naively we might have expected to get something like $\left(1+\tfrac{1}{Y_{k-1|vw}}\right)\left(1+Y_{k-1|vw}\right)$ in the above equation. Because it is exactly analogous to what happens in the equations for $vw$ strings it is very natural that we get a doubled contribution instead, though it arises in the far less obvious fashion shown in appendix \ref{app:sTBAYk}.

\subsection{Y-system}

\subsubsection*{$w$ and $vw$ strings}

Deriving the Y-system for $w$-strings from the simplified TBA equation is now a simple matter; applying $s^{-1}$ to the equations above\footnote{As usual $f \circ s^{-1}(u) \equiv \displaystyle{\lim_{\epsilon\to 0} }[f(u+i/g-i \epsilon) + f(u-i/g+i \epsilon)]$.}  immediately yields
\begin{align}
\label{eq:YsysMw}
Y_{1|w}^+ Y_{1|w}^- & =(1+Y_{2|w})\left(\frac{1-Y_-^{-1}}{1-Y_+^{-1}}\right)^{\theta(u_b-|u|)}  \\
Y_{M|w}^+ Y_{M|w}^- & =(1+Y_{M+1|w})(1+Y_{M-1|w}) \, \, , \, \, \, M=2,\ldots,k-3 \, ,\\
Y_{k-2|w}^+ Y_{k-2|w}^- & = (1+Y_{k-3|w})(1+Y_{k-1|w})^2 \, ,\\
Y_{k-1|w}^+ Y_{k-1|w}^- & = (1+Y_{k-2|w}) \, .
\end{align}
For $vw$-strings we similarly get
\begin{align}
\label{eq:YsysMvw}
Y_{1|vw}^+ Y_{1|vw}^- & =\left(\frac{1+Y_{2|vw}}{1+Y_2}\right)\left(\frac{1-Y_-}{1-Y_+}\right)^{\theta(u_b-|u|)}  \\
Y_{M|vw}^+ Y_{M|vw}^- & =(1+Y_{M+1|vw})(1+Y_{M-1|vw})\left(\frac{1}{1+Y_{M+1}}\right) \, ,\\
Y_{k-2|vw}^+ Y_{k-2|vw}^- & = (1+Y_{k-3|vw})(1+Y_{k-1|vw})^2 \left(\frac{1}{1+Y_{k-1}}\right)\, ,\\
Y_{k-1|vw}^+ Y_{k-1|vw}^- & = (1+Y_{k-2|vw}) \left(\frac{1}{1+Y_{k}}\right)\, .
\end{align}
Away from $M=1$ we indeed see precisely the XXZ structure advertised in figure \ref{fig:XXXtoXXZ} in the introduction for both the $w$ and $vw$ functions independently. The two Y-functions in the lime-green bubble there are $Y_{k-1|(v)w}$ and $Y_{0|(v)w}$ and are inverse to each other. We implemented their relation at the level of canonical equations already, so that we precisely get the doubled contribution of $Y_{k-1|(v)w}$ on the right hand side of the Y-system for the $k-2$th Y-function. Note also how the $Y_Q$ particles are nicely compatible with this structure by existing precisely one length beyond the $Y_{M|vw}$ functions. Dropping the $Y_Q$ terms in the above now clearly gives precisely the structure of figure \ref{fig:HubbtoDefHubb}.

\subsubsection*{$y$-particles}

For $y$-particles only the Y-system for $Y_-$ is of a nice local form. As in the undeformed case we can derive them by applying $s^{-1}$ to the equation for $Y_-$, and subtracting the canonical TBA equations for $1|w$ and $1|vw$ strings to get
\begin{equation}
Y_{-}^{(\alpha)+}Y_{-}^{(\alpha)-} = \frac{1+Y_{1|vw}^{(\alpha)}}{1+Y_{1|w}^{(\alpha)}}\frac{1}{1+Y_1}\, .
\end{equation}

\subsubsection*{$Q$ particles}

Taking $|u|<u_b$ for $Q=1$ we have
\begin{align}
\label{eq:YsysY1}
\frac{Y_1^+ Y_1^-}{Y_2} = &\frac{\displaystyle{\prod_{\alpha=1,2}} \left(1-\tfrac{1}{Y_{-}^{(\alpha)}}\right)}{1+Y_{2}}\, ,
\end{align}
while for $Q=2,\ldots,k-1$ the Y-system is
\begin{align}
\label{eq:YsysYQ}
\frac{Y_Q^+ Y_Q^-}{Y_{Q+1}Y_{Q-1}} = \frac{\displaystyle{\prod_{\alpha=1,2}} \left(1+\tfrac{1}{Y_{Q-1|vw}^{(\alpha)}}\right)}{(1+Y_{Q-1})(1+Y_{Q+1})}\, .
\end{align}
Finally at $Q=k$ we have
\begin{align}
\label{YsysYk}
\frac{Y_k^+ Y_k^-}{Y_{k-1}^2} = & \frac{\displaystyle{\prod_{\alpha=1,2}}\left(1+\tfrac{1}{Y_{k-1|vw}^{(\alpha)}}\right)^2}{1+Y_{k-1}}\, .
\end{align}
Taking the above equations together, we have the full the Y-system for the quantum deformed mirror model illustrated in figure \ref{fig:DefYsys}. The four Y-functions coupling to $Y_k$ at the top of the diagram are precisely the four $Y_{k-1|vw}$ terms - two for each $\alpha$ - in the above equation. As in the undeformed case the Y-system admits analytic continuation onto the complex $u$-plane with cuts where the TBA equations unambiguously fix the corresponding jump discontinuities of the Y-functions. Both the TBA equations and the Y-system presented here manifestly tend to their undeformed cousins in the limit $k\to \infty$.

\section{Ground state energy}
\label{sec:groundsolution}

In this section we will find the constant solution of the ground state TBA equations and show that the ground state has zero energy. To do so we follow \cite{Frolov:2009in} where this was shown for the undeformed ground state. The idea is that the ground state equations are solved by
\begin{equation}
Y_Q = 0\, , \, \, \, \, Y_\pm = 1\, ,
\end{equation}
with constant $Y_{M|(v)w}$ functions. However, the TBA equations for $Q$ particles are singular at the point $Y_Q = 0$ and so we need to regularize the solution and take a limit to obtain our result. To this end we write
\begin{equation}
Y_\pm^{(\alpha)} = 1 + h A_\pm^{(\alpha)} + \ldots\, .
\end{equation}
As in the undeformed case, the small $h$ expansion of the other equations is consistent with $Y_+ = Y_- $ and constant $Y_{M|(v)w}$ functions.  Then the ground state TBA equation for $Q$ particles, eq. (\ref{eq:cTBAQ0}), shows that
\begin{equation}
\log Y_Q = -2 \log h \star (K^{yQ}_-+K^{yQ}_+) + \mbox{finite terms}\, .
\end{equation}
Since $K^{yQ}_-+K^{yQ}_+$ is normalized to one, this shows that $Y_Q \sim h^2 B_Q$ where $B_Q$ is left to be determined. We then need to solve the TBA equations for $vw$ and the completely decoupled $w$ strings with constant Y-functions, and use the resulting $vw$ functions to fix $B_Q$. The equations for $w$ and $vw$ strings take the same form, namely for $Y_{M|(v)w} = A_M$ the simplified TBA equations reduce to the Y-system
\begin{align}
A_M^2 &= (1+A_{M-1})(1+A_{M+1})\,,\\
A_{k-2}^2 &= (1+A_{k-3})(1+A_{k-1})^2 \, ,\\
A_{k-1}^2 &= (1+A_{k-2})\, ,
\end{align}
with $A_0=0$, which in principle has many solutions. However, requiring the solution of this Y-system to solve the canonical TBA equations singles out the solution
\begin{align}
A_M & = M(M+2) \, , \\
A_{k-1} & = k-1\, ,
\end{align}
as can be verified by explicit computation. To determine $B_Q$ we need to solve
\begin{equation}
\log{B_Q} = - L \mathcal{E}_Q + 2 \log \left(1+\tfrac{1}{A_{M}} \right)\star  K^{MQ}_{vwx} + 2\log\left(1+A_{k-1}\right) \star  K^{k-1,Q}_{vwx}\, .
\end{equation}
Noting that $K^{Mk}_{vwx}$ is normalized to one, we easily find
\begin{equation}
\log{B_k} = - L \mathcal{E}_k + 2 \sum_{M=1}^{k-2} \log \left(1+\tfrac{1}{M(M+2)} \right) + 2\log \tfrac{k^2}{k-1} = -L \mathcal{E}_k + 2\log{2k}\, .
\end{equation}
Rather than computing the other terms explicitly, we can use the small $h$ expansion of the Y-system (\ref{eq:YsysY1}-\ref{YsysYk}) to conclude
\begin{equation}
Y_Q \sim 4 h^2 Q^2 e^{-L \mathcal{E}_Q} \, , \, \, \, \mbox{for}\, \,  Q =1,\ldots,k\, .
\end{equation}
The associated energy is then given by
\begin{equation}
E \sim 4 h^2 \sum_{Q=1}^k Q^2 \int \frac{d \tilde{p}}{2\pi} e^{-L \mathcal{E}_Q(\tilde{p})} \, .
\end{equation}
Since we sum a finite number of convergent integrals, in the limit $h\to0$ the ground state energy is manifestly zero\footnote{The divergence of this expression for $L=2$ observed in the undeformed case \cite{Frolov:2009in} does not occur here since we only have a finite number of integrals. The sum does grow with $k$ however, and for $L=2$ the limit $k\to \infty$ and $h\to 0$ do not commute.}.

\section{Conclusion}

In the present paper we have derived the ground state TBA equations of the quantum deformed $\AdS$ mirror model, based on a $\psu_q(2|2)^2$ invariant S-matrix. As we have seen, these equations show a rich structure, summarized in figure \ref{fig:DefYsys}. The next natural step is to generalize these equations to excited states along the lines of \cite{KP,DT96,BLZe,Arutyunov:2011uz,AFS09}, and as indicated we will provide the necessary tools to do so - an asymptotic solution - in an upcoming publication \cite{qdeftbapart2}.

The main important feature of the equations we have found is that there is a finite number of them. This will hopefully allow us to gain further valuable insight into the $\AdS$ mirror model by considering our equations in an appropriate limit. Having only a finite number of TBA equations may help avoid technical obstacles which arise in the strict undeformed limit. As one potential example, when the above equations are generalized to excited states an obvious question will be that of wrapping corrections at small coupling, for example for the deformed analogue of the Konishi operator. With a finite number of TBA equations, it might be possible to obtain an analytic expression for wrapping corrections at a fixed value of $k$\footnote{The previous obstacle of inversion of an infinite dimensional matrix of integral operators now at least becomes finite dimensional.} without relying on the clever but unfortunately limited approach of \cite{BH10a,BH10b} which appears to work specifically just beyond first wrapping in the $\alg{sl}(2)$ sector. The idea would then be to find a sequence of wrapping corrections at fixed values of $k$, and hope that this naturally extrapolates to the undeformed case, giving higher order wrapping corrections for SYM operators in the limit $k\to \infty$, previously unaccessible analytically.

Coming to more general questions, as we saw above the quantum deformed model at roots of unity shows qualitatively rather dramatic differences to the undeformed case. In contrast we can expect deformations with real $q$ to be `smoother' and can ask whether the resulting model has interesting properties. Considering the amount of material already presented, we decided against trying to incorporate this in the present paper. However, we should note that the answer to the above question is positive, though admittedly less dramatic than was the case here. We would like to briefly elaborate on this. The XXZ model considered as a real deformation of the XXX model has $|\Delta|>1$, and in this case the TBA equations \cite{Gaudin:1971,Takahashi:book}  are simply modified by being appropriately defined on an interval rather than a whole line, while otherwise their structure remains intact. Of course a similar thing happens in our model when $q$ is real, but the fact that we have an additional coupling constant in the game results in something special. Firstly, for fixed $q$ there is a natural bound to the coupling constant $g$ above which the model is no longer well defined, and secondly the support of $Y_\pm$ functions varies with respect to that of the other Y-functions and becomes equal to it exactly on the boundary for $g$, which seems to result in a Y-system without discontinuities provided this limit is well defined. These comments of course also apply to the deformed Hubbard model subsystem. We hope to address this in more detail in the near future.

Also, let us note that the thermodynamics of the deformed Hubbard model itself might have interesting features from a condensed matter point of view, and apparently provides the next simplest setting beyond the XXZ model to study the TBA of quantum deformed systems. It would be interesting to look into  the thermodynamics of the quantum deformed Hubbard and associated models\footnote{While in undeformed Hubbard-like models we can directly go from the `mirror' type model (Hubbard) to `string' type models (corresponding to noncompact support for the $Y_\pm$ functions in the standard parametrization) it appears that with the root of unity deformation the string hypothesis for the `stringy' type deformed Hubbard model requires more than the obvious modifications.} in more detail along the lines of {\it e.g.} \cite{Frolov:2011wg}.

\section*{Acknowledgements} We would like to thank Niklas Beisert for fruitful discussions and suggestions, for early-stage collaboration on this project and for comments on the manuscript. Moreover we thank Sergey Frolov for interesting discussions and Ben Hoare, J. Luis Miramontes and Sergey Frolov for valuable comments on the paper. G.A. acknowledges support by the Netherlands Organization for Scientific Research (NWO) under the VICI grant 680-47-602. The work by M.L. is supported by the SNSF under project number 200021-137616. The work by G.A and S.T. is also part of the ERC Advanced grant research programme No. 246974,  {\it ``Supersymmetry: a window to non-perturbative physics"}. G.A. would like to thank IPMU and the Yukawa Institute of Kyoto university, where part of this work was done, for their kind hospitality.

\appendix

\section{The S-matrix}
\label{app:matrixSmatrix}

In this appendix we summarize the properties of the $q$-deformed S-matrix. We use $E_{ij}$ to denote the $4\times 4$ $(i,j)$ matrix unity, {\it i.e.} a matrix with a one in the $(i,j)$th entry and zeroes everywhere else. Next, we introduce the following definition
\begin{equation}
E_{kilj}=(-1)^{\epsilon(l)\epsilon(k)}E_{ki}\otimes E_{lj}\, ,
\end{equation}
where $\epsilon(i)$ denotes the parity of the index, equal to $0$ for $i=1,2$ (bosons) and to $1$ for $i=3,4$ (fermions). The matrices $E_{kilj}$
can be used to write down invariance with respect to the action of two copies of $U_q(\su(2))$. If we introduce
\bea
\Lambda_1&=&E_{1111}+\frac{q}{2}E_{1122}+\frac{1}{2}(2-q^2)E_{1221}+\frac{1}{2}E_{2112}+\frac{q}{2}E_{2211}+E_{2222}\, ,\nonumber\\
\Lambda_2&=&\frac{1}{2}E_{1122}-\frac{q}{2}E_{1221}-\frac{1}{2q}E_{2112}+\frac{1}{2}E_{2211}\, , \nonumber \\
\Lambda_3&=&E_{3333}+\frac{q}{2}E_{3344}+\frac{1}{2}(2-q^2)E_{3443}+\frac{1}{2}E_{4334}+\frac{q}{2}E_{4433}+E_{4444} \, , \nonumber\\
\Lambda_4&=&\frac{1}{2}E_{3344}-\frac{q}{2}E_{3443}-\frac{1}{2q}E_{4334}+\frac{1}{2}E_{4433}\, , \nonumber\\
\Lambda_5&=&E_{1133}+E_{1144}+E_{2233}+E_{2244}\, ,\\
\Lambda_6&=&E_{3311}+E_{3322}+E_{4411}+E_{4422}\, , \nonumber\\
\Lambda_7&=&E_{1324}-qE_{1423}-\frac{1}{q}E_{2314}+E_{2413}\, , \nonumber\\
\Lambda_8&=&E_{3142}-qE_{3214}-\frac{1}{q}E_{4132}+E_{4231}\, , \nonumber\\
\Lambda_9&=&E_{1331}+E_{1441}+E_{2332}+E_{2442}\, , \nonumber\\
\Lambda_{10}&=&E_{3113}+E_{3223}+E_{4114}+E_{4224}\, , \nonumber
\eea
the S-matrix of the $q$-deformed model is given by
\bea
S_{12}(p_1,p_2)=\sum_{k=1}^{10}a_k(p_1,p_2)\Lambda_k\, ,
\eea
where the coefficients are
\bea
a_1&=&1\, ,  \nonumber \\
a_2&=&-q+\frac{2}{q}\frac{x^-_1(1-x^-_2x^+_1)(x^+_1-x^+_2)}{x^+_1(1-x^-_1x^-_2)(x^-_1-x^+_2)}\nonumber \\
a_3&=&\frac{U_2V_2}{U_1V_1}\frac{x^+_1-x^-_2}{x^-_1-x^+_2}\nonumber \\
a_4&=&-q\frac{U_2V_2}{U_1V_1}\frac{x^+_1-x^-_2}{x^-_1-x^+_2}+\frac{2}{q}\frac{U_2V_2}{U_1V_1}\frac{x^-_2(x^+_1-x^+_2)(1-x^-_1x^+_2)}{x^+_2(x^-_1-x^+_2)(1-x^-_1x^-_2)}\nonumber \\
a_5&=&\frac{x^+_1-x^+_2}{\sqrt{q}\, U_1V_1(x^-_1-x^+_2)}
\\
\nonumber
a_6&=&\frac{\sqrt{q}\, U_2V_2(x^-_1-x^-_2)}{x^-_1-x^+_2} \\
a_7&=&\frac{ig}{2}\frac{(x^+_1-x^-_1)(x^+_1-x^+_2)(x^+_2-x^-_2)}{\sqrt{q}\, U_1V_1(x^-_1-x^+_2)\gamma_1\gamma_2}
\nonumber \\
a_8&=&\frac{2i}{g}\frac{U_2V_2\,  x^-_1x^-_2(x^+_1-x^+_2)\gamma_1\gamma_2}{q^{\frac{3}{2}} x^+_1x^+_2(x^-_1-x^+_2)(x^-_1x^-_2-1)}\nonumber \\
a_9&=&\frac{(x^-_1-x^+_1)\gamma_2}{(x^-_1-x^+_2)\gamma_1} \nonumber \\
\nonumber
a_{10}&=&\frac{U_2V_2 (x^-_2-x^+_2)\gamma_1}{U_1V_1(x^-_1-x^+_2)\gamma_2}\, .
\eea
Here the central charges are given by
\bea
U_i^2=\frac{1}{q}\frac{x^+_i+\xi}{x^-_i+\xi}\, , ~~~~V^2_i=q\frac{x^+_i}{x^-_i}\frac{x^-_i+\xi}{x^+_i+\xi} \, ,
\eea
and the parameters $\gamma_i$ are
\bea
\gamma_i=q^{\frac{1}{4}}\sqrt{\frac{ig}{2}(x^-_i-x^+_i)U_iV_i}\, .
\eea
The dependence of the S-matrix on the variables $\gamma_i$, $i=1,2$, is gauge-like. Indeed, introducing the diagonal matrix
$\Gamma_i={\rm diag}(1,1,\gamma_i,\gamma_i)$, we find
\bea
\Big[ \Gamma_1\otimes \Gamma_2\Big]\,  S_{12}^{\gamma_i=1}(\z_1,\z_2)\, \Big[\Gamma_1^{-1}\otimes \Gamma_2^{-1}\Big]=S_{12}(\z_1,\z_2)\, ,
\eea
where $S_{12}^{\gamma_i=1}$ is the S-matrix where $\gamma_1$ and $\gamma_2$ are set to one.

\medskip

Let us summarize the most important properties of the S-matrix. The S-matrix satisfies
\begin{itemize}
\item The Yang-Baxter equation;

\item The unitarity condition
\bea
S_{21}(\z_2,\z_1)S_{12}(\z_1,\z_2)=1;
\eea

\item The transposition property
\bea
S^t(\z_1,\z_2)=\mI^g \Omega \, S(\z_1,\z_2) \, \Omega^{-1} \mI^g\, ,
\eea
where $\mI^g=(-1)^{\epsilon_i\epsilon_j}E_i^i \otimes E_j^j$ is the graded identity and $\Omega$ is given by
\bea
\Omega=\exp \frac{i\pi}{2}\Big(E_1\otimes F_1+F_1\otimes E_1+E_3\otimes F_3+F_3\otimes E_3\Big)\, .
\eea
Here $E_i,F_i$ are positive and negative roots of two $\su(2)$'s. The graded identity commutes with $\Omega$. In the limit
$q\to 1$ one finds
$$
\lim_{q\to 1}\Big[\Omega\, S(\z_1,\z_2) \, \Omega^{-1}\Big]= \lim_{q\to 1}S(\z_1,\z_2) \, .
$$
Note also the formula $x^+(\z,q)=-x^{-}(-\z,1/q)$.

\item Generalized physical pseudo-unitarity

\begin{equation}
\label{eq:pseudounitarity}
S(\z_1,\z_2)^\dagger = B S^{-1}(\z_1^*,\z_2^*) B^{-1}
\end{equation}
where\footnote{Our matrix $B$ is Hermitian as is required for pseudo-unitarity \cite{Mostafazadeh:2001A}.}
\begin{equation}
\label{eq:pseudounitaritylocal}
B = A\otimes A \, , \, \, \mbox{with} \, \, A = \mbox{diag}(\sigma_1,\sigma_1)\,
\end{equation}
and $\sigma_1 = \left(\begin{array}{cc} 0 & 1 \\ 1& 0\end{array}\right)$ is the first Pauli matrix.

In addition, the S-matrix has a unitary spectrum on the real line of the string and mirror theory, but is generically not unitarizable by a local basis transformation. In fact the many body S-matrix on the real string line has non-unitary eigenvalues. Nonetheless, on the mirror line the many body S-matrix appears to have unitary eigenvalues always\footnote{We have checked this numerically for the first few $n$-body S-matrices.}. This is equivalent to being quasi-unitary, where quasi-unitarity means that the matrix $B$ above is positive definite; $B=O O^\dagger$ \cite{Mostafazadeh:2001B}. However, there does not appear to be a matrix $B$ of this form which is also factorizable over the one-particle basis, and as such we have not rigorously proved quasi-unitarity on the mirror line.

\item For coincident arguments the S-matrix reduces to the permutation.

\item It is compatible with crossing symmetry. If we introduce the following
charge conjugation matrix
{\small
$$
C=\left(\begin{array}{cccc}  0 & -i q^{1/2} & 0 & 0\\
i q^{-1/2} & 0 & 0 & 0 \\
0 & 0 & 0 & q^{1/2} \\
0 & 0 & -q^{-1/2} & 0
\end{array}\right)\, ,
$$
}
the crossing relation reads\footnote{Under the crossing transformation $x^{\pm}\to 1/x^{\pm}$ so that the central charges transform as
$U^2\to 1/U^2$ and $V^2\to 1/V^2$.}
\bea
S_{12}(\z_1,\z_2)C_2 S_{12}^{t_2}(\z_1,\z_2-w_2)C_2^{-1}=\frac{1}{q}
\frac{(x_1^+-x_2^-)\Big(1-\frac{1}{x_1^+x_2^+}\Big)}{(x_1^--x_2^-)\Big(1-\frac{1}{x_1^-x_2^+}\Big)}\, .
\eea
\end{itemize}

\vskip 0.3cm

The complete S-matrix comprising two copies of $S$ can be written in the form\footnote{In \cite{Hoare:2011wr} the factor $\frac{x_1^+ x_2^-}{x_1^-x_2^+}$ in the S-matrix was replaced by $\frac{U_1^2}{U_2^2}=\frac{x_1^++\xi}{x_1^-+\xi}\frac{x_2^-+\xi}{x_2^++\xi}$. This is a minor change which leads to the corresponding modification of the crossing equation for $\sigma$.}
\bea
&&\hspace{-1.5cm}
{\mathbf S}=S_{\su(2)} S\hat{\otimes} S\, , ~~~S_{\su(2)}=\frac{1}{\sigma(\z_1,\z_2)^2}\frac{x_1^+}{x_1^-}\frac{x_2^-}{x_2^+}\cdot
\frac{x_1^--x_2^+}{x_1^+-x_2^-}\frac{1-\frac{1}{x_1^-x_2^+}}{1-\frac{1}{x_1^+x_2^-}}\, ,
\eea
where $\hat{\otimes} $ stands for the graded tensor product and $\sigma$ is the dressing phase \cite{Arutyunov:2004vx,Beisert:2006ez}. Substituting this representation for ${\mathbf S}$  into the crossing equation \cite{Janik:2006dc}, we deduce that the dressing phase must obey the following equation
\bea
\label{cross1}
\sigma(\z_1,\z_2)\sigma(\z_1,\z_2-\omega_2)=q^{-1}\frac{x_1^+}{x_1^-}\frac{x_1^--x_2^+}{x_1^--x_2^-}\frac{1-\frac{1}{x_1^+x_2^+}}{1-\frac{1}{x_1^+x_2^-}}\, .
\eea
Using the unitarity relation the previous formula also implies that
\bea
\label{cross2}
\sigma(\z_1+\omega_2,\z_2)\sigma(\z_1,\z_2)=q^{-1}\frac{x_2^-}{x_2^+}\cdot \frac{x_1^--x_2^+}{x_1^--x_2^-}\frac{1-\frac{1}{x_1^+x_2^+}}{1-\frac{1}{x_1^+x_2^-}}\, .
\eea
The formulae (\ref{cross1}) and (\ref{cross2})
are $q$-deformed analogue of eqs. (2.8), (2.9) and (2.10) from \cite{Arutyunov:2009kf}. A natural solution of this equation has been obtained in \cite{Hoare:2011wr}, where an explicit expression for $\sigma$ can be found, generalizing the DHM representation of the undeformed dressing phase \cite{Dorey:2007xn}.

\smallskip

The dressing factor $\sigma^{PQ}$ which describes scattering of $P$- and $Q$-particle bound states can be obtained from the dressing factor
for fundamental particles by means of fusion. The corresponding crossing equations have almost the same form as in the undeformed case, {\it cf.} eq.(2.14) in \cite{Arutyunov:2009kf},
\bea
\sigma^{PQ}(z_1,z_2)\sigma^{PQ}(z_1-\omega_2,z_2)&=&q^{-PQ}\left(\frac{x_1^+}{x_1^-}\right)^Q h^{PQ}(z_1,z_2)\, ,\\
\sigma^{PQ}(z_1,z_2)\sigma^{PQ}(z_1+\omega_2,z_2)&=&q^{-PQ}\left(\frac{x_2^-}{x_2^+}\right)^P h^{PQ}(z_1,z_2)\, .
\eea
Here $x_1$ and $x_2$ solve the bound state conditions
\bea
\frac{1}{q^P}\left(x^+_1+\frac{1}{x^+_1}\right)-q^P\left(x^-_1+\frac{1}{x^-_1}\right)&=&\left(q^P-\frac{1}{q^P}\right)\left(\xi+\frac{1}{\xi}\right)\, \, ,\\
\frac{1}{q^Q}\left(x^+_2+\frac{1}{x^+_2}\right)-q^Q\left(x^-_2+\frac{1}{x^-_2}\right)&=&\left(q^Q-\frac{1}{q^Q}\right)\left(\xi+\frac{1}{\xi}\right)\, , \eea
and we have introduced the crossing function
\bea
\label{crossfunc}
h^{PQ}=\frac{x_1^--x_2^+}{x_1^--x_2^-}\frac{1-\frac{1}{x_1^+x_2^+}}{1-\frac{1}{x_1^+x_2^-}} \prod_{j=1}^{P-1} S_{Q-P+2j}\, ,
\eea
where $S_Q$ is defined in \eqref{eq:SM}.

\section{String hypothesis for the deformed Hubbard model}
\label{app:Stringhypo}

As explained in the main text, in the limit $R \rightarrow \infty$ we obtain $Q$-particle bound states. As we will show here, going down one level and considering the limit $K^{\mathrm{I}} \rightarrow \infty$ yields $vw$-strings, while going down two levels and considering the limit $K^{\mathrm{II}} \rightarrow \infty$ gives $w$-strings. While not in the strictly logical order, we will start by discussing the string hypothesis for $w$-strings because it gives the clearest illustration of the effects of the deformation.

\subsection{$y$-particles, $w$-strings, and the XXZ model}
\label{subsec:StringHypo-w}

To describe the string hypothesis at the level of the second auxiliary Bethe equation, we need to know which type of $y$-roots can appear there. This is derived in the section below, where we will see that $|y| \neq 1$ necessarily leads to the emergence of a $vw$-string so that $y$-particles are characterized by the condition $|y|=1$. With this condition, the equations for $w$ particles become nothing but the Bethe equations of the inhomogeneous XXZ spin chain, where the inhomogeneities are unitary defects. As such the results here closely mimic the string hypothesis for the XXZ spin chain \cite{Takahashi:1972,Takahashi:book}, but the derivation is slightly different in form.

The discussion is simplest if we consider the second auxiliary Bethe equation in the form (\ref{eq:auxw})
\begin{eqnarray}
-1= \prod_{i=1}^{K^{\mathrm{II}}} \frac{q \W_k - \V_i}{\W_k - q \V_i} \prod_{i=1}^{K^{\mathrm{III}}}\frac{\W_k - q^2 \W_i}{ q^2 \W_k - \W_i }\, ,
\end{eqnarray}
where in this notation $|y_i|=1$ implies that the $\V_i$ are positive. We want to consider the types of solutions these equations admit in the limit $K^{\mathrm{II}} \rightarrow \infty$. Now for $q$ on the upper half unit circle and $\V$ positive we have
\begin{equation}
\left | \frac{q \W - \V}{\W - q \V}  \right | >1 ~~~\mbox{for}~~~\mbox{Im}\, \W>0
\end{equation}
and vice versa. Thus, the product over $y$-particles diverges or goes to zero unless $\W$ is real. Without loss of generality we will start with a $\W$ root, say $\W_1$, for which the product in the bracket below diverges
\begin{equation}
\label{div1}
-1=\left(\prod_{i=1}^{K^{\mathrm{II}}}\frac{q\W_1-\V_i}{\W_1-q\V_i}\right) \prod_{i=1}^{K^{\mathrm{III}}}\frac{\W_1-q^2\W_i}{q^2\W_1-\W_i}\, .
\end{equation}
The only way this root can be part of a solution of the Bethe equations is if this divergence is compensated by a zero in the second product in the Bethe equation, say
\begin{equation}
\label{w2string}
\W_1 - q^2 \W_2 = 0 \, .
\end{equation}
However we should now also ensure that there are no problems in the equation for $\W_2$
\begin{equation}
\label{div2}
-1=\left(\prod_{i=1}^{K^{\mathrm{II}}}\frac{q\W_2-\V_i}{\W_2-q\V_i}\right)\prod_{i= 1}^{K^{\mathrm{III}}}\frac{\W_2-q^2\W_i}{q^2\W_2-\W_i}\, .
\end{equation}
This can be determined by multiplying the two equations (\ref{div1}) and (\ref{div2}), giving
\begin{equation}
\label{div3}
1=\prod_{i=1}^{K^{\mathrm{II}}}\frac{q\W_1-\V_i}{\W_1-q\V_i}\frac{q\W_2-\V_i}{\W_2-q\V_i}
\prod_{i\neq 1,2}^{K^{\mathrm{III}}}\frac{\W_1-q^2\W_i}{q^2\W_1-\W_i}\frac{\W_2-q^2\W_i}{q^2\W_2-\W_i}\, ,
\end{equation}
whereby the divergent term in (\ref{div2}) originating from (\ref{w2string}) cancels out. At this point there are two possibilities. Either the equation for $\W_2$ is finite in the limit $K^{\mathrm{II}}\to\infty$, which is equivalent to
\begin{equation}
\left|\frac{q \W_1 - \V}{\W_1 - q \V}\frac{q \W_2 - \V}{\W_2 - q \V}\right| = \left|\frac{q^3 \W_2 - \V}{\W_2 - q \V}\right| = 1 \, ,
\end{equation}
giving a string of length two, or this product is greater than one and the process continues to give a longer string. In the latter case we necessarily have to involve another root to find
\begin{equation}
\W_2 - q^2 \W_3 = 0 \, ,
\end{equation}
and analyze its equation in turn. Continuing along these lines we see that a general string configurations of $w$-roots are given by
\begin{equation}
\{\W\}=\{q^{M+1-2j}\W\}\, ,~~~~j=1,\ldots, M\, .
\end{equation}
where $M$ denotes the length of the string. Consistency of the resulting string solution requires that the center of the string is real, as can be seen via fusion, see appendix \ref{app:Smatrices}. This is not the only constraint however.

As we saw above, in order for our string configuration to keep growing, we must keep a divergent product at every stage of the derivation. This means that in order to obtain a string of length $M$ we must have
\begin{equation}
\prod_{k=1}^j \left|\frac{q \W_k - \V}{\W_k - q \V}\right|=\left |  \frac{q^j(\V-q^M \W)}{q^{2j}\V-q^M \W}\right | > 1 \, , \, \, \mbox{for} \, \, j=1,...,M-1 \, .
\end{equation}
Taking into account that $\V$ is positive and $q=e^{i \frac{\pi}{k}}$, we find that the last condition is equivalent to the following requirement
\begin{equation}
\W \cos{\Big(M \frac{\pi}{k}\Big)}  < \W \cos{\Big((M-2j) \frac{\pi}{k}\Big)}  \, , \, \, \mbox{for} \, \, j=1,...,M-1 \, .\end{equation}
Given these conditions, we can have strings of length one through $k-1$ for a positive center ($\W>0$) and a single type of string of length one with negative center ($\W<0$). We will refer to the positivity or negativity of the string center as positive and negative parity respectively. In terms of the $u$-plane parametrization, positive parity strings have real rapidity $w$, while the rapidities of negative parity strings lie on the line $+i k/g$.

Note that the positive parity string of length $k$ lies just on the edge of these conditions and is not allowed; there would have been no divergence to begin with. However, were we to perturb slightly away from $k$ being an integer and let $k$ be $m +\epsilon$, of course it could become an allowed string configuration; correspondingly in the limit where $k$ is exactly an integer this string configuration has fixed momentum $\pi$ and trivial scattering with all other roots, which is another way to see it plays no role in thermodynamics.

The constraints on the length of strings in the XXZ model were first postulated in \cite{Takahashi:1972} and later shown to be equivalent to normalizability of the Bethe wave function of the associated string configuration \cite{hida:1981}. The argument of normalizability of the wave function would of course also apply here.

\subsection{$vw$-strings}

\label{subsec:StringHypo-vw}

Now that we have seen how to analyze the string solutions of the equations for $w$ roots, let us take a look at the equations for $y$ roots in the limit $K^{\mathrm{I}} \rightarrow \infty$. To analyze this limit, we need to analyze the product over $Q$ particle excitations in \eqref{eq:auxy}. This product consists of terms of the form
\begin{equation}
S^{Qy}=q^{Q/2}\,\frac{x^--y}{x^+-y}\sqrt{\frac{x^+}{x^-}}\,,
\end{equation}
where $x^{\pm}$ describes a $Q$-particle bound state with real rapidity $u$. Due to the conjugation properties in the mirror theory (\ref{cr}) the ratio $x^+/x^-$ is real. Hence the modulus of $S^{Qy}$ is given by
\begin{equation}
\left|S^{Qy}\right|^2=\frac{y^*-\frac{1}{x^+}}{y^*-\frac{1}{x^-}}\frac{y-x^-}{y-x^+}\frac{x^+}{x^-}=
\frac{|y|^2\frac{x^+}{x^-}+1-y^*x^+-\frac{y}{x^-}}{|y^2|+\frac{x^+}{x^-}-y^*x^+-\frac{y}{x^-}}\, .
\end{equation}
Hence this modulus is smaller than one provided
\begin{equation}
\left(1-\frac{x^+}{x^-}\right)(1-|y|^2)<0\, .
\end{equation}
Now for any bound state number $Q=1,\ldots, 2k-1$ and real $u$ we have $x^+/x^-<1$. So $|S^{Qy}|<1$ for $|y|>1$ and vice versa. As in the undeformed case, this implies that in the limit $K^{\mathrm{I}} \rightarrow \infty$ the product over the momentum carrying particles in \eqref{eq:auxy} diverges when $|y_k|<1$ and goes to zero when $|y_k|>1$. Auxiliary roots that have $|y_k|=1$ are identified as $y$-particles and $w$-strings exist in their presence as we discussed above; here we assume that  $|y_k|\neq1$.

\smallskip

Without loss of generality we take $|y_1|<1$. Then in the limit $K^{\mathrm{I}} \rightarrow \infty$ eq.\eqref{eq:auxy} becomes
\begin{equation}
\label{eq:auxyzero}
1  =\frac{1}{0} \times \prod_{i=1}^{K^{\mathrm{III}}}\frac{\sinh{\frac{\pi g}{2k}(v_k-w_i-\frac{i}{g})}}{\sinh{\frac{\pi g}{2k}( v_k-w_i + \frac{i}{g})}} \,,
\end{equation}
which requires one of the numerators in the product on the right hand side of the equation to vanish. This can be achieved by taking for instance
\begin{equation}
\label{eq:1stVWcondition}
v_1 - w_1 -\frac{i}{g} = 0 \, .
\end{equation}
We should also satisfy the equation for $w_1$ however, which now gives
\begin{equation}
1  =0 \times \prod_{i=2}^{K^{\mathrm{II}}} \frac{\sinh{\frac{\pi g}{2k}(w_1 - v_i+\frac{i}{g})}}{\sinh{\frac{\pi g}{2k}(w_1 - v_i - \frac{i}{g})}} \prod_{j=1}^{K^{\mathrm{III}}}\frac{\sinh{\frac{\pi g}{2k}(w_1 - w_j-\frac{2i}{g})}}{\sinh{\frac{\pi g}{2k}(w_1 - w_j+\frac{2i}{g})}}\, ,
\end{equation}
so that we should take\footnote{Taking a denominator in the product over $w$ roots to vanish would give $v_1 + \frac{i}{g}= w_2$ which contradicts the original considerations in \eqref{eq:auxyzero}, since the zero in the equation for $w_1$ by assumptions scales as the one in \eqref{eq:auxyzero}.}
\begin{equation}
\label{eq:1vwstring}
w_1 -v_2 - \frac{i}{g} = 0\, .
\end{equation}
Now $y_2$ has become involved in the game, and there are two options. If $|y_2|>1$ the equation for $y_2$ is `satisfied' and we get a $1|vw$ string.

\smallskip

In case $|y_2|<1$ we get further conditions which we explicitly work out one further level before giving the general pattern. Assuming $|y_2| <1$, analogous to \eqref{eq:auxyzero} and \eqref{eq:1stVWcondition} we take
\begin{equation}
\label{eq:2ndWinVW}
v_2 - w_2 - \frac{i}{g}=0\, .
\end{equation}
However we now have the equation for $w_2$ as well, which tells us
\begin{equation}
1  = 0 \times \prod_{i\neq 2}^{K^{\mathrm{II}}} \frac{\sinh{\frac{\pi g}{2k}(w_2 - v_i+\frac{i}{g})}}{\sinh{\frac{\pi g}{2k}(w_2 - v_i - \frac{i}{g})}} \prod_{j=1}^{K^{\mathrm{III}}}\frac{\sinh{\frac{\pi g}{2k}(w_2 - w_j-\frac{2i}{g})}}{\sinh{\frac{\pi g}{2k}(w_2 - w_j+\frac{2i}{g})}}\, ,
\end{equation}
so that $w_2 - v_3 -\frac{i}{g} = 0$. However we should not forget that from \eqref{eq:1vwstring} and \eqref{eq:2ndWinVW} we now also get
\begin{equation}
w_2 =  w_1 -\frac{2i}{g}\, ,
\end{equation}
meaning we should additionally have a numerator vanish\footnote{Note that we have  $w_2 = w_1 - \frac{2i}{g}$ only `after' imposing \eqref{eq:2ndWinVW}; it cannot fix the original vanishing denominator.} by imposing a condition
\bea
w_2 - v_4 + \frac{i}{g} = 0\, .
\eea
At this point we have the configuration
\begin{equation}
\label{eq:2vwstring}
v_1 - \frac{2i}{g} =  v_2 = w_1 -\frac{i}{g} =  w_2 + \frac{i}{g} = v_3 + \frac{2i}{g}= v_4 \, ,
\end{equation}
The configuration \eqref{eq:2vwstring} forms a $2|vw$ string if $|y_{3,4}|<1$. If $|y_3|>1$ or $|y_4|>1$ we get more conditions, and in general we can end up with an $M|vw$ string given by the following configuration of roots
\begin{eqnarray}
\{ w \}&=&\left\{ v+(M-1)\frac{i}{g},\ldots ,  v -(M-1)\frac{i}{g}\right\}\, ,
\nonumber \\
\label{vwsol}
\{v\}&=&\left\{v +M\frac{i}{g},\,  \underbrace{v+(M-2)\frac{i}{g},\ldots , v-(M-2)\frac{i}{g}}_{M-1},\, v-M\frac{i}{g}\right\}\, .
\end{eqnarray}
Note that there are two $y$-roots $y_j,1/y_j$ associated to each of the underbraced rapidities
$$v_j=v+(M-2j),~~~~ j=1,\ldots\, , M-1$$ in a $M|vw$-string. We denote by
$y_M$ and $y_{-M}$ the roots associated with the first and the last rapidities in (\ref{vwsol}), respectively.
According to our construction, $|y_M|<1$ and $|y_{-M}|>1$. In total there are $2M$ $y$-roots.

\smallskip

As for $w$-strings, the reality conditions for the string centers follow by fusing the S-matrices and insisting that the S-matrix for the total $M|vw$ string is unitary. Fusing the S-matrix $S^{Qy}$ over the constituents $y$-roots of an $M|vw$ string gives
\begin{align}
\prod_{i\in M|vw} S^{Qy}(u,v) & = q^{QM}\frac{x^--y_M}{x^+-y_M}\frac{x^--y_{-M}}{x^+-y_{-M}}\frac{x^+}{x^-}
\prod_{i=1}^{M-1} \frac{x^--y_i}{x^+-y_i}\frac{x^--\frac{1}{y_i}}{x^+-\frac{1}{y_i}}\frac{x^+}{x^-}  \nonumber \\
& = q^{QM}\frac{x^--y_M}{x^+-y_M}\frac{x^--y_{-M}}{x^+-y_{-M}}\frac{x^+}{x^-}
\prod_{i=1}^{M-1} \frac{x^-+\frac{1}{x^-}-y_i-\frac{1}{y_i}}{x^++\frac{1}{x^+}-y_i-\frac{1}{y_i}}\nonumber \\
& = q^{QM}\frac{x^--y_M}{x^+-y_M}\frac{x^--y_{-M}}{x^+-y_{-M}}\frac{x^+}{x^-}
\prod_{i=1}^{M-1} \frac{q^{-Q}e^{\frac{\pi g u}{k} }-q^{M-2i}e^{\frac{\pi g v}{k} }}{q^Q e^{\frac{\pi g u}{k} }-q^{M-2i}e^{\frac{\pi g v}{k} }}  \nonumber \\
& = q^{QM}\frac{x^--y_M}{x^+-y_M}\frac{x^--y_{-M}}{x^+-y_{-M}}\frac{x^+}{x^-}
\prod_{i=1}^{M-1} q^{M-Q-2i} S_{Q+M-2i}(u-v) \\
& = q^{Q}\frac{x^--y_M}{x^+-y_M}\frac{x^--y_{-M}}{x^+-y_{-M}}\frac{x^+}{x^-}
\prod_{i=1}^{M-1} S_{Q+M-2i}(u-v)\, .
\end{align}
Requiring this S-matrix to be unitary requires that $y_M^*=\frac{1}{y_{-M}}$ and $u$ to be real modulo $i k/g$.

At this point we do not have a restriction on the string length yet. This is because at every stage there are two $y$ roots associated to each rapidity, which means we can always choose the one with the desired properties. However, going beyond length $k$ it is not hard to see that we must necessarily have coincident roots. For $w$ roots however, we had no length $k$ string either. Again here the peculiarities of $y$-roots do not allow us to dismiss the length $k$ string; we can obtain a bona-fide divergence and repeat the arguments above. However, similarly to the case of $w$-strings the would-be $k|vw$ string has fixed momentum and scatters trivially with everything else. This means our strings can have length up to $k-1$.

Next, coming back to the location of the center, it is easy to see that by translating a given root configuration by $2ik/g$ the set of all but the outer two $y$-roots is immediately invariant. Now since we have the constraints $|y_M|<1$ and $y_M^*=\frac{1}{y_{-M}}$, also the outermost $y$ roots cannot change either, and hence the center of the string can be taken to lie either on the real line or on the line $ik/g$.

At this point it appears we can have negative parity $vw$ strings of length up to $k-1$, while for $w$ strings this was not the case. Also this strange fact arises due to the curious property of the $y$-roots mentioned earlier; we can naively choose the correct $y$ roots to make our string hypothesis work. However, the root configurations that lead to negative parity strings of length greater than one are not consistent at the at the level of $w$ roots, where they violate the conditions we derived above\footnote{Of course the conditions on the $w$ roots immediately imply the conditions on the $vw$ strings derived above independently.}. This in particular implies that the Bethe wavefunction associated to a negative parity $vw$ string of length greater than one in the deformed Hubbard model subsystem would not be normalizable. Hence we are led to exclude negative parity string of length greater than one, and end up with a string content exactly analogous to the second auxiliary level.

\subsection*{Parametrization of $y$-roots}

As mentioned earlier, $x(u)$ runs over the lower half unit circle as $u$ runs from $-u_b$ to $u_b$. Therefore, a $y$-particle with negative imaginary part is parametrized by $x(u)$, while a $y$-particle with positive imaginary part is parametrized by $\frac{1}{x(u)}$, with $-u_b \leq u \leq u_b$. In other words
\begin{equation}
y^+(v) = \frac{1}{x(v)}\,, \,\,\,y^-(v) = x(v) \, .
\end{equation}
Next, in an $M|vw$ string of positive parity the outermost $y$ roots are $y_{\pm M}$, associated to the rapidity $v \pm i M/g$, with the requirement that $|y_M|<1$ and $|y_{-M}|>1$. These roots are therefore parametrized as
\begin{equation}
y^+_{\pm M} = x(v \pm i M/g)\, .
\end{equation}
From the properties of $x(u)$, see {\it e.g.} figure \ref{fig:xmirror}, it follows that $\mbox{Im}\, y_M^{\pm}<0$ for any $M=1,\ldots, k-1$. The negative parity string of length one has two $y$-roots associated to $v + i(k\pm1)/g$, with $|y_{1}|<1$ and $|y_{-1}|>1$. These roots are therefore parametrized as
\begin{equation}
y^-_{1} = x(v+i(k+1)/g) \, , \, \, \, y^-_{-1} = \frac{1}{x(v+i(k-1)/g)} \, .
\end{equation}
Here we have $\mbox{Im}\, y_{\pm 1}^{-}>0$.

\section{Derivation of the TBA equations}

\label{app:TBAderivation}

\subsection*{The Bethe-Yang equations in the thermodynamic limit}

As indicated in the main text, given a string hypothesis and the corresponding set of Bethe-Yang equations the derivation of the TBA equations becomes a text book story. For completeness we present it in this appendix. The first step is to find the integral equations for the hole and particle densities, the analogue of the Bethe-Yang equations in the thermodynamic limit. In the present context there is a subtlety regarding the definition of counting functions, so for this reason as well as general continuity let us quickly go through the general story.

The string hypothesis classifies the types of solutions that can arise to the Bethe-Yang equations in the thermodynamic limit. This means that the products over elementary excitations in the Bethe-Yang equations should arrange themselves as products running over the allowed string complexes containing the relevant excitation, where each string type of complex can occur a given number of times. In other words we have a product over the ($r$) allowed types of strings labeled by $M$, the number of strings of type denoted $N_{M}$, and a product over the constituents of individual strings, that is
\begin{equation}
\prod_{i=1}^{K} S (u_l, u_i)\rightarrow \prod_{M=1}^{r} \prod_{m=1}^{N_{M}} \prod_{j \in M_m} S (u_l, u_j)\, .
\end{equation}
Due to the structure of the string complexes, the S-matrices involved in the products over individual strings fuse to give the S-matrix that describes scattering of the entire complex in one go
\begin{equation}
\prod_{j \in M_m} S (u_l, u_j) \rightarrow S^M(u_l,u_m) \, ,
\end{equation}
where we labeled the central rapidity of the string $M_m$ by $u_m$. This gives us the Bethe-Yang equations for scattering of elementary excitations with string configurations. To describe scattering of string complexes between each other, we take a product of these Bethe-Yang equations over the external elementary excitations making up a string of length $N$
\begin{equation}
e^{-i p_l L} = \prod_{M=1}^{r} \prod_{m=1}^{N_{M}} S^M(u_l,u_m) \, \rightarrow \, e^{-i p_N L} = \prod_{M=1}^{r} \prod_{m=1}^{N_{M}} S^{NM}(u_N,u_m) \, ,
\end{equation}
where we labeled the momentum and rapidity of the string as $p_N$ and $u_N$ respectively. For the auxiliary equations, the analogue of $e^{i p_l}$ is the S-matrix for scattering of the higher level root with the root under consideration. The fused $S$-matrices implicitly used here are explicitly given in the next appendix, section \ref{app:Smatrices}.

Next, we take the logarithm of the Bethe-Yang equations. Through a fixed choice of branch for the logarithm this introduces an integer $I$ in each equation which labels the possible solutions of the equation
\begin{equation}
2 \pi I_N = L p_N  - i \sum_{M=1}^{r} \sum_{m=1}^{N_{M}} \log{S^{NM}(u_N,u_m)} \, ,
\end{equation}
Now in the thermodynamic limit the solutions of the Bethe-Yang equations become dense and it makes sense to generalize the integer $I$ to a function of the relevant momentum, the counting function, interpolating between the integers corresponding to different solutions of the Bethe-equations. It is important that this counting function is monotonic so that if increasing, the derivative of this counting function gives us the density of possible solutions as a function of momentum. If the derivative is negative, since we would like to think of densities as positive, we should change the sign of the definition of the counting function. By definition, the total solution density is then a sum of the particle and hole densities, corresponding to solutions occupied and left vacant respectively.

\smallskip

Simultaneously, on the right hand side of the logarithmic Bethe-Yang equations the sums over excitations solving the Bethe equations turn into integrals over particle densities
\begin{equation}
\sum_{M=1}^{r} \sum_{m=1}^{N_{M}}\rightarrow \sum_{M=1}^{r} \int d u \rho_M (u)
\end{equation}
In short, we take the logarithmic derivative of the Bethe equations and equate the result to the sum of the particle and hole density for the excitations under consideration. This gives us integral equations for particle and hole densities; the thermodynamic analogue of the Bethe-Yang equations
\begin{equation}
\label{eq:integralBYgeneral}
\mbox{sign}\left(\frac{d p^N}{du}\right)(\rho_N (v) + \bar{\rho}_N (v)) = \frac{L}{2\pi} \frac{d p^N }{du}  + \sum_{M=1}^{r} \int du \, K^{NM}(v,u)\, \rho_M (u) \, ,
\end{equation}
where $K^{NM} (v,u)\equiv \frac{1}{2 \pi i} \frac{d}{dv} \log{S^{NM}(v,u)}$. The above sign is precisely such that we have a monotonically increasing counting function in accordance with the string hypothesis, see {\it e.g.} \cite{Faddeev:1996iy}. This sign is especially important in the present context; it comes in due to remarkably nice properties of the scattering kernels following from the special nature of the deformation. Namely for $w$ and $vw$ strings we have
\begin{align}
\label{eq:Kproperties11}
\frac{d}{du} \log S_\chi^{0N}(u,v) & = - \frac{d}{du} \log S_\chi^{k-1,N}(u,v) \, ,\\
\label{eq:Kproperties12}
\frac{d}{du} \log S_\chi^{M0}(u,v) & = - \frac{d}{du} \log S_\chi^{M,k-1}(u,v) \, ,\\
\label{eq:Kproperties13}
\frac{d}{du} \log S^{00}(u,v) & = \frac{d}{du} \log S^{11}(u,v) \, ,
\end{align}
where the last equation holds simply by the definition of type $0$ $w$ and $vw$ strings, and $\chi$ is a generic label denoting any relevant S-matrix. These properties immediately show that negative parity strings have counting functions defined with a relative minus sign with respect to their positive parity cousins.

\smallskip

In light of the above properties it should be clear that the canonically defined kernels for negative parity $w$ and $vw$ strings are given by minus the kernels for their length $k-1$ positive parity cousins. Hence we introduce the following \emph{temporary} notation: repeated indices of type $M$ ($M$, $N$ and possibly $L$) do not quite denote the standard summation, but rather $a^M b_M = \sum_{p=1}^{k-1} a^p b_p -a^0 b_{k-1}$. Repeated indices of type $Q$ ($Q$, $P$ and possibly $R$) denote a standard sum with $Q$ running from one to $k$. Let us also define three types of convolution
\bea
f\star h(u,v)&=&\int_{-\infty}^{\infty}\, dt\, f(u,t)h(t,v) \, , \label{eq:stdconv}\\
f\, \hat{\star}\,  h(u,v)&=&\int_{-u_b}^{u_b}\, dt\, f(u,t)h(t,v)\, , \label{eq:hatconv}\\
f\, \check{\star}\,  h(u,v)&=&\int_{-\infty}^{-u_b}\, dt\, f(u,t)h(t,v)+\int_{u_b}^{\infty}\, dt\, f(u,t)h(t,v)\label{eq:checkconv}\, .
\eea
We mention again that all $S$-matrices are derived and defined in section \ref{app:Smatrices}. With these definitions we explicitly have the following equations

\paragraph{Q-particles}

\begin{equation}
\label{eq:intBYQ}
\rho_Q + \bar{\rho}_Q = \frac{R}{2\pi} \frac{d \tilde{p}}{du} + K_{sl(2)}^{QP} \star \rho_P + \sum_{\alpha=1}^2 \left( \sum_{\beta=\pm} K^{Qy}_{\beta} \hat{\star} \rho^{(\alpha)}_{y^\beta} +
 K^{QM}_{xv}\star \rho^{(\alpha)}_{M|vw} \right)\, ,
\end{equation}
where generically we have $K^A (u,v) \equiv \frac{1}{2\pi i} \frac{d}{du} \log S^A (u,v)$, except of course for $K^{QM}_{xv}$ at $M=0$.

\paragraph{$y$-particles}

\begin{equation}
\label{eq:intBYy}
\mp (\rho^{(\alpha)}_\pm + \bar{\rho}^{(\alpha)}_\pm) = \pm K^{yQ}_{\pm} \star \rho_Q + K^M \star \rho^{(\alpha)}_{M|vw} +
K^N\star \rho^{(\alpha)}_{N|w}\, ,
\end{equation}
where $K^{yQ}_{\beta}$ is special; $K^{yQ}_{\beta}(u,v) \equiv \beta \frac{1}{2\pi i} \frac{d}{du} \log S^{Qy}_{\beta} (v,u)$.

\paragraph{$w$-strings}

\begin{align}
\label{eq:intBYw}
\rho^{(\alpha)}_{M|w} + \bar{\rho}^{(\alpha)}_{M|w}& =  K^M \hat{\star} (\rho^{(\alpha)}_+ + \rho^{(\alpha)}_-) + K^{MN} \star \rho^{(\alpha)}_{N|w} \, , \, \,  M=1,\ldots,k-1,\\
\rho^{(\alpha)}_{0|w} + \bar{\rho}^{(\alpha)}_{0|w}& =  \rho^{(\alpha)}_{k-1|w} + \bar{\rho}^{(\alpha)}_{k-1|w}
\end{align}
where we made use of \eqref{eq:Kproperties11} to rewrite the kernels in the second equation, and of the fact that the counting function for a $0|w$ string is defined oppositely from those of the other $k-1$ to ensure positive particle densities.

\paragraph{$vw$-strings}

\begin{align}
\nonumber
\rho^{(\alpha)}_{M|vw} + \bar{\rho}^{(\alpha)}_{M|vw} & = K^{MQ}_{vwx} \star \rho_Q
-  K^M \hat{\star}(\rho^{(\alpha)}_+ + \rho^{(\alpha)}_-) -
 K^{MN} \star \rho^{(\alpha)}_{N|vw} \, , \, \, M=1,\ldots, k-1,\\
\label{eq:intBYvw}
\rho^{(\alpha)}_{0|vw} + \bar{\rho}^{(\alpha)}_{0|vw}& =\rho^{(\alpha)}_{k-1|vw} + \bar{\rho}^{(\alpha)}_{k-1|vw} \, ,
\end{align}
where $K^{MQ}_{vwx}(u,v) \equiv - \frac{1}{2\pi i} \frac{d}{du} \log S^{QM}_{xv}(v,u)$. Also here the counting functions for strings of type $0$ are opposite to those of the others.

\subsection*{Free energy and TBA equations}

To derive the TBA equations using the above integral equations for densities, we need to minimize the free energy at temperature $T=1/L$ given by
\bea
\label{eq:FreeEnergyGeneral1}
&&\mathcal{F}_\gamma (L)=\int {\rm d}u\,\left[ \sum_{Q=1}^{2k-1} \widetilde{{\cal E}}^Q(u)\rho_Q(u) - \frac{i\gamma}{L}\, \sum_{\alpha=1}^2(-1)^\alpha( \rho_{y^-}^{(\alpha)}(u)+ \rho_{y^+}^{(\alpha)}(u))  - \frac{S}{L} \right]\,.~~~
\eea
First we write the free energy compactly as
\bea
\label{eq:FreeEnergyGeneral2}
\mathcal{F}_\gamma(L)
=\int {\rm d}u\, \sum_{i}\left[ \widetilde{{\cal E}}_{i}\, \rho_i - \frac{i\gamma_i}{L}\, \rho_i  - \frac{1}{L} \alg{s}(\rho_i) \right]\,,
\eea
where $\widetilde{{\cal E}}_{j}$ and $\gamma_j$ are nonzero only for $Q$- and $y$-particles respectively. Next, varying the Bethe equations \eqref{eq:intBYQ}-\eqref{eq:intBYvw} gives
\begin{eqnarray}
\label{eq:BYvariation}
\delta\rho_i(u) + \delta\bar{\rho}_i(u) = K_{ij}\star\delta\rho_j \,,~~~~
\end{eqnarray}
where we note that the implicit sum over $j$ can be of the special kind introduced above. Let us explicitly mention that for $w$ and $vw$ strings we have the special case
\begin{eqnarray}
\label{eq:BYvariationspecial}
\delta\rho_{0|(v)w}(u) + \delta\bar{\rho}_{0|(v)w}(u) = \delta\rho_{k-1|(v)w}(u) + \delta\bar{\rho}_{k-1|(v)w}(u) = K_{k-1 j}\star\delta\rho_j \,,~~~~
\end{eqnarray}
which we will come back to shortly. Varying the entropy function then gives
\bea
\label{eq:EntropyVariation}
\delta \alg{s}(\rho_i) =(\epsilon_i-i\gamma_i)\delta\rho_i + \log\left(1+e^{i\gamma_i-\epsilon_i} \right)K_{ij}\star\delta \rho_j\,,
\eea
where we note again that the sum over $j$ can be special, while in the sum over $i$ we need to take into account \eqref{eq:BYvariationspecial} but that aside from this it is a regular sum. The pseudo-energies $\epsilon_i$ are defined as
\bea
e^{i\gamma_i-\epsilon_i} = \frac{\rho_i}{\bar{\rho}_i}\,.
\eea
Upon imposing $\delta\mathcal{F}_\gamma(L)=0$, we get the TBA equations
\bea
\label{eq:TBAgeneral}
\epsilon_j = L\, \widetilde{{\cal E}}_{j} - \log\left(1+e^{i\gamma_i-\epsilon_i} \right)\star K_{ij}\,.
\eea
For $0|w$ or $0|vw$ strings, we should take into account the special sum we introduced above. This means the concise notation of $K_{i0}$ actually stands for $-K_{i,k-1}$, and since $w$ and $vw$ strings carry no energy\footnote{Or viewed differently, their `energies' are related via (\ref{eq:Kproperties11}-\ref{eq:Kproperties13}).}, we then immediately see that
\begin{equation}
\label{eq:inverseYfunctions}
\epsilon_{0|(v)w} = -\, \epsilon_{k-1|(v)w}
\end{equation}
so that we have just $k-1$ independent pseudo-energies to consider for $w$ and $vw$ strings\footnote{Their precise relationship is different if we were to include chemical potentials, we may come back to this in more detail in our upcoming publication \cite{qdeftbapart2}.}. Note also that $\epsilon_{y^\pm}$ is defined only for $u \in (-u_b,u_b)$.

\smallskip

Taking into account that the entropy function can be written in the form
\bea
\label{eq:entropy2}
\alg{s}(\rho_i) =\frac{R}{2\pi}\frac{d\tilde{p}_i}{du}\log\left(1+e^{i\gamma_i-\epsilon_i} \right)
+ (\epsilon_i-i\gamma_i)\rho_i + \log\left(1+e^{i\gamma_i-\epsilon_i} \right)K_{ij}\star \rho_j\,,~~~~~
\eea
at the extremum \eqref{eq:TBAgeneral} the free energy is given by
\bea
\label{eq:FreeEnergy} \mathcal{F}_\gamma(L) =-\frac{R}{L}\int {\rm d}u\,
\sum_{i}\frac{1}{2\pi}\frac{d\tilde{p}_i}{du}\log\left(1+e^{i\gamma_i-\epsilon_i}
\right)\,.
\eea
Finally, since the ground state energy of the $q$-deformed `string' theory is related to the free energy of the q-deformed mirror model as
\bea
E_\gamma(L)= \lim_{R\rightarrow\infty}\frac{L}{R}\mathcal{F}_\gamma(L)\,,
\eea
we get
\bea
E_\gamma(L) &=&-\int {\rm d}u\, \sum_{Q=1}^{k}\frac{1}{2\pi}\frac{d\tilde{p}^Q}{du}\log\left(1+e^{-\epsilon_Q} \right)\,.
\eea

The explicit form of the TBA equations \eqref{eq:TBAgeneral} depends on the particle type; we give them in full detail in the main text. There we put them in a form where we have introduced $Y$-functions as
\begin{align}
e^{-\epsilon_Q}& = Y_Q \, ,\\
e^{\epsilon_{M|(v)w}} & = Y_{M|(v)w} \, ,\\
e^{\epsilon_{y^\pm}} & = Y_\pm \, ,
\end{align}
and note that $\gamma = i \pi$ for $y$-particles \cite{AF09b}. We would like to emphasize that the $k-1$ $w$ and $vw$ $Y$-functions enter the equations in a special way because of \eqref{eq:inverseYfunctions}. From this point onward in the main text, the indices $M$,$N$ and $L$ run from one to $k-1$, while $Q$,$P$ and $R$ run from one to $k$, and repeated indices always indicate a standard sum.

\section{S-matrices and kernels}

\subsection{Fusion of S-matrices}

\label{app:Smatrices}

In what follows, unless otherwise indicated the indices $M$, $N$ and $L$ run from one to $k-1$ while the indices $Q$,$P$ and $R$ run from one to $k$.

\subsection*{$\boldsymbol S_M$ and $\boldsymbol S_{MN}$}

We will encounter the fusion of the basic S-matrix in the equations for $y$-particles, $w$, and $vw$ strings, so let us discuss it first. The basic S-matrix $S_1$
\begin{equation}
S_1(u-v) \equiv \frac{\sinh{\frac{\pi g}{2k}(u - v - i/g)}}{\sinh{\frac{\pi g}{2k}(u - v + i/g)}} \, ,
\end{equation}
fuses as
\begin{equation}
\label{eq:SM}
S_M (u-v) \equiv \prod_{i \in (v)w}^M S_1(u_i-v) = \prod_{j\in (v)w}^M S_1(u-v_j) = \frac{\sinh{\frac{\pi g}{2k}(u - v - M i/g)}}{\sinh{\frac{\pi g}{2k}(u - v + M i/g)}}\,,
\end{equation}
where the product is over the $w$ roots making up an $M|(v)w$ string. This S-matrix describes the scattering of the $w$ part of an $M|(v)w$ string with a $y$-particle. Alternately, fusing in $u$ or $v$ over the $v$ roots making up a $M|vw$ string gives
\begin{equation}
\prod_{i \in v(w)}^M S_1(u_i-v) = \prod_{j\in v(w)}^M S_1(u-v_j) = S_{M+1}(u-v)S_{M-1}(u-v) = \prod_{i\in w}^M S_2(u_i-v) \, ,
\end{equation}
which as we noted in the last equality, also coincides with fusion of the $w$-particle scattering matrix over the $w$ roots making up an $M|(v)w$ string. By definition negative parity particles scatter with
\begin{equation}
S_0 (u-v)\equiv S_1(u+i k/g-v)\, ,
\end{equation}
with positive parity particles, and regularly between themselves as we will come back to shortly. We would like to note that the fundamental scattering matrix for negative parity particles is almost inverse to the one for a $k-1$ positive parity string, namely
\begin{equation}
\label{eq:negparSM}
S_0 (u-v) S_{k-1} (u-v) = -1\,.
\end{equation}
Next, to describe scattering of strings with one another this S-matrix is to be fused over the previously untouched argument. Note that there are two relevant cases for our considerations; either we fuse $S_M$ over the $v$ roots in a $vw$ string, or we have to fuse $S_{M+1}S_{M-1}$ over the $w$ roots in a $(v)w$ string. Of course either fusion gives the same result, as is immediately clear when interchanging the order of fusion, upon noting that the result is symmetric in the interchange of $M$ and $N$. Let us define $S_{MN}$ as
\begin{equation}
S_{MN} (u-v) \equiv \prod_{j\in v(w)}^N S_M(u-v_j) = \prod_{j\in v(w)}^N \prod_{i\in (v)w}^M S_1(u_i-v_j) \, ,
\end{equation}
where the first product runs over the constituents of an $N|(v)w$ string. This product can be rewritten as
\begin{eqnarray}
S^{MN}(u-v)=S_{M+N}(u-v)S_{|M-N|}(u-v)\prod_{m=1}^{\min{(M,N)}-1}S_{|M-N|+2m}^2(u-v)\, .
\end{eqnarray}
manifestly showing its symmetry under interchange of $M$ and $N$. Negative parity particles scatter with positive parity strings with
\begin{equation}
S_{0M} (u-v) \equiv S_{1M} (u+ik/g-v) = S^{-1}_{M+1} (u-v) S^{-1}_{M-1} (u-v)  \,,
\end{equation}
while between themselves they scatter as positive parity particles meaning
\begin{equation}
S_{00} (u-v) \equiv S_{11} (u-v) \,.
\end{equation}
Note that here we have
\begin{equation}
\label{eq:negparSMN}
S_{0M} (u-v) S_{k-1,M} (u-v) = 1\,.
\end{equation}
Finally, note that these S-matrices are trivial when one of the indices is equal to $k$
\begin{equation}
S_k = -1 \, , \, \, \, S_{Mk} = S_{kM} =1\,.
\end{equation}
This shows that at $k$ these solutions to the discrete Laplace equation have a natural boundary.

\subsection*{$\boldsymbol S^{yQ}$, $\boldsymbol S^{Qy}$ and $\boldsymbol S^{QM}_{xv}$}

Fusing the scattering matrix of $y^\pm$ particles with fundamental particles over a $Q$-particle bound state directly gives
\begin{align}
S_-^{yQ}(u,v)&= q^{Q/2} \, \frac{x(u) -x^-(v)}{x(u)-x^+(v)}\sqrt{\frac{x^+(v)}{x^-(v)}} \, , \\
S_+^{yQ}(u,v) &= q^{Q/2} \, \frac{\frac{1}{x(u)} -x^-(v)}{\frac{1}{x(u)}-x^+(v)}\sqrt{\frac{x^+(v)}{x^-(v)}} \, ,
\end{align}
where $x^\pm$ are the parameters for a $Q$-particle bound state; $x^\pm(v) = x(v \pm i Q/g)$, and the subscript $\pm$ in $S^{Qy}_\pm$ denotes the sign of the imaginary part of the $y$-particle under consideration. Analogously we define the S-matrices for scattering of bound states with $y$-particles as
\begin{align}
S_-^{Qy}(u,v)&= q^{Q/2} \, \frac{x^-(u)-x(v)}{x^+(u)-x(v)}\sqrt{\frac{x^+(u)}{x^-(u)}} \, , \\
S_+^{Qy}(u,v) &= q^{Q/2} \, \frac{x^-(u)-\frac{1}{x(v)}}{x^+(u)-\frac{1}{x(v)}}\sqrt{\frac{x^+(u)}{x^-(u)}} \, .
\end{align}
As shown above in section \ref{subsec:StringHypo-vw}, the scattering matrix of a $Q$-particle bound state with an $M|vw$ string of positive parity is given by
\begin{equation}
\label{eq:SQMxv}
S^{QM}_{xv}(u,v) \equiv q^{Q}\frac{x^-(u)-x^+(v) }{x^+(u)-x^+(v) }\frac{x^-(u)-x^-(v) }{x^+(u)-x^-(v)}\frac{x^+(u)}{x^-(u)}
\prod_{i=1}^{M-1} S_{Q+M-2i}(u-v)
\end{equation}
where $x^\pm(v) = x(v \pm i M/g)$, while for length one $vw$ strings with negative parity we have simply
\begin{equation}
S^{Q0}_{xv}(u,v) \equiv q^{Q}\frac{x^-(u)-x(v+i(k+1)/g) }{x^+(u)-x(v+i(k+1)/g) }\frac{x^-(u)-\frac{1}{x(v+i(k-1)/g)} }{x^+(u)-\frac{1}{x(v+i(k-1)/g)}}\frac{x^+(u)}{x^-(u)} \, .
\end{equation}
Again there is a special relation between these S-matrices, namely
\begin{equation}
\label{eq:negparSQMxv}
S^{Q0}_{xv}(u,v) S^{Qk-1}_{xv}(u,v) = (-1)^Q \, .
\end{equation}

\subsection*{$\boldsymbol S_{\mathfrak{sl}(2)}^{QP}$}

In the present paper we will not derive the expression for the fused dressing factor, but assume structural analogy to the undeformed case and the effects of the deformation on the other kernels. Introducing the same split as in the undeformed case we write
\begin{equation}
S_{\mathfrak{sl}(2)}(x_1,x_2) = S_2^{-1} \Sigma(x_1,x_2)^{-2}\, ,~~
\end{equation}
where we have introduced the improved dressing factor $\Sigma$
\begin{equation}
\Sigma(x_1,x_2)\equiv \frac{1-\frac{1}{x_1^+x_2^-}}{1-\frac{1}{x_1^-x_2^+}}\sigma(x_1,x_2)\, .
\end{equation}
We then assume these factors to fuse similarly to the undeformed case, to give
\begin{equation}
S^{QM}_{\mathfrak{sl}(2)}= S_{QM}^{-1} \Sigma_{QM}^{-2}\, ,~~
\end{equation}
where now $\Sigma_{QM}$ should be a solution of the discrete Laplace equation in the bulk, and on the boundary at one with a source term analogous to the undeformed case \cite{AF09d}.

\subsection{Kernels and their properties}

\label{app:kernels}

\subsection*{$\boldsymbol K_M$ and $\boldsymbol K_{MN}$}

As in the main text, we have
\begin{equation}
\label{eq:KM}
K_M(u) \equiv \frac{1}{2\pi i} \frac{d}{du} \log{S_M(u)} = \frac{g}{2k} \frac{\sin{\frac{M\pi}{k}}}{\cosh{\frac{\pi g u}{k}}-\cos{\frac{M\pi}{k}}}\, ,
\end{equation}
and
\begin{equation}
K_{MN}(u) \equiv \frac{1}{2\pi i} \frac{d}{du} \log{S_{MN} (u)} =\mathbf{K}_{M+N}+\mathbf{K}_{|M-N|} + 2 \sum_{j=1}^{\min{(M,N)}-1}\mathbf{K}_{|M-N|+2j}\, .
\end{equation}
The kernels involving scattering of the negative parity particles can be defined in terms of the above via \eqref{eq:negparSM} and \eqref{eq:negparSMN}; note that \emph{we do not define them independently}. For values of $M$ greater than $2k$ we would have to take into account the $2k$ periodicity of the S-matrix and take $M \mod 2k$ in the above formulae, but we will not encounter this case. In order to simplify the TBA equations we would like to understand the integral identities satisfied by these kernels.

For real rapidities the kernel $K_M$ is real and positive and has the following Fourier transform
\begin{equation}
\hat{K}_M(\omega) \equiv \int_{-\infty}^{\infty} du e^{i g \omega u} K_M(u)= \frac{\sinh{(k-M)\omega }}{\sinh{k\omega}}\, ,
\end{equation}
where we defined the Fourier transform with an unconventional factor of $g$. The Fourier transform of the kernel $K_{MN}$ is then
\begin{equation}
\hat{K}_{MN}(\omega)= \frac{\coth{\omega}}{\sinh{k\omega}} \left(\cosh{(|M-N|-k)\omega}-\cosh{(M+N-k)\omega}\right) - \delta_{MN} \, .
\end{equation}
These kernels satisfy the following properties
\begin{align}
& \hat{K}_N (\delta_{N,M} - I_{NM} \hat{s}) = \hat s \delta_{M,1} \, , \label{eq:simpKM} \, ,\\
& \hat{K}_{ML}(\delta_{L,N} - I_{LN} \hat{s}) = \hat{s}I_{M,N} \, .\label{eq:simpKMN}\, ,
\end{align}
where $\hat{s}(\omega)=\frac{1}{2\cosh{\omega}}$ is the Fourier transform of $s$
\begin{equation}
s(u)=\frac{g}{2\pi}\int_{-\infty}^{\infty} d\omega \frac{e^{-i g \omega u}}{2\cosh{\omega}}=\frac{g}{4 \cosh\frac{g \pi u}{2}}\, ,
\end{equation}
and $I_{MN} = \delta_{M,N-1} + \delta_{M,N+1}$ is the incidence matrix, to be appropriately interpreted on the boundary as $I_{N,k-1} = \delta_{N,k-2}$\footnote{Note that we have $K_{k}=K_{Mk}=0$.}. Since we will need it again, let us define
\begin{equation}
\label{eq:Kp1inv}
(K+1)^{-1}  \equiv 1 - I \star s\, .
\end{equation}

\subsection*{$\boldsymbol K^{yQ}$, $\boldsymbol K^{Qy}$, $\boldsymbol  K^{QM}_{xv}$ and $\boldsymbol K^{MQ}_{vwx}$}

In line with the conventions in the undeformed case, in the main text we defined
\begin{align}
K^{QM}_{xv}(u,v) & \equiv \frac{1}{2\pi i} \frac{d}{du} \log S^{QM}_{xv} (u,v)\, ,\\
K^{MQ}_{vwx}(u,v) & \equiv - \frac{1}{2\pi i} \frac{d}{du} \log S^{QM}_{xv}(v,u)\, ,\\
K^{Qy}_{\beta}(u,v) & \equiv \frac{1}{2\pi i} \frac{d}{du} \log S^{Qy}_{\beta} (u,v) \, ,\\
K^{yQ}_{\beta}(u,v) & \equiv \beta \frac{1}{2\pi i} \frac{d}{du} \log S^{Qy}_{\beta} (v,u) \, .
\end{align}
The scattering kernels for negative parity strings can be defined in terms of the above kernels via \eqref{eq:negparSQMxv}. We will mainly work with linear combinations of the last two kernels, namely
\begin{align}
K^{Qy}_{-}(u,v) - K^{Qy}_{+}(u,v) & \equiv K_{Qy}(u,v) = K(u+iQ/g,v) - K(u-iQ/g,v)\, , \\
K^{Qy}_{-}(u,v) + K^{Qy}_{+}(u,v) & = K_Q(u,v)\, , \\
K^{yQ}_{-}(u,v) - K^{yQ}_{+}(u,v) & = K_Q(u,v)\, , \label{eq:KyQKQdef}\\
K^{yQ}_{-}(u,v) + K^{yQ}_{+}(u,v) & \equiv K_{yQ}(u,v) = K(u,v+iQ/g) - K(u,v-iQ/g)\, ,\label{eq:KyQKyQdef}
\end{align}
where
\begin{equation}
K(u,v) = \frac{1}{2\pi i} \frac{d}{du} \log \frac{x(u)-\frac{1}{x(v)}}{x(u)-x(v)} \, ,
\end{equation}
and $K_Q$ is defined in \eqref{eq:KM}. These kernels satisfy the following properties under application of the operator $(K+1)^{-1}$
\begin{align}
K^{QN}_{xv}(\delta_{N,M}-I_{NM}\star s) & = \delta_{Q-1,M} s + \delta_{M,1} K_{Qy}\, \hat{\star}\,  s  \, , \label{eq:simpKQMxv} \\
K^{MP}_{vwx}(\delta_{P,Q}-I_{PQ}\star s) & = \delta_{M+1,Q} s + \delta_{Q,1}\check{K}_M\,  \check{\star} \, s \, ,\label{eq:simpKMQvwx} \\
K_{yP}(\delta_{P,Q}-I_{PQ}\star s)& = \delta_{Q,1}(2\check{K}\, \check{\star}\,  s + s)\, ,\label{eq:simpKyQ}  \\
K_{P}(\delta_{P,Q}-I_{PQ}\star s)& = \delta_{Q,1} s\, ,\label{eq:simpKQ}
\end{align}
where we note again that for $M$ (not $Q$-particle) type indices $I_{N,k-1} = \delta_{N,k-2}$ while for $Q$ type indices this incidence matrix is fine for $Q=k-1$, but at $Q=k$ we need slightly different identities. Namely
\begin{align}
K^{MP}_{vwx}(\delta_{P,k}-2\delta_{P,k-1}\star s) & = \delta_{M,k-1} s - K_{M,k-1} \star s \, ,\label{eq:simpKMQvwxbndry} \\
K_{yP}(\delta_{P,k}-2\delta_{P,k-1}\star s)& = 0\, .\label{eq:simpKyQbndry}
\end{align}
The kernels entering in the above identities are
\begin{equation}
\check{K}(u,v) = \theta(|u|-u_b) \frac{1}{2\pi i} \frac{d}{du} \log \frac{x(u)-\frac{1}{x_s(v)}}{x(u)-x_s(v)} \, ,
\end{equation}
and
\begin{equation}
\check{K}_M(u,v) \equiv \check{K}(u+i M/g,v)+ \check{K}(u-i M/g,v)\, .
\end{equation}
Let us also define
\bea
\label{eq:checkE}
\check{\cal E}=\log\frac{x_s+\xi}{\frac{1}{x_s}+\xi}\, .
\eea
In the above formulas $x_s$ is given by
\begin{equation}
x_s(u)=\frac{ \left(e^{\frac{\pi g u}{2 k}} \sinh\frac{\pi  g u}{2 k}-g^2 \sin^2\frac{\pi}{k}\right) \left(1+\sqrt{1-\frac{e^{\frac{\pi g u}{k}} g^2 \sin^2\frac{\pi }{k}  \left(1+g^2\sin^2\frac{\pi}{k}\right)}{\left(e^{\frac{\pi  g u}{2k}}\sinh \frac{\pi g u}{2 k}-g^2 \sin^2\frac{\pi }{k}\right)^2}}\right)}{\sqrt{\left(1+g^2 \sin^2\frac{\pi }{k}\right)g^2\sin^2\frac{\pi }{k}}}\, .
\end{equation}

\subsection*{$\boldsymbol K_{\mathfrak{sl}(2)}^{QP}$}

The main kernel $K_{\mathfrak{sl}(2)}$ has the following structure
\begin{equation}
K_{\mathfrak{sl}(2)}^{QP}(u,v) = - K_{QP}(u-v) - 2 K^\Sigma_{QP}(u,v)\,.
\end{equation}
Proceeding by analogy to the undeformed case \cite{AF09b}, we \emph{assume} that the following identity \cite{AF09d} is valid
\begin{equation}
K^\Sigma_{QP}\star (K + 1)^{-1}_{PR}=\delta_{1,R}\check{K}_{Q}^\Sigma \check{\star} s \,,
\end{equation}
where the kernel $\check{K}_{Q'}^\Sigma(u,v)$ vanishes for $|v|<u_b$, and $(K+1)^{-1}$ is defined in (\ref{eq:Kp1inv}).

\section{The simplified TBA equation for $Y_k$}

\label{app:sTBAYk}

To derive the simplified TBA equation for $Y_k$ we act with $\delta_{Q,k} - 2 \delta_{Q,k-1} s$ on the canonical TBA equations for $Q$-particles and make use of the special identities \eqref{eq:simpKMQvwxbndry} and \eqref{eq:simpKyQbndry} and definitions \eqref{eq:KyQKQdef} and \eqref{eq:KyQKyQdef} to get
\begin{align}
\label{eq:sTBAQktemp}
\log Y_k =\, & 2\log{Y_{k-1}}\star s + \log\left(1+\tfrac{1}{Y_{k-1|vw}^{(\alpha)}}\right)^2 Y_{k-1|vw}^{(\alpha)} \star s \nonumber \\
& - \log\left(1 + \tfrac{1}{Y^{(\alpha)}_{M|vw}}\right) \star K_{M,k-1}\star s - \log\left(1 + Y^{(\alpha)}_{k-1|vw}\right) \star K_{k-1,k-1}\star s\\
& -\log \frac{1-\tfrac{1}{Y_-^{(\alpha)}}}{1-\frac{1}{Y_+^{(\alpha)}}}\,
 \hat{\star}\,  K_{k-1} \star s  +\log \left(1+Y_P \right)\star (K_{\mathfrak{sl}(2)}^{Pk}- 2 K_{\mathfrak{sl}(2)}^{Pk-1}\star s)  \nonumber\, .
\end{align}
Note that the energy $\tilde{\cal E}_Q$ is annihilated by the operator $\delta_{Q,k} - 2 \delta_{Q,k-1} s\,$ because $\mathcal{E}_{k+1} = \mathcal{E}_{k-1}$ and it satisfies the discrete Laplace equation by construction. Examining eq.(\ref{eq:sTBAQktemp}) we immediately recognize a large part of the canonical TBA equation for $Y_{k-1|vw}$, albeit integrated with $s$ and without the $Y_Q$ term, for each
$\alpha$. Rewriting then gives
\begin{align}
\label{eq:sTBAQktemp2}
\log Y_k =\, & 2\log{Y_{k-1}}\star s + \log\left(1+\tfrac{1}{Y_{k-1|vw}^{(\alpha)}}\right)^2\star s \\
& \quad + \log \left(1+Y_P \right)\star (K_{\mathfrak{sl}(2)}^{Pk}- 2 K_{\mathfrak{sl}(2)}^{Pk-1}\star s - 2 K^{Pk-1}_{xv} \star s)  \nonumber\, .
\end{align}
The kernel $K_{\mathfrak{sl}(2)}$ satisfies the discrete Laplace equation, meaning that if we introduce the would-be fused $k+1$st kernel  we should have
\begin{equation}
\nonumber
K_{\mathfrak{sl}(2)}^{Pk}- 2 K_{\mathfrak{sl}(2)}^{Pk-1}\star s = K_{\mathfrak{sl}(2)}^{PQ}(K+1)^{-1}_{Q,k} - K_{\mathfrak{sl}(2)}^{Pk-1}\star s + K_{\mathfrak{sl}(2)}^{Pk+1}\star s = (K_{\mathfrak{sl}(2)}^{Pk+1}-K_{\mathfrak{sl}(2)}^{Pk-1}-\delta_{Pk-1})\star s\,.
\end{equation}
In this way we reduce the equation for $Y_k$ to the following form
\begin{align}
\label{eq:sTBAQktemp2int}
\log Y_k =\, & 2\log{Y_{k-1}}\star s + \log\left(1+\tfrac{1}{Y_{k-1|vw}^{(\alpha)}}\right)^2\star s -\log(1+Y_{k-1})\star s\\
&\quad \quad + \log \left(1+Y_P \right)\star (K_{\mathfrak{sl}(2)}^{Pk+1}-K_{\mathfrak{sl}(2)}^{Pk-1} - 2 K^{Pk-1}_{xv} ) \star s \nonumber\, .
\end{align}
Finally, the combination of the kernels in eq.(\ref{eq:sTBAQktemp2int}) can be evaluated as follows. Let us consider the corresponding combination of S-matrices
\bea
\nonumber
\frac{S_{\sls(2)}^{Pk+1}(u,v)}{S_{\sls(2)}^{Pk-1}(u,v)}\Big(S_{xv}^{Pk-1}(u,v)\Big)^{-2}=S_{\mathfrak{sl}(2)}^{P1}\Big(u,v+\tfrac{ ik}{g}\Big)S_{\mathfrak{sl}(2)}^{P1}\Big(u,v-\tfrac{ik}{g}\Big)\left(S^{Pk-1}_{xv}(u,v)\right)^{-2} .\eea
Here we have used the fact that $S^{PQ}_{\sls(2)}$ is obtained by fusing the S-matrices $S_{\sls(2)}^{P1}$. Now
we immediately realize that the first two terms are related by crossing. Using fusion in the first argument
it is not difficult to show that
\bea
&&S_{\mathfrak{sl}(2)}^{P1}\Big(u,v+\tfrac{ ik}{g}\Big)S_{\mathfrak{sl}(2)}^{P1}\Big(u,v-\tfrac{ik}{g}\Big)=\\
\nonumber
&&\hspace{1cm}=\left(\frac{x_1^+}{x_1^-}\right)^2\Big(\sigma^{P1}\Big(u,v+\tfrac{ ik}{g}\Big)\sigma^{P1}\Big(u,v-\tfrac{ ik}{g}\Big)\Big)^{-2}=
q^{2P}h^{P1}\Big(u,v+\tfrac{ ik}{g}\Big)^{-2}\, ,
\eea
where $\sigma^{P1}$ is the dressing factor and we have used the crossing equation describing scattering of a $P$-particle bound state
with the fundamental particle. Here
\begin{equation}
h^{PQ}(u,v) = \frac{x(u-i P/g) - x(v+i Q/g)}{x(u-i P/g) - x(v-i Q/g)}\frac{1- \frac{1}{x(u+i P/g)x(v+i Q/g)}}{1- \frac{1}{x(u+i P/g)x(v-i Q/g)}}
\prod_{j=1}^{P-1} S_{Q-P+2j}(u-v)\,,  \nonumber
\end{equation}
as defined in appendix A.
Thus, we are left to compute the quantity
\begin{equation}
\label{eq:crossingtoSQMxv}
q^{-P}h^{P1}(u,v+ik/g) S^{Pk-1}_{xv}(u,v) \, .
\end{equation}
To proceed we will denote $x(u\pm iP/g)$ and $x(v\pm i(k-1)/g)$ as $x^\pm_1$ and $x^\pm_2$. Note that $x(v\pm i/g + ik/g) = x(v\mp i(k-1)/g)^{\mp1} = (x_2^\mp)^{\mp1}$. Working out the product (see also eq. \eqref{eq:SQMxv}), immediately removing the shift on $v$  in $h$ as just indicated, we get
\begin{align}
& q^{-P}h^{P1}(u,v+ik/g) S^{Pk-1}_{xv}(u,v)  = \nonumber \\
& \hspace{40pt} =
\frac{x^-_1 - \tfrac{1}{x^-_2}}{x^-_1 - x^+_2}\frac{1- \frac{x^-_2}{x^+_1}}{1- \frac{1}{x^+_1 x^+_2}} \frac{x^-_1-x^+_2 }{x^+_1-x^+_2 }\frac{x^-_1-x^-_2 }{x^+_1-x^-_2}\frac{x^+_1}{x^-_1} \nonumber \\
& \hspace{60pt} \times \prod_{j=1}^{P-1} S_{1-P+2j}(u,v+i k/g) \prod_{m=1}^{k-2} S_{P+1-k+2m}(u,v) \nonumber\\
& \hspace{40pt} =  \frac{1 - \tfrac{1}{x^-_1x^-_2}}{1- \frac{1}{x^+_1 x^+_2}}\frac{x^-_1-x^-_2 }{x^+_1-x^+_2} \prod_{j=1}^{P-1} S_{1-P+2j}(u,v+i k/g) \prod_{m=1}^{k-2} S_{P+1-k+2m}(u,v) \nonumber \\
& \hspace{40pt} =  -q^{-(P-1)} S_{P+1-k} (u,v)\prod_{j=1}^{P-1} S_{1-P+2j}(u,v+i k/g) \prod_{m=1}^{k-2} S_{P+1-k+2m}(u,v) \nonumber \\
& \hspace{40pt} =  (-1)^{P}q^{-(P-1)}\, .
\end{align}
In the third equality we used the general identity
\begin{equation}
\frac{1 - \tfrac{1}{x^-_i x^-_j}}{1- \frac{1}{x^+_i x^+_j}}\frac{x^-_i - x^-_j}{x^+_i - x^+_j} = q^{-(P+Q)}S_{P-Q} \, ,
\end{equation}
where $P$ and $Q$ refer to the bound state numbers of $i$ and $j$ respectively. In the fourth we used
\begin{align}
& \prod_{j=1}^{M-1} S_{1-P+2j} = S_{P-1}\, , \\
& S_P(u\pm i k/g) = -S_{P-k}(u)\, \end{align}
and
\begin{equation}
\prod_{j=1}^k S_{P+2j} = (-1)^{P+k+1}\, .
\end{equation}
Thus, we have proved that the combination of the S-matrices we are interested in is constant and therefore the corresponding combination of the kernels vanishes.
With this result eq. \eqref{eq:sTBAQktemp2} simplifies further, now properly deserving the name, giving
\begin{equation}
\log Y_k =\, 2\log{Y_{k-1}}\star s -\log(1+Y_{k-1})\star s + \log\left(1+\tfrac{1}{Y_{k-1|vw}^{(\alpha)}}\right)^2 \star s \, .
\end{equation}
Note that the contribution of $Y_{k-1|vw}$ has doubled up here, exactly as it did in the equations for $vw$ strings, though admittedly in a less obvious fashion.

\section{Representations of $\su_q(2|2)$ at roots of unity}
\label{app:reptheory}

Here we discuss some elements of representation theory of the centrally extended $\su_q(2|2)\equiv U_q(\mathfrak {su}(2|2))$ at even roots of unity. We follow the construction of \cite{deLeeuw:2011jr}.

%%%%%%%%%%%%%%%%%%%%%%%%%%%%%%%%%%%%%%%%%%%%%%%%%%%%%%%%%%%%%%%%%%%
\subsection*{Quantum oscillators}
Let us first introduce quantum oscillators. We will use those to build our representations.

The $q$-oscillator ($q$-Heisenberg-Weyl algebra) $U_q(\alg{h}_4)$ is the associative unital algebra consisting of the generators $\{\ad,\a,w,w^{-1}\}$ that satisfy the following relations
\begin{align}\label{eqn;DefRelQosc}
w\,\ad &= q\,\ad w, && q w\,\a = \a\,w,\\
w w^{-1} &=  w^{-1} w =1, &&\a\,\ad - q\,\ad \a = w^{-1}.\nonumber
\end{align}
From the defining relations one can see that the element $w^{-1}(\ad \a -\frac{w-w^{-1}}{q-q^{-1}})$ is central. As such, we will set it to zero in the remainder. The defining relations then imply that
\begin{align}\label{eqn;DefRelQosc2}
&\ad \a = \frac{w-w^{-1}}{q-q^{-1}}, &&\a\,\ad = \frac{qw-q^{-1}w^{-1}}{q-q^{-1}}.
\end{align}
Since we are interested in representations of supergroups, we will also need to consider the fermionic version of the $q$-oscillator. The above notion is extended to include fermionic operators by adjusting the defining relations in the following way (we keep the same notation for bosonic and fermionic $\a,\,\ad$ for now)
\begin{align}
w\,\ad &= q\,\ad w, && q w\,\a  = \a\,w,\\
w w^{-1} &=  w^{-1} w =1, &&\a\,\ad + q\,\ad \a = w.\nonumber
\end{align}
In this case, the central element is $w(\ad \a -\frac{w-w^{-1}}{q-q^{-1}})$. Again we set this element to zero, resulting in the following identities
\begin{align}\label{eqn;oscferm}
&\ad \a = \frac{w-w^{-1}}{q-q^{-1}}, &&\a\,\ad = \frac{qw^{-1} - q^{-1}w}{q-q^{-1}}.
\end{align}
Of course in the fermionic case the operators $\a,\ad$ square to zero. From \eqref{eqn;oscferm} it then follows that this only is consistent if $w^2 = 1,q^2$. Below we will identify $w\equiv q^N$, where $N=0,1$ is the number of fermions making it indeed compatible.

\subsection*{Fock space}

Let us first build the Fock representation of $U_q(\alg{h}_4)$. For this purpose consider a vacuum state $|0\rangle$ such that
\begin{align}
&\a|0\rangle = 0,  && w|0\rangle = |0\rangle,
\end{align}
then the Fock vector space $\cal{F}$ generated by the states of the form
\begin{align}
|n\rangle = (\ad)^n|0\rangle\,,
\end{align}
is an irreducible module of $U_q(\alg{h}_4)$. Let us first consider the bosonic $q$-oscillators. With the help of the defining relations \eqref{eqn;DefRelQosc} and \eqref{eqn;DefRelQosc2} one finds that the action of the oscillator algebra generators on this module is
\begin{align}\label{normalFock}
&\ad|n\rangle = |n+1\rangle, && \a|n\rangle = [n]_q|n-1\rangle,&& w|n\rangle = q^n|n\rangle.
\end{align}
This makes it natural to identify $w \equiv q^N$, where $N$ is understood as a number operator. Analogously, fermionic generators are found to act as
\begin{align}
\label{eq:fermgenact}
&\ad|n\rangle = |n+1\rangle, && \a|n\rangle = [2-n]_q|n-1\rangle,&& w|n\rangle = q^n|n\rangle.
\end{align}
However, due to the fermionic nature, $n$ can only take the values $0$ and $1$ and thus in \eqref{eq:fermgenact} we can identify $[n-2]_q$ with $[n]_q$.

\subsection*{Fock space at roots of unity}

Let $q=e^{\frac{i\pi}{k}}$ \footnote{All the results are straightforwardly generalized to the case where $q=e^{\frac{2\pi i}{k}}$, with $k$ an odd integer.}, then it is easy to see that $\a^{2k},(\ad)^{2k},w^{2k},w^{-2k}$ are central elements. Of course, for fermionic oscillators, this is trivial and holds always. Because of this, the Fock space is reducible and has a finite number of weight spaces. Indeed, since
\begin{align}
&\a|n\rangle = [n]_q |n-1\rangle =0,  && w|n\rangle = q^n|n\rangle
\end{align}
we see that only the weights $\{1,\ldots,q^{2k-1}\}$ are present. In other words, we can restrict our states in the Fock space to $n=0,\ldots,k-1$ only and doing so gives a irreducible module again.

This allows us to introduce the class of so-called semi-cyclic representations of the quantum oscillator algebra. Let $(\lambda,\mu) \in (\mathbb{C}\times\{0\}) \cup (\{0\}\times \mathbb{C} )$, then we define
\begin{align}
&\a |n\rangle = [n]_q |n-1\rangle, &&\mathrm{for}~n=1,\ldots k-1 && \a |0\rangle = \mu |k-1\rangle \\
&\ad |n\rangle = |n+1\rangle,&&\mathrm{for}~ n=0,\ldots k-2 && \ad |k-1\rangle = \nu |0\rangle \\
&w|n\rangle = q^n|n\rangle, &&\mathrm{for}~ n=0,\ldots k-1.
\end{align}
The idea behind this is easy; we identify $|k\rangle \sim |0\rangle$. Putting $\mu=\nu=0$ simply returns us to the normal restricted Fock space.

\subsection*{Representations of $U_q(\alg{sl}(2))$}

We will now construct representations of $U_q(\alg{sl}(2))$. Consider two copies of bosonic $q$-oscillators $\a_i,\ad_i,w_i=q^{N_i}$ which mutually commute. Then the Fock space is naturally spanned by vectors of the form
\begin{align}
|m,n\rangle = (\ad_1)^m (\ad_2)^n |0\rangle.
\end{align}
Define the subspace $\mathcal{F}_M = \mathrm{span}\{\,|m,M-m\rangle\;|\; m=0,\ldots,M\,\}$. We will refer to $M$ as the total number of particles and $\dim\mathcal{F}_M =M+1$. It is easy to see that under the identification
\begin{align}
&E = \ad_2 \a_1, && F = \ad_1 \a_2, && H = N_2-N_1,
\end{align}
we obtain a presentation of $U_q(\alg{sl}(2))$. Then depending on the representation of the q-oscillators on the Fock space (normal or semi-cyclic) this will define an explicit representation.

\subsection*{Irreducible representations}

\paragraph{Generic $q$.} Let $q$ be a complex number that is not a root of unity. In this case the q-oscillators have the standard representation on the Fock space \eqref{normalFock}. Moreover, the subspace $\mathcal{F}_M = \mathrm{span}\{\,|m,M-m\rangle\;|\; m=0,\ldots,M\,\}$ is an irreducible $\mathcal{U}_q(\alg{sl}(2))$-representation of dimension $M+1$. Explicitly we have
\begin{align}\label{genericqoscSU2}
&E|n,M-n\rangle = [n]_q |n-1,M-n+1\rangle, && F|n,M-n\rangle = [M-n]_q |n+1,M-n-1\rangle, \nonumber\\
& H|n,M-n\rangle = (M-2n)|n,M-n\rangle.
\end{align}
It is easy to check that indeed satisfies all the defining relations of $U_q(\alg{sl}(2))$.

\paragraph{Root of unity $q=e^{\frac{i\pi}{k}}$.} Now we have to use the semi-cyclic presentation of the algebra. Let $\mu_1,\nu_1$ and $\mu_2,\nu_2$ be the parameters describing the representation of the first and second set of oscillators respectively. Notice that the only when the total number of particles is $k-1$ there is a possibility of seeing the semi-cyclicity. Thus, when restricting to states with total $M<k-1$ number of particles, representations coincide trivially with those for generic $q$ given in \eqref{genericqoscSU2}.

However, when $M=k-1$ we find a one-parameter family of representations. This is obtained by taking $\mu_1=\nu_2=0$. Then apart from the ordinary relations for $n=1,\ldots,k-1$ \eqref{genericqoscSU2} we find
\begin{align}
& E|0,k-1\rangle = \nu_1\mu_2 |k-1,0\rangle.
\end{align}
Alternatively, we can set $\nu_1=\mu_2=0$ and get a special relation for $F$ acting on $|0,k-1\rangle$.

The special point where $\mu_i=\nu_i=0$ is most natural in the sense that the representations simply correspond to the restriction of the normal q-oscillator representations like for $M<k-1$. All these representations are irreducible.

\subsection*{Reducible representations at roots of unity}

By construction, any representation based on the normal oscillator representation on $\mathcal{F}_M$ with $M>k-1$ for $q=e^{\frac{i\pi}{k}}$ is reducible. At $M=k$, the representation contains two singlets $|0,k\rangle,|k,0\rangle$. In general, the representations will be indecomposable. More precisely, let $M = n k+m$ with $0\leq m<k$ then the representation is reducible and contains
\begin{align}
\bigoplus_{i=0}^{n} \mathcal{F}_m \subset \mathcal{F}_M.
\end{align}
Only when $m=k-1$, the representation is decomposable. The above can be readily proven by using that the copies of $ \mathcal{F}_m$ are generated by $|a k,M -ak\rangle,|ak+1,M-1-ak\rangle,\ldots$ where $a=0,\ldots,n$.

\subsection*{Representations of centrally extended $\su_q(2|2)$.}

We will now construct the bound state representation for centrally extended $\su_q(2|2)$ in the $q$-oscillator language. We need to consider two copies of $U_q(\alg{sl}(2))$, a bosonic and a fermionic one. Thus we need four sets of $q$-oscillators $\a_i,\ad_i,w_i=q^{N_i}$, where the index $i=1,2$ denotes bosonic oscillators and $i=3,4$ -- fermionic ones. Using these we write
\begin{align}
&E_1 = \ad_2 \a_1, && F_1 = \ad_1 \a_2, && H_1 = N_2-N_1,\\
&E_2 = a~ \ad_4 \a_2 + b~ \ad_1 \a_3 && F_2 = c~ \ad_3 \a_1 + d~ \ad_2 \a_4 , && H_2 =-C +\frac{N_1+N_3-N_2-N_4}{2},\\
&E_3 = \ad_3 \a_4, && F_3 = \ad_4 \a_3, && H_3 = N_4-N_3,
\end{align}
where $C$ is central. Let $V=q^C$. We first consider the case where $q$ is not a root of unity. It is then straightforward to check that this set of generators forms a representation of $\su_q(2|2)$ on the Fock space when restricting to the subspace of total particle number $M$ upon setting
\begin{align}\label{eqn;CentralOsc}
  &ad = \frac{[C+{\textstyle \frac{M}{2}}]_q}{[M]_q}, && bc =\frac{[C-{\textstyle \frac{M}{2}}]_{q}}{[M]_q}, && ab =g\alpha \frac{1-U^2V^2}{[M]_q}, && cd = \frac{g}{\alpha}\frac{V^{-2}-U^{-2}}{[M]_q}.
\end{align}
The dimension of this representation is $4M$.

Let us spell out the representation more explicitly.  The bound state representation is defined on vectors
\begin{align}
|m,n,k,l\rangle=(\mathsf{a}_{3}^{\dag})^{m}(\mathsf{a}_{4}^{\dag})^{n}(\mathsf{a}_{1}^{\dag})^{k}(\mathsf{a}_{2}^{\dag})^{l}\,|0\rangle,
\end{align}
where the indices $1$, $2$ denote bosonic oscillators and $3$, $4$ - fermionic and the total number of excitations $k+l+m+n=M$ is the bound state number.

The triples corresponding to the bosonic and fermionic $\mathfrak{sl}_{q}(2)$ in this representation are given by
\begin{align}
 & H_{1}|m,n,k,l\rangle=(l-k)|m,n,k,l\rangle, &  & H_{3}|m,n,k,l\rangle=(n-m)|m,n,k,l\rangle,\el
 & E_{1}|m,n,k,l\rangle=[k]_{q}\,|m,n,k-1,l+1\rangle, &  & E_{3}|m,n,k,l\rangle=|m+1,n-1,k,l\rangle,\el
 & F_{1}|m,n,k,l\rangle=[l]_{q}\,|m,n,k+1,l-1\rangle, &  & F_{3}|m,n,k,l\rangle=|m-1,n+1,k,l\rangle.
\end{align}
The supercharges act on basis states as
\begin{align}
H_{2}|m,n,k,l\rangle= & ~-\left\{ C-\frac{k-l+m-n}{2}\right\} |m,n,k,l\rangle,\el
E_{2}|m,n,k,l\rangle= & ~a~(-1)^{m}[l]_{q}\,|m,n+1,k,l-1\rangle+b~|m-1,n,k+1,l\rangle,\el
F_{2}|m,n,k,l\rangle= & ~c~[k]_{q}\,|m+1,n,k-1,l\rangle+d~(-1)^{m}\,|m,n-1,k,l+1\rangle.
\end{align}
For generic $q$ these representations are irreducible.

\paragraph{Root of unity.} At the root of unity, we can once again use the semi-cyclic representation of the quantum oscillators. This will again yield a special representation at $M=k-1$, which upon setting $\mu_i=\nu_i=0$ reduces to the one coming from the normal representation on the Fock space.

\paragraph{Reducible representations at roots of unity.} In the case where $\mu_i=\nu_i=0$  one finds reducible representations for $M\geq k$. Since this representation contains a bosonic $\alg{su}(2)$, the discussion of the reducibility of those representations also applies here. In fact, we immediately see that we get indecomposable representations again. Let $V_M$ be the $M$-particle bound state representation and let $M = n k+m$ with $0\leq m<k$. Then, it is not difficult to see that we again find
\begin{align}
\bigoplus_{i=0}^{n} V_m \subset V_M.
\end{align}
The central charges obviously are the same in all components and consequently each of the sub representations share the same central charges $U,V$.

%%%%%%%%%%%%%%%%%%%%%%%%%%%%%%%%%%%

\end{document}